\documentclass[trackchanges,twocolumn]{aastex7}
\usepackage{subcaption} 

\newcommand{\meoh}{CH$_{3}$OH}
\newcommand{\mecn}{CH$_{3}$CN}
\newcommand{\hcho}{H$_{2}$CO}

\newcommand{\ppn}{HC$_{3}$N}
\newcommand{\hac}{CH$_{3}$CHO}

\newcommand{\mf}{CH$_{3}$OCHO}

\newcommand{\hcop}{HCO$^+$}
\newcommand{\htcop}{H$^{13}$CO$^+$}

\begin{document}

\title{The ALMA-ATOMS-QUARKS survey: Resolving a chemically rich massive protostellar outflow}

\correspondingauthor{Jiahang Zou; Tie Liu}
\email{zojh1000@163.com; liutie@shao.ac.cn}

\author[orcid=0009-0000-9090-9960]{Jia-Hang Zou}
\affiliation{School of Physics and Astronomy, Yunnan University, Kunming 650091, People's Republic of China}
\affiliation{Shanghai Astronomical Observatory, Chinese Academy of Sciences, 80 Nandan Road, Shanghai 200030, People's Republic of China}
\email{zojh1000@163.com}

\author[orcid=0000-0002-5286-2564]{Tie Liu} 
\affiliation{Shanghai Astronomical Observatory, Chinese Academy of Sciences, 80 Nandan Road, Shanghai 200030, People's Republic of China}
\email{liutie@shao.ac.cn}

\author[orcid=0000-0001-5950-1932]{Fengwei Xu} 
\affiliation{Max Planck Institute for Astronomy, Königstuhl 17, 69117 Heidelberg, Germany}
\affiliation{Kavli Institute for Astronomy and Astrophysics, Peking University, 5 Yiheyuan Road, Haidian District, Beijing 100871, People's Republic of China}
\email{fengweilookuper@gmail.com}

\author[orcid=0000-0002-4154-4309]{Xindi Tang}
\affiliation{Xinjiang Astronomical Observatory, Chinese Academy of Sciences, 830011 Urumqi, People's Republic of China}
\email{tangxindi@xao.ac.cn}

\author[orcid=0009-0000-5764-8527]{Dezhao Meng}
\affiliation{Xinjiang Astronomical Observatory, Chinese Academy of Sciences, 830011 Urumqi, People's Republic of China}
\affiliation{University of Chinese Academy of Sciences, Beijing 100080, People’s Republic of China}
\affiliation{Shanghai Astronomical Observatory, Chinese Academy of Sciences, 80 Nandan Road, Shanghai 200030, People's Republic of China}
\email{mengdezhao@xao.ac.cn}

\author[orcid=0000-0001-7817-1975]{Yan-Kun Zhang}
\affiliation{Shanghai Astronomical Observatory, Chinese Academy of Sciences, 80 Nandan Road, Shanghai 200030, People's Republic of China}
\email{zhangyankun@shao.ac.cn}


\author[0000-0003-4546-2623]{Aiyuan Yang}
\affiliation{National Astronomical Observatories, Chinese Academy of Sciences, Beijing 100101, People's Republic of China}
\affiliation{Key Laboratory of Radio Astronomy and Technology, Chinese Academy of Sciences, A20 Datun Road, Chaoyang District, Beijing, 100101, People's Republic of China}
\email{yangay@bao.ac.cn}

\author[]{Tapas Baug}
\affiliation{S. N. Bose National Centre for Basic Sciences, Block-JD, Sector-III, Salt Lake City, Kolkata 700106, India}
\email{tapas.polo@gmail.com}

\author[0000-0002-4154-4309]{Chang-Won Lee}
\affiliation{University of Science and Technology, Korea (UST), 217 Gajeong-ro, Yuseong-gu, Daejeon 34113, Republic of Korea}
\affiliation{Korea Astronomy and Space Science Institute, 776 Daedeokdaero, Yuseong-gu, Daejeon 34055, Republic of Korea}
\email{cwl@kasi.re.kr}

\author[0000-0002-5310-4212]{L. Viktor T\'oth} 
\affiliation{Institute of Physics and Astronomy, E\"otv\"os Lor\`and University, P\'azm\'any P\'eter s\'et\'any 1/A, H-1117 Budapest, Hungary}
\affiliation{Faculty of Science and Technology, University of Debrecen, H-4032 Debrecen, Hungary}
\email{toth.laszlo.viktor@ttk.elte.hu}

\author[orcid=0009-0003-6633-525X]{Ariful Hoque}
\affiliation{S. N. Bose National Centre for Basic Sciences, Block-JD, Sector-III, Salt Lake City, Kolkata 700106, India}
\email{arifulh882@gmail.com}

\author[orcid=0000-0002-8697-9808]{Sami Dib}
\affiliation{Max Planck Institute for Astronomy, K\"{o}nigstuhl 17, 69117, Heidelberg, Germany}
\email{sami.dib@gmail.com}

\author[orcid=0000-0002-8586-6721]{Pablo Garcia}
\affiliation{Chinese Academy of Sciences South America Center for Astronomy, National Astronomical Observatories, CAS, Beijing 100101, China}
\affiliation{Instituto de Astronom\'ia, Universidad Cat\'olica del Norte, Av. Angamos 0610, Antofagasta, Chile}
\email{astro.pablo.garcia@gmail.com}

\author[orcid=0000-0003-3343-9645]{Hong-Li Liu}
\affiliation{School of Physics and Astronomy, Yunnan University, Kunming 650091, People's Republic of China}
\email{hongliliu2012@gmail.com}

\author[]{Prasanta Gorai}
\affiliation{Rosseland Centre for Solar Physics, University of Oslo, PO Box 1029 Blindern, 0315 Oslo, Norway}
\affiliation{Institute of Theoretical Astrophysics, University of Oslo, PO Box 1029 Blindern, 0315 Oslo, Norway}
\email{prasanta.astro@gmail.com}

\author[0000-0001-7151-0882]{Swagat R. Das}
\affiliation{Departamento de Astronomia, Universidad de Chile, Las Condes, 7591245 Santiago, Chile}
\email{swagat@das.uchile.cl}

\author[]{Guido Garay}
\affiliation{Chinese Academy of Sciences South America Center for Astronomy, National Astronomical Observatories, CAS, Beijing 100101, China}
\affiliation{Departamento de Astronomia, Universidad de Chile, Las Condes, 7591245 Santiago, Chile}
\email{guido@das.uchile.cl}

\author[orcid=0000-0002-7125-7685]{Patricio Sanhueza}
\affiliation{Department of Astronomy, School of Science, The University of Tokyo, 7-3-1 Hongo, Bunkyo, Tokyo 113-0033, Japan}
\email{patosanhueza@gmail.com}

\author[orcid=0009-0009-8154-4205]{Li Chen}
\affiliation{School of Physics and Astronomy, Yunnan University, Kunming 650091, People's Republic of China}
\email{li.chen@mail.ynu.edu.cn}

\author[]{Di Li}
\affiliation{Department of Astronomy, Tsinghua University, Beijing 100084, People's Republic of China}
\email{dili@bao.ac.cn}

\author[0000-0001-7866-2686]{Jihye Hwang}
\affiliation{Institute for Advanced Study, Kyushu University, Japan}
\affiliation{Department of Earth and Planetary Sciences, Faculty of Science, Kyushu University, Nishi-ku, Fukuoka 819-0395, Japan}
\email{hwang.jihye.514@m.kyushu-u.ac.jp}

\author[]{Dongting Yang}
\affiliation{School of Physics and Astronomy, Yunnan University, Kunming 650091, People's Republic of China}
\email{dongting@mail.ynu.edu.cn}


\collaboration{all}{The ATOMS-QUARKS collaboration}

\begin{abstract}

We present a comprehensive study on the physical and chemical structures of a chemically rich bipolar outflow in a high-mass star forming region IRAS 16272$-$4837 (SDC335), utilizing high-resolution spectral line data at 1.3 mm and 3 mm dual-bands from the ALMA ATOMS and QUARKS surveys. The high-velocity jet is enveloped by a lower-velocity outflow cavity, containing bright knots that show enhanced molecular intensities and elevated excitation temperatures. Along the outflow, we have identified 35 transitions from 22 molecular species. By analyzing the spatial distribution and kinematics of these molecular lines, we find that the molecular inventory in the outflow is regulated by three processes: (i) direct entrainment from the natal molecular core by the outflow; (ii) shock-induced release of molecules or atoms from dust grains; and (iii) thermal desorption and gas-phase reactions driven by shock heating. These results confirm that outflows are not only dynamical structures but also active chemical factories, where entrainment, shocks, and thermal processing jointly enrich the molecular content. Our findings confirmed that outflow chemistry has multi-origin nature, and provide critical insights into chemical evolution during high-mass star formation.

\end{abstract}



\section{Introduction} 

Protostellar outflows represent one of the most prominent dynamical phenomena in the early stages of star formation \citep{2018ARA&A..56...41M,2020A&ARv..28....1L}. Outflows are spectroscopically identified by broad high-velocity wings in molecular lines, and spatially by bipolar red- and blue-shifted lobes centered on the protostar. Some outflows exhibit exceptionally rich molecular line emission, including SiO, CH$_3$OH, and various sulfur- or nitrogen-bearing molecules. The enhanced abundances of these molecules likely result from non-equilibrium processes induced by shocks, such as ice mantle desorption and grain sputtering \citep{1989ApJ...342..306H,1994ASPC...58..332F,1995Ap&SS.233..111D}. The chemically enriched regions in these ``chemically active outflows'' provide a unique laboratory for studying shock-driven chemistry in protostellar environments \citep{2011IAUS..280...88T,2017A&A...605L...3C,2024MNRAS.531.2653C,2025ApJ...980..263O}.

According to current observational and experimental studies, there are two primary mechanisms that can release complex organic molecule (COMs) from the ice mantles into the gas phase -- (i) thermal desorption and (ii) dust grain sputtering. In thermal desorption, the ices sublimate at temperatures around 100 K, fully releasing the trapped molecules into the gas phase \citep{2022A&A...665A..96B,2021A&A...655A..65T}. In such cases, the observed rotational temperatures are typically comparable to or higher than the sublimation threshold ($\sim$100 K). In the latter scenario, sputtering of dust grains requires shock events, which typically occur when the outflowing material collides with the surrounding dense medium at a high speed. These collisions can liberate molecules from the ice mantles into the gas phase \citep{1997A&A...322..296C}. These shock-induced collisions not only liberate molecules, but also produce line profiles with broad widths and/or high-velocity emission wings. In addition, shocks can heat the ambient material and induce thermal desorption, further enhancing molecular abundances \citep{2017ApJ...839...47M}. As the gas cools post-shock, the thermal signature may diminish \citep{2020A&A...634A..17J}. Therefore, it is essential to conduct detailed chemical analysis within molecular outflows to distinguish from these two different mechanisms. 

In recent years, extensive studies have focused on the chemical characteristics of outflows in low-mass protostars, often inferring excitation mechanisms from velocity-resolved molecular line profiles and revealing that their shock regions can also be chemically rich \citep{2008ApJ...681L..21A,2011ApJ...740...14O,2014MNRAS.445..151M,2020A&A...640A..75D,2020ApJ...896...37F,2022ApJ...933L..35F,2024ApJ...976...29H,2024A&A...690A.205C,2025MNRAS.539.2380B}. For instance, the ASAI survey toward L1157 \citep{2018MNRAS.477.4792L} shows that its two shocked knots (B1 and B2), located $\sim$1.5–2$'$ away from the driving source, exhibit a molecular inventory even richer than the protostar itself, including various COMs \citep{2017MNRAS.469L..73L}. In particular, B1 and B2 are among the richest Galactic sources of HNCO and NH$_2$CHO, two molecules of prebiotic interest \citep{2014MNRAS.445..151M}. However, previous studies of high-mass protostellar outflows have mostly relied on limited molecular species (e.g., CO and SiO) and often lack spatially resolved kinematic information, which hinders a comprehensive understanding of shock-induced chemistry within these sources \citep{2017MNRAS.467.2723P,2018ApJS..235....3Y,2020ApJ...901...31L,2021ApJ...909..177L,2022A&A...659A..23C,2022A&A...658A.160Y,2022ApJS..262...13R,2023ApJ...949...89X,2024A&A...681A.104B,2024A&A...683A.249F,2024AJ....167..285X,2024MNRAS.527.2110R,Hoque_2025}.

\begin{figure*}
\gridline{
  \fig{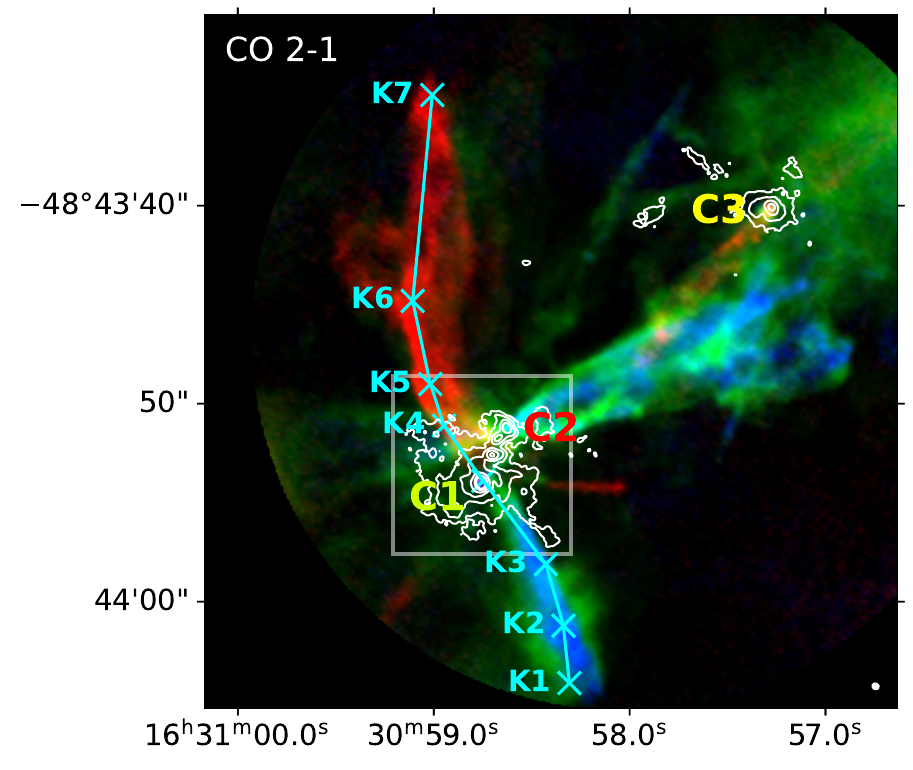}{0.43\textwidth}{(a)}
  \fig{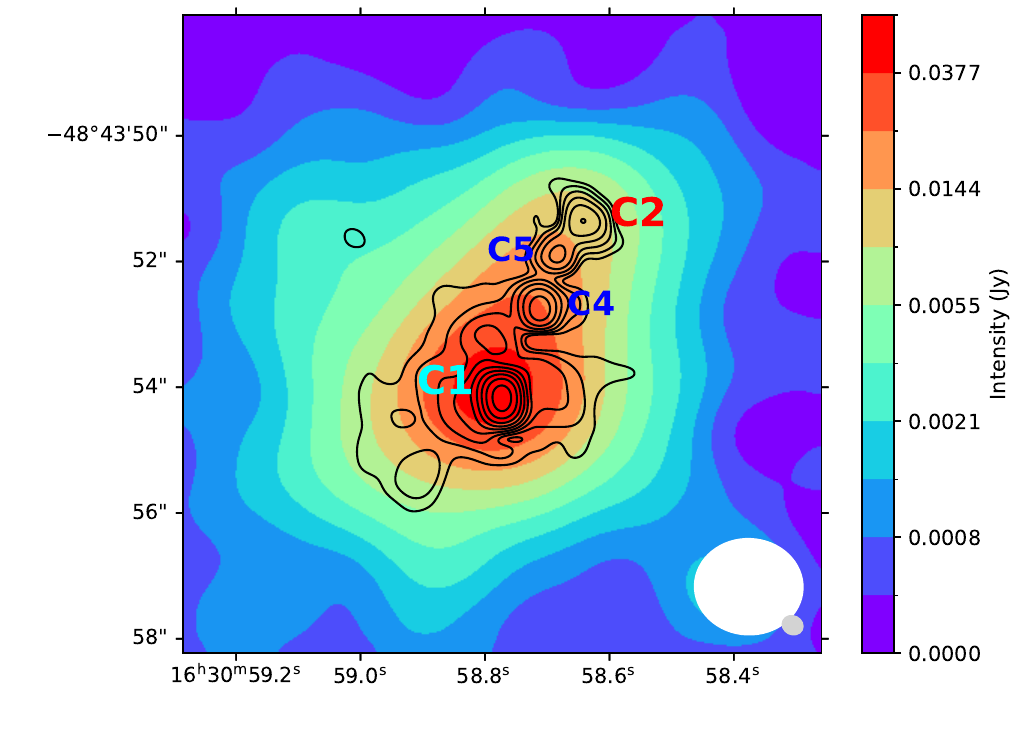}{0.53\textwidth}{(b)}
}
\caption{
\textbf{(a)} CO\,(2--1) RGB image constructed using the ACA+TM1+TM2 combined data. 
Blue- and red-shifted components are integrated over velocities of $-42$ to $-12$ and $+12$ to $+42~\mathrm{km\,s^{-1}}$ relative to the systemic velocity, while the green channel traces emission within $\pm12~\mathrm{km\,s^{-1}}$.
The cyan line marks the jet axis. Hot cores C1--C3 and several emission knots are visible. 
\textbf{(b)} Zoom-in of the central region outlined in \textbf{(a)}, with 3~mm continuum in colorscale and 1.3~mm contours highlighting compact sources. 
Continuum contours start at 21\,mJy\,beam$^{-1}$ and increase in steps of a factor of 1.618 (golden ratio). 
The beams of 3~mm (white) and 1.3~mm (grey) are shown in the lower-right corner.}
\label{fig:cont}
\end{figure*}

To overcome these limitations, we resolve a chemically rich massive protostellar outflow in the high-mass star-forming region IRDC SDC335.579$-$0.292 (IRAS 16272$-$4837, also G335.59-0.29,  hereafter SDC335), which is located at a distance of 3.54 kpc. This clump contains a total gas mass of $\sim$3700~M$\odot$ \citep{2008AJ....136.2391C,2021MNRAS.508.2964A}.
Within the central region corresponding to the area shown in Figure~\ref{fig:cont}~(b), which encompasses the massive core SDC335-C1, the mass is $\sim$400~M$\odot$ \citep{2023MNRAS.520.3259X}, exhibits high density, and shows chemical richness in COMs such as CH$_3$CN, CH$_3$OH, CH$_3$OCHO, and C$_2$H$_5$CN \citep[]{2022MNRAS.511.3463Q,2025A&A...697A.190L,Xu2024Assemble1}. It is also associated with a clearly defined bipolar high-velocity outflow \citep{2021A&A...645A.142A}, making it an ideal observational platform for exploring non-thermal chemical evolution in high-mass star formation.

In this letter, we report the detection of chemically enriched regions associated with jet-driven shocks in the bipolar molecular outflow of the high-mass protostellar core SDC335 C1, utilizing the wide-band molecular line data at both 3 mm and 1.3 mm bands from the ATOMS-QUARKS survey. Section \ref{sec:observations} introduces the two ALMA observations used in this study. Section \ref{sec:result} presents the integrated intensity maps of selected molecular transitions, which reveal chemically enriched zones throughout the bipolar outflow. In Section \ref{sec:discuss}, we further explore the chemical origin of each species within these local regions. Finally, Section \ref{sec:conclusions} summarizes our main findings and conclusions. 

\section{Observations} \label{sec:observations}

\subsection{ALMA 3-mm data}

We utilized ALMA Band 3 data from the ATOMS survey \citep[Project ID: 2019.1.00685.S; PI: Tie Liu;][]{Liu2020b}, which targets 146 massive star-forming clumps using both the 12-m array in C3 configuration and the ACA 7-m array. SDC335 was observed in single-pointing mode, with a spectral setup covering multiple dense gas and shock tracers. In this study, we use the 3~mm continuum map obtained with the 12-m array, which offers higher angular resolution than the ACA-combined data,  with a synthesized beam of $1.7'' \times 1.5''$ and a noise rms of 10 mJy beam$^{-1}$. We also analyze selected molecular line data from the 12-m $+$ 7-m combined cubes, including transitions of \hcop (1--0), \htcop (1--0), SiO (2--1), and \meoh, with velocity resolutions of 0.1--0.2~km~s$^{-1}$. More details on the observations and data reduction of this source can be found in \citealt{2021MNRAS.505.2801L,2023MNRAS.520.3259X}.

\subsection{ALMA 1.3-mm data}

We also used ALMA Band 6 data from the QUARKS survey \citep[Project ID: 2021.1.00095.S; PIs: Lei Zhu, Guido Garay, and Tie Liu;][]{2024RAA....24b5009L}, which is a high-resolution 1.3~mm follow-up survey of the ATOMS survey. SDC335 was observed using both the 12-m C5 (TM1), 12-m C2 (TM2) and the ACA 7-m arrays, with the best resolution of $0.3"$ and the maximum recoverable scale of $\sim28''$ \citep{Xu2024Quarks2,2025ApJS..280...33Y}. In this work, we also used combined TM2 $+$ ACA data for lines because of the higher signal-to-noise ratios at lower resolution. The achieved resolutions of continuum image and line data cube from the TM2 $+$ ACA combined data are approximately $1.3'' \times 1.1''$. 
The noise rms of the continuum image is $\sim0.8$~mJy beam$^{-1}$. The four spectral windows at 217--221 and 229--233 GHz spectral ranges cover numerous transitions from species such as SiO (5--4), SO (6--5), HNCO, \meoh, and other molecules of interest. The velocity resolution of the line data is about 1.3~km~s$^{-1}$.

\begin{sidewaysfigure*}
\centering
\includegraphics[width=\textwidth]{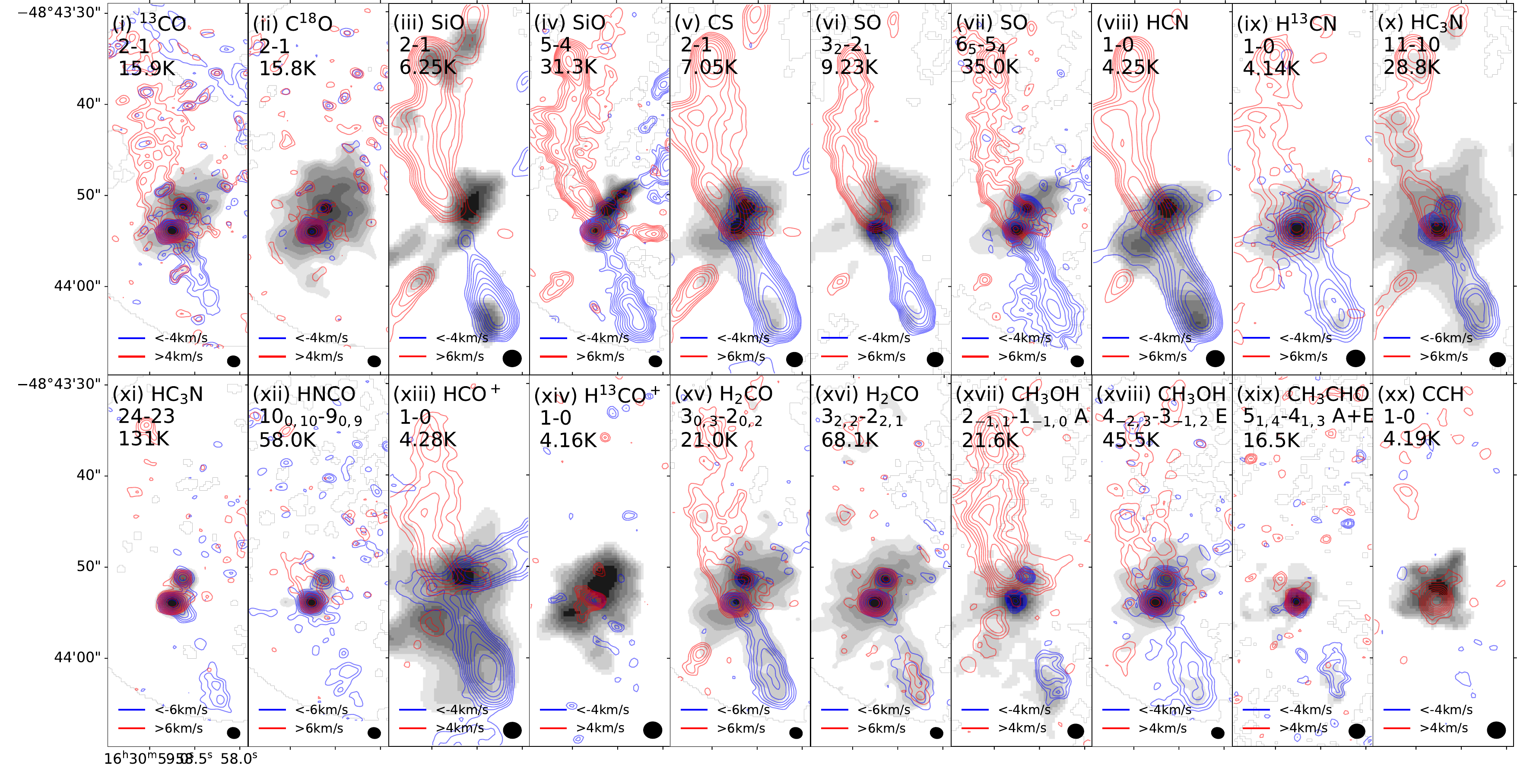}
\caption{
Integrated intensity maps of selected molecular transitions detected in SDC335. 
\textbf{Red/Blue contours:} red-shifted and blue-shifted components, with velocity ranges labeled in each panel.
\textbf{Background:} emission integrated over the systemic velocity gap between the blue- and red-shifted ranges.
Contours are drawn at 10 levels logarithmically spaced from 3$\sigma$ to the map maximum.  
The typical 1$\sigma$ rms in the integrated-intensity maps is 2.6\,K\,km\,s$^{-1}$ for the 3~mm data and 0.7\,K\,km\,s$^{-1}$ for the 1.3~mm data.
Each panel lists the corresponding transition and upper energy level.}
\label{fig:moment0}
\end{sidewaysfigure*}

\section{results} \label{sec:result}

\subsection{The outflow and driven source} \label{sec:result-cont}

Fig.~\ref{fig:cont}(a) presents the RGB composite image of CO J=2-1 outflows in SDC335, with blue, green, and red colors representing gas velocities blue-shifted by more than 10~km~s$^{-1}$, near the systemic velocity, and red-shifted by more than 7~km~s$^{-1}$, respectively. A bipolar jet-like outflow originates from the core C1, which is identified as the driving source based on its position and chemical richness. The 1.3~mm continuum is overlaid as white contours, and the jet axis, spanning $\sim$37$''$ (0.63~pc at the distance to the source), is indicated by a cyan line. Seven knots marked with “X” correspond to positions along the jet where CH$_3$OH and SiO emission are significantly enhanced in 1.3~mm band (panels iv and xviii of Fig.~\ref{fig:moment0}). These knots, along with the jet axis, serve as the basis for subsequent analysis and discussion.

Fig.~\ref{fig:cont}(b) shows a zoomed-in view of the white box in panel (a), with the background displaying the 3mm continuum. A low threshold is adopted to highlight the extended warm envelope, while high-threshold black contours of the 1.3mm continuum emphasize compact internal structures. In addition to the three main cores C1, C2, and C3, two earlier-phase cores (C4 and C5) are detected between C1 and C2, consistent with the cores detected in ALMA-DIHCA survey \citep{2021ApJ...909..199O}.
C1 is a chemically rich core with a deconvolved size of $1.14''\times1.01''$. In the ATOMS 3mm band survey, 18 COMs were detected toward C1, including c-C$_2$H$_4$O, C$_2$H$_5$CN, CH$_3$CHO, and CH$_3$OCHO~\citep{2022MNRAS.511.3463Q,2025A&A...697A.190L}, indicating that it is a typical hot molecular core. The average line width is $\sim$4 km~s$^{-1}$. The non-detection of radio recombination lines (e.g., H${30\alpha}$ or H${40\alpha}$) is suggestive of the absence of strong ionized gas emission.

\begin{sidewaysfigure*}
\centering
\includegraphics[width=1\textwidth]{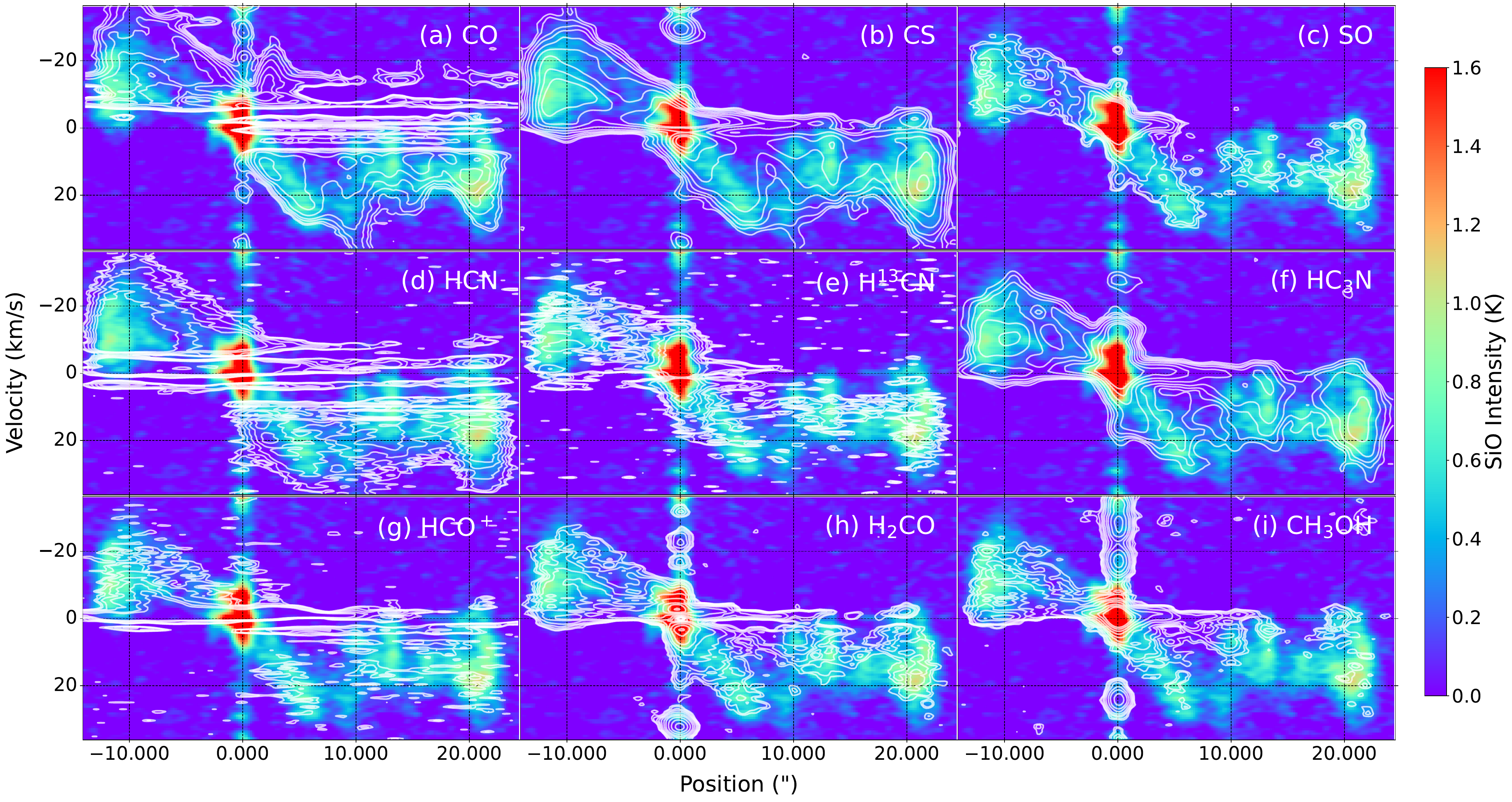}
\caption{
PVDs extracted along the jet axis of SDC335. The background color scale in each panel shows SiO emission, tracing the full spatial and velocity extent of the outflow. 
Overlaid contours correspond to representative transitions of other molecules, as labeled in each panel. 
Negative offsets indicate the southern (blue-shifted) lobe, and positive offsets the northern (red-shifted) lobe, with 0$''$ marking the C1.
}
\label{fig:pvds}
\end{sidewaysfigure*}

\subsection{Molecular lines detected along the outflow} \label{sec:result-pvdchmap}

We first extracted line spectra at the seven bright knots located along the walls of the jet-driven outflow cavity, where \meoh~and SiO emission are significantly enhanced (see Sect.~\ref{sec:result-cont} and Figs.~\ref{fig:cont} and \ref{fig:moment0}). To examine their velocity structure at higher spatial resolution, we employed the ACA+TM1+TM2 combined data (see Figs.~\ref{fig:spectra-3} and \ref{fig:spectra-1.3} ). Owing to the relatively low signal-to-noise ratio of these data, we integrated every 3~km~s$^{-1}$ channel to enhance the visibility of extended structures. Given the potential for significant missing flux in the interferometric maps, we did not use these data for quantitative analysis. The resulting channel maps reveal that emission from CO~$2\text{--}1$, SiO~$5\text{--}4$, SO~$6_5$–$5_4$, H$_2$CO~$3_{0,3}$–$2_{0,2}$, and \meoh~$4_{-2,3}$–$3_{-1,2}$~is morphologically separated into two kinematic components: a high-velocity ($>12$~km~s$^{-1}$) component concentrated along the collimated outflow axis, and a low-velocity (6–12~km~s$^{-1}$) component forming a hollow-shell cavity structure. The seven knots correspond to prominent features in the low-velocity cavity shell, while their high-velocity counterparts trace localized regions along the axis. At a given velocity interval, the spatial distributions of these molecules are broadly similar, with differences mainly in emission strength and the maximum velocity reached—for example, the high-velocity emission of \meoh~extends only to $\sim$21~km~s$^{-1}$, H$_2$CO to $\sim$24~km~s$^{-1}$, SO to $\sim$30~km~s$^{-1}$, and SiO and CO to $\sim$33~km~s$^{-1}$. These differences likely reflect variations in excitation conditions or shock chemistry sensitivity among species. In total, 35 transitions from 22 molecular species were identified at the 3~mm and 1.3~mm bands (see Appendix Table~\ref{tab:transitions}).

Fig.~\ref{fig:moment0} presents the integrated intensity maps of 20 representative transitions, with color coding corresponding to blue-shifted, systemic, and red-shifted components. As can be seen in the figure, while most transitions are closely associated with outflow structure, the molecular species exhibit diverse spatial distributions and kinematic properties. For example, H$^{13}$CO$^+$ and CCH are confined to the quiescent envelope surrounding C1 and C2, characterized by narrow single-peaked profiles (FWHM~$\leq$~2~km~s$^{-1}$), whereas SiO emission is enhanced at jet-impacted regions and shows broad line wings typical of fast shocks. Similarly, SO displays a spatial and velocity distribution akin to SiO, suggesting both species trace shocked gas. COMs including CH$_3$CHO are mainly detected near the outflow termini, indicating sensitivity to local heating or mild shocks. \mf~and \mecn~ are marginally detected and exhibit low S/N as shown in Fig.~\ref{fig:spectra-1.3}. CH$_3$OH emission, although widespread, exhibits relatively weak high-velocity components and is concentrated near the systemic velocity, likely originating from sublimated ices induced by shocks in outflow and the cavity walls.

To further examine the gas kinematics, Fig.~\ref{fig:pvds} presents position–velocity diagrams (PVDs) of several important molecules extracted along the outflow-jet axis (K1–K7, cyan line in Fig.~\ref{fig:cont}(a)). Consistent with the channel maps (Sect.~\ref{sec:channelmap}), the PVDs confirm the morphological separation between high-velocity emission along the axis and low-velocity components in the cavity walls. In addition, the PVDs highlight localized high-velocity peaks at shock positions traced most clearly by SiO and SO, as well as discrete, spatially confined knots in CS, SO, HC$_3$N, H$_2$CO, and CH$_3$OH. These kinematic patterns, combined with the spatial distributions in Fig.~\ref{fig:moment0}, reflect diverse excitation mechanisms and physical origins, which will be further analyzed in the following section.


\begin{figure*}
\hspace*{-0.07\textwidth} 
\begin{minipage}{1.1\textwidth} 
\gridline{\fig{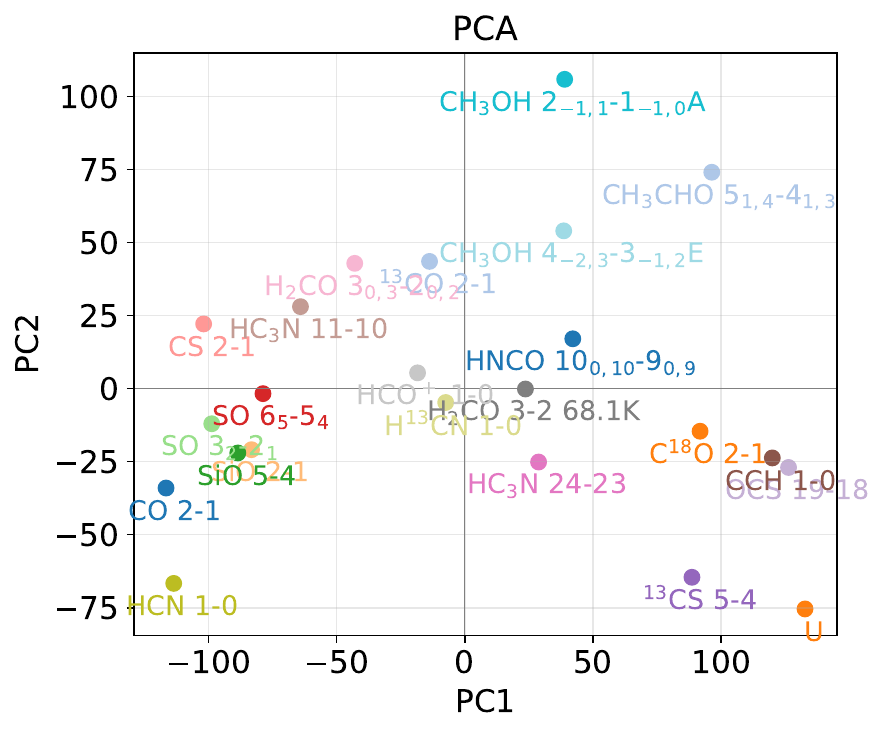}{0.5\textwidth}{(a)}
  \fig{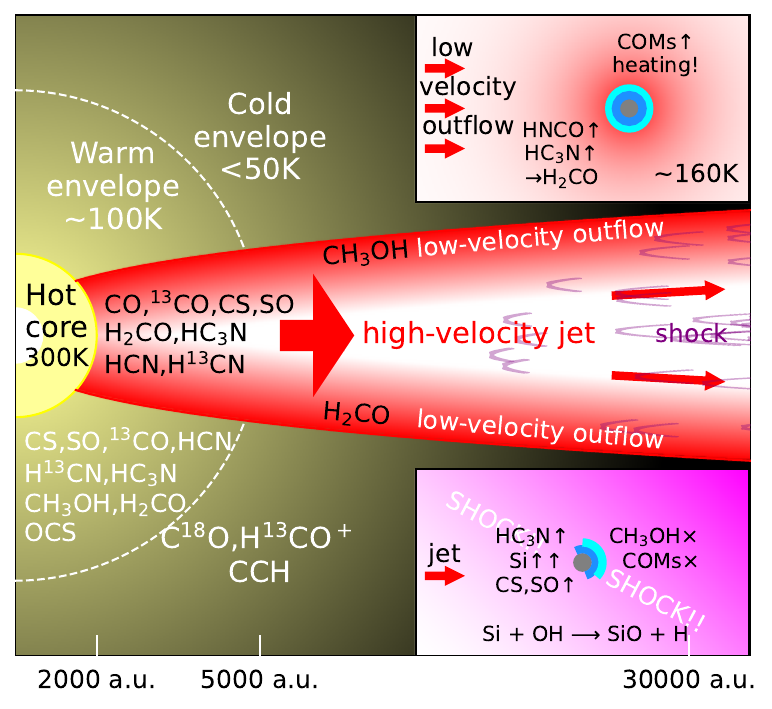}{0.5\textwidth}{(b)}}
\end{minipage}
\caption{
(\textbf{a}) PCA of molecular emission along the jet axis (data within $\pm2^{\prime\prime}$ of C1 excluded). Molecules are separated into four quadrants reflecting dominant chemical–kinematic behavior:
Quadrant I: COMs enhanced in shocked regions;
Quadrant II: outflow tracers with low- and high-velocity components;
Quadrant III: shock/high-velocity tracers;
Quadrant IV: systemic-velocity or low S/N lines.
(\textbf{b}) Schematic view of the physical and chemical structure of the bipolar outflow in SDC335. 
The main panel shows the spatial distribution of key molecular tracers associated with the hot core, warm envelope, and jet-driven outflows. 
The upper inset illustrates post-shock low-velocity outflows heating the gas and sublimating molecules from icy mantles. 
The lower inset depicts strong shocks destroying dust grains, releasing ice-phase material and driving Si atoms from grain cores into the gas phase to form SiO.
}
\label{fig:pca_model}
\end{figure*}


\begin{figure*}
\centering
\includegraphics[width=1.05\textwidth]{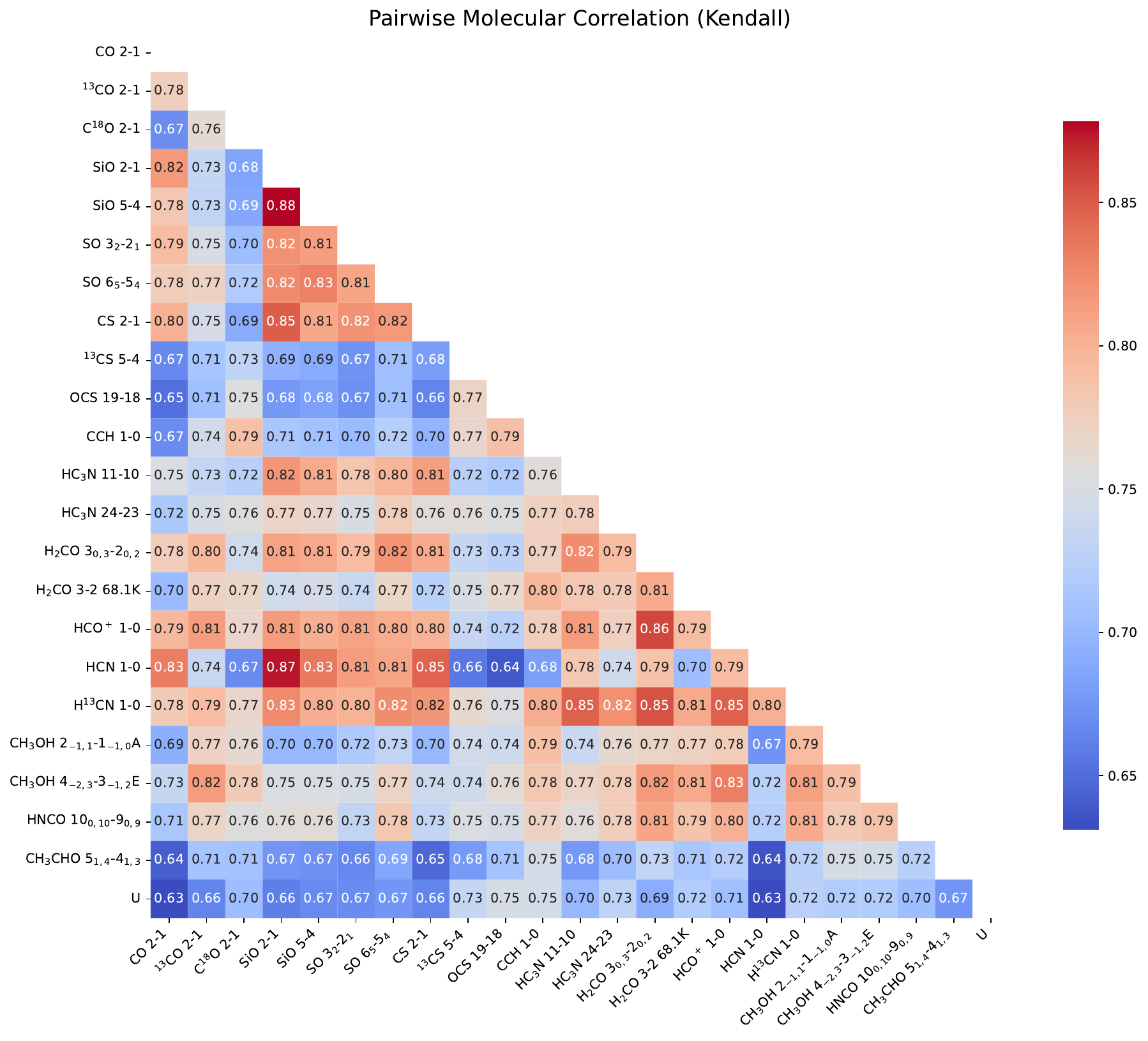}
\caption{Kendall rank correlation coefficients between integrated intensities of molecular lines, shown as a lower-triangular matrix to avoid duplication.}
\label{fig:kendall}
\end{figure*}

\section{Analysis and Discussion} \label{sec:discuss}

\subsection{Quantitative Analysis and Global Results} \label{sec:4.1}

Benefiting from the large number of molecular lines detected across both bands (Section~\ref{sec:result-pvdchmap}), we focus on a subset of transitions with sufficiently high S/N for reliable kinematic and excitation analysis. This selection enables a systematic investigation of the physical and chemical origins of the detected species, which we develop in this section.

The channel maps (Section~\ref{sec:channelmap}) reveal a clear separation between high-velocity emission confined to the jet axis and a surrounding low-velocity cavity, with molecule-dependent variations in extent and maximum velocity. Consistently, the PVD (Fig.\ref{fig:pvds}) and moment0 map (Fig.\ref{fig:moment0}) show that the bipolar jet driven by C1 is remarkably symmetric within $\pm$14$''$ in both morphology and velocity, reaching line-of-sight velocities of up to $\pm$40~km~s$^{-1}$ at $\sim$$\pm$10$''$ from C1. The broadest line widths (up to 20~km~s$^{-1}$) and the highest gas temperatures are found at knots K1, K6, and K7, coinciding with peaks in molecular emission. Independent constraints from H$_2$CO rotational analysis (Fig.~\ref{fig:trot-sio}) reveal $T{\rm rot} \gtrsim 200$~K at several knots, consistent with strong shock heating by the protostellar jet and suggesting that multiple chemical processes—direct entrainment, shock sputtering, and thermal desorption—operate simultaneously along the outflow.

To quantify chemical variations, we computed position-integrated intensities of multiple transitions along the jet axis (Fig.\ref{fig:B1}). Assuming LTE and optically thin conditions, these serve as proxies for column densities. Most molecules, except $^{13}$CO and C$^{18}$O, show enhancements near K1–K7. Using $^{13}$CO as a tracer of H$_2$ column density, relative intensity ratios (Fig.\ref{fig:ni-13co}) highlight localized abundance peaks at the knots.

Spectral profiles provide complementary constraints (Fig.~\ref{fig:spectra1}). Molecules linked to shocks, such as SiO, SO, and HCN, display broad wings and strong high-velocity components, whereas \hac~and \mecn~peak near the systemic velocity with weak wings. These differences reflect distinct excitation and formation pathways, consistent with the spatial–kinematic separation observed in the channel maps and PVDs.

Building on these diagnostics—integrated intensity (Fig.\ref{fig:ni-13co}), line profile (Fig.\ref{fig:spectra1}), and the ratio of systemic- to non-systemic-velocity components (Fig.\ref{fig:moment0})—we performed PCA on the molecular PVDs (Fig.\ref{fig:pca_model}(a)). This separates molecules into four groups, consistent with Section~\ref{sec:4.2}:
\begin{itemize}
\item Quadrant I (Sec.\ref{sec:4.2.3}): COMs, concentrated near systemic velocity but enhanced in terminal knots, reflecting ice-mantle sublimation and gas-phase reactions;
\item Quadrant II (Sec.\ref{sec:4.2.1}): outflow tracers spanning low- and high-velocity components, with stable intensities from the hot core outward, indicative of entrained ambient gas;
\item Quadrant III (Sec.~\ref{sec:4.2.2}): species associated with shocks or high-velocity jets, peaking at terminal knots where collisions produce strong shocks;
\item Quadrant IV: molecules dominated by systemic-velocity emission or with low S/N, weakly affected by jet-induced shocks or of uncertain origin.
\end{itemize}

Finally, PCA on the full 18,684-dimensional vectors (173 positions × 108 high-velocity channels) shows that PC1 and PC2 account for 36.2\% and 10.3\% of the variance (46.5\% combined), with nearly all variance driven by spatial differences in high-velocity ($>4$km~s$^{-1}$) line intensities. 

Complementing this, pairwise Kendall correlations (Fig.~\ref{fig:kendall}) reveal that CS, SO, HCN, and SiO are tightly correlated ($>0.8$), with the strongest correlation (0.88) between two SiO transitions J=2–1 and  J=5–4. H$_2$CO and HCO$^+$ correlate uniformly with most species (0.77–0.83) and with each other (0.86), indicating strong chemical coupling. Notably, \meoh~$2_{-1,1}$–$1_{-1,0}$ A and $4_{-2,3}$–$3_{-1,2}$ E correlate at only 0.79, and their correlations with other molecules show larger differences, possibly related to nuclear spin isomers (Zou et al., in prep).

The results are also illustrated in a schematic cartoon (Fig.~\ref{fig:pca_model} (b)). We emphasize that the PCA quadrants indicate the dominant origin of each molecule along the jet, but many molecules can have contributions from more than one source. This motivates the detailed classification and discussion in the following subsections.

\subsection{Classification of Molecules} \label{sec:4.2}

\subsubsection{Molecules Entrained by the Outflow} \label{sec:4.2.1}

This group comprises molecules that are abundant in the dense core and are primarily carried outward by the jet-driven outflow, including CO, $^{13}$CO, HCO$^+$, H$_2$CO, CS, and HC$_3$N. As shown in the channel maps (Fig.~\ref{fig:channelCO}, \ref{fig:channelH2CO}) and PVDs (Fig.~\ref{fig:pvds}), these species are detected both along the collimated jet axis and within the surrounding cavity-like low-velocity structures. Their line profiles (Appendix Fig.~\ref{fig:spectra1}) generally display broad wings together with systemic or intermediate-velocity components, indicating that they mainly trace gas physically entrained and dynamically processed by the jet–ambient interaction \citep{2010A&A...522A..91T,2020A&A...636A..60T,2021A&A...648A..45P,2019A&A...632A.101T}.

Relative intensity ratios to $^{13}$CO (Appendix Fig.~\ref{fig:ni-13co}, \ref{fig:ni-13co-static}) remain nearly flat in regions lacking strong shocks, but show moderate enhancements at bright knots. Within this set, CO and HCO$^+$ are canonical tracers of entrained gas; H$_2$CO can be supplied by gas-phase formation and/or non-thermal release from grains \citep{2019A&A...632A.101T}; CS and HC$_3$N follow the same spatial–kinematic pattern and predominantly trace dense gas swept up by the outflow.

\subsubsection{Molecules Released into the Gas Phase by Jet-Induced Shocks} \label{sec:4.2.2}

A second group of molecules shows clear evidence of shock-related enhancement, including SiO, SO, HCN (and its isotopologues H$^{13}$CN and DCN), \ppn, and HC$_3$N. Channel maps and PVDs (Figs.~\ref{fig:channelSiO}, \ref{fig:channelSO}, and \ref{fig:pvds}) reveal that their emissions are concentrated along the jet axis and peak at bright knots, with broad wings and high-velocity components comparable to SiO (Appendix Fig.~\ref{fig:spectra1}). Their intensity ratios relative to $^{13}$CO also rise sharply at these knots (Appendix Fig.~\ref{fig:ni-13co}), confirming that they trace material chemically reprocessed in shocks.

In cold dense clouds, Si is locked in dust grains and SiO is therefore heavily depleted \citep{1992A&A...265L..49G,1997A&A...325..758D}. Jet-induced shocks liberate Si atoms via non-thermal sputtering or grain fragmentation, which then react with OH or O to form SiO \citep{1997A&A...322..296C,1997A&A...321..293S,2008A&A...482..809G,2008A&A...490..695G}. Such localized SiO emission has been widely observed in sources such as TMC1, SMM3, and HH 211 \citep{2015A&A...581A..85P,2014Natur.507...78S,2019A&A...632A.101T}.  

SO chemistry is likewise tied to shocks: sulfur released from grains can react with OH or O to form SO \citep{1980ApJ...236..182H,1993dca..book.....M}, while high-temperature reactions involving H$_2$S further enhance SO abundance in fast shocks where OH and O abundances increase sharply \citep{2008A&A...490..695G}. Although slower shocks can also contribute, the strongest SO emission is typically found in energetic jet–ambient collisions.

HCN and its isotopologues (H$^{13}$CN, DCN) exhibit SiO-like kinematics in our data. Their enhancement can be explained by gas-phase reactions of CN with H$_2$ in warm post-shock gas \citep{2009A&A...503L..13B,2018A&A...615A..75V}. In HH211, H$^{13}$CN has been observed tracing cavity walls, supporting a scenario where UV irradiation and high-temperature chemistry act in tandem \citep{2019A&A...632A.101T}.

Although HC$_3$N is often regarded as a more complex N-bearing molecule, it likewise shows signs of shock enhancement \citep{2021ApJ...909..177L,Hoque_2025}. Possible formation pathways include polymerization of CN intermediates or release of atomic N in nitrogen-rich, low C/O environments \citep{2004A&A...415.1021J,2009A&A...507L..25C,2010A&A...518L.112C}. An additional efficient gas-phase route is the reaction $\mathrm{C_2H_2 + CN \rightarrow HC_3N + H}$ \citep{1997ApJ...489..113F,2009MNRAS.394..221C,2018MNRAS.475.5501M}, which becomes more active in shocked gas.

Overall, the enhancement of these molecules is best understood through two complementary processes:
\begin{itemize}
  \item[(i)] \textbf{sputtering of atoms or molecules} (e.g., Si, S, HCN) from dust grains into the gas phase, and
  \item[(ii)] \textbf{rapid activation of high-temperature chemistry} at shock fronts, producing additional N- and O-bearing species.
\end{itemize}

\subsubsection{Molecules Thermally Excited by Shock Heating in the Jet} \label{sec:4.2.3}

A third group of molecules shows enhanced abundances that are best explained by shock-induced heating rather than direct sputtering. Channel maps and PVDs (Figs.~\ref{fig:channelMeOH}, \ref{fig:pvds}) reveal that their emissions peak at the systemic velocity of the outflow, while intensity ratios relative to $^{13}$CO still increase at bright knots (Appendix Fig.~\ref{fig:ni-13co}). Line profiles (Appendix Fig.~\ref{fig:spectra1}) confirm that CH$_3$OH and HNCO sometimes display additional broad wings tracing the outflow, whereas CH$_3$CHO, CH$_3$CN, and CH$_3$OCHO are confined to narrow systemic components. This pattern, combined with the correlation between SiO gradients and H$_2$CO rotational temperatures derived with RADEX \citep{2017A&A...598A...7M} following methods similar to \citet{2024ApJ...963..163I} (Fig.~\ref{fig:trot-sio}), indicates excitation primarily driven by local thermal heating in shocked gas.

The observed enhancements can be attributed to high-temperature gas-phase chemistry coupled with thermal desorption of icy mantles. CH$_3$OH, for example, forms efficiently on grain surfaces at low temperatures and is readily released into the gas phase by heating \citep{2004ApJ...616..638W,2009A&A...504..891O}. Other species such as CH$_3$CHO and HNCO are likewise desorbed or reprocessed through high-temperature reactions in shocked zones \citep{2012A&A...541L..12B,2020ApJ...891...73S}. Comparisons with hot cores, such as S68N \citep{2020A&A...639A..87V,2021A&A...650A.150N}, show similar abundance ratios, suggesting shared thermal histories or transient regeneration in shocks. Observations in L1157-B1 and SMM3 further support this scenario, where localized heating can mimic hot-core conditions and temporarily enable the release or synthesis of complex organics \citep{2020A&A...640A..75D,2021A&A...655A..65T}.

In summary, the elevated abundances of CH$_3$OH, HNCO, and several COMs are mainly attributed to high temperatures induced by jet–ambient shocks. Thermal desorption and rapid gas-phase chemistry work in concert, enhancing molecular complexity in the outflow beyond that expected from the hot core alone.

Overall, our classification highlights the predominant behaviors of different molecular groups—entrainment from the core, shock-induced release, and thermal excitation by heating. However, these categories should not be regarded as mutually exclusive. As shown in Fig.~\ref{fig:kendall} and Fig.~\ref{fig:pca_model}, some molecules exhibit stronger correlations across groups than within their assigned category, underscoring the complexity of chemical processes in jet–outflow systems. This scheme therefore provides a useful framework for interpretation, but the role of each molecule may vary with local physical conditions, and caution is warranted when drawing sharp boundaries between categories.

\section{Conclusions} \label{sec:conclusions}
We have presented ALMA observations of the chemically rich jet and outflow system in SDC335, detecting 35 transitions from 22 molecular species. By extracting channel maps, PVDs, integrated intensity ratios, and line profiles, we conducted a quantitative analysis of the spatial–kinematic and excitation properties of different molecules. 

Principal component analysis (PCA) of these parameters shows that line profiles provide the largest variance among species, serving as the dominant feature distinguishing their behaviors. The PCA distribution naturally separates molecules into three quadrants, corresponding to three predominant classes: (1) species entrained from the dense core by the outflow (e.g., CO, HCO$^+$, CS, HC$_3$N); (2) molecules released into the gas phase or newly formed in jet-induced shocks (e.g., SiO, SO, HCN, DCN); and (3) molecules thermally excited by shock heating and subsequent gas-phase or desorption processes (e.g., CH$_3$OH, CH$_3$CHO, CH$_3$CN, HNCO, CH$_3$OCHO). These groups display distinct spatial–kinematic signatures, with entrained tracers extended along the cavity walls, shock tracers peaking at high-velocity knots, and thermally excited species confined to systemic-velocity components with local enhancements at knots. However, individual species may participate in multiple processes—entrainment, sputtering, or thermal heating—depending on local conditions. The observed chemistry of the outflow is therefore shaped by the interplay of all three mechanisms rather than by a single pathway.

In summary, our study highlights the multi-origin nature of molecular emission in a massive protostellar outflow, combining entrainment from the core, shock-driven release, and thermal excitation by shock heating. Future observations with higher angular resolution and sensitivity, together with laboratory and modeling efforts on grain-surface and gas-phase chemistry under shock conditions, will be essential to disentangle the relative contributions of these processes and to quantify the physical–chemical environment of massive outflows.

\begin{acknowledgments}

Tie Liu acknowledges the supports by the National Key R\&D
Program of China (No. 2022YFA1603101), National Natural Science Foundation of China (NSFC) through grants No.12073061 and No.12122307, the PIFI program of Chinese Academy of Sciences through grant No. 2025PG0009, and the Tianchi Talent Program of Xinjiang
Uygur Autonomous Region.\\


H.-L. Liu is supported by Yunnan Fundamental Research Project (grant No. 202301AT070118, 202401AS070121), and by Xingdian Talent Support Plan--Youth Project. \\
\end{acknowledgments}

A.H. thank the support of the S. N. Bose National Centre for Basic Sciences under the Department of Science and Technology, Govt. of India and the CSIR-HRDG, Govt. of India for funding the fellowship.

SRD acknowledges support from the Fondecyt Postdoctoral fellowship (project code 3220162) and ANID BASAL project FB210003.





%
\facilities{ALMA}

\bibliography{sample7}

@ARTICLE{2008ApJ...681L..21A,
       author = {{Arce}, H{\'e}ctor G. and {Santiago-Garc{\'\i}a}, Joaqu{\'\i}n and {J{\o}rgensen}, Jes K. and {Tafalla}, Mario and {Bachiller}, Rafael},
        title = "{Complex Molecules in the L1157 Molecular Outflow}",
      journal = {\apjl},
         year = 2008,
        month = jul,
       volume = {681},
       number = {1},
        pages = {L21},
          doi = {10.1086/590110},
archivePrefix = {arXiv},
       eprint = {0805.2550},
 primaryClass = {astro-ph},
       adsurl = {https://ui.adsabs.harvard.edu/abs/2008ApJ...681L..21A}
}

@ARTICLE{2025ApJ...980..263O,
       author = {{Oya}, Yoko and {Saiga}, Eri and {Miotello}, Anna and {Koutoulaki}, Maria and {Johnstone}, Doug and {Ceccarelli}, Cecilia and {Chandler}, Claire J. and {Codella}, Claudio and {Sakai}, Nami and {Bianchi}, Eleonora and {Bouvier}, Mathilde and {Charnley}, Steven and {Cuello}, Nicolas and {De Simone}, Marta and {Francis}, Logan and {Hanawa}, Tomoyuki and {Jim{\'e}nez-Serra}, Izaskun and {Loinard}, Laurent and {Menard}, Francois and {Sabatini}, Giovanni and {Vastel}, Charlotte and {Zhang}, Ziwei and {Aikawa}, Yuri and {Alves}, Felipe O. and {Balucani}, Nadia and {Busquet}, Gemma and {Caselli}, Paola and {Caux}, Emmanuel and {Choudhury}, Spandan and {Dulieu}, Francois and {Dur{\'a}n}, Aurora and {Evans}, Lucy and {Fedele}, Davide and {Feng}, Siyi and {Fontani}, Francesco and {Hama}, Tetsuya and {Herbst}, Eric and {Hirano}, Shingo and {Hirota}, Tomoya and {Isella}, Andrea and {Kahane}, Claudine and {Lefloch}, Bertrand and {Le Gal}, Romane and {Liu}, Hauyu Baobab and {L{\'o}pez-Sepulcre}, Ana and {Maud}, Luke T. and {Maureira}, Mar{\'\i}a Jos{\'e} and {Mercimek}, Seyma and {Moellenbrock}, George and {Mori}, Shoji and {Nomura}, Hideko and {Oba}, Yasuhiro and {O'Donoghue}, Ross and {Ohashi}, Satoshi and {Okoda}, Yuki and {Ospina-Zamudio}, Juan and {Pineda}, Jaime and {Podio}, Linda and {Rimola}, Albert and {Sakai}, Takeshi and {Segura-Cox}, Dominique and {Shirley}, Yancy and {Svoboda}, Brian and {Testi}, Leonardo and {Viti}, Serena and {Watanabe}, Naoki and {Watanabe}, Yoshimasa and {Zhang}, Yichen and {Yamamoto}, Satoshi},
        title = "{Evidence for Jet/Outflow Shocks Heating the Environment around the Class I Protostellar Source Elias 29: FAUST XXI}",
      journal = {\apj},
         year = 2025,
        month = feb,
       volume = {980},
       number = {2},
          eid = {263},
        pages = {263},
          doi = {10.3847/1538-4357/adabe7},
archivePrefix = {arXiv},
       eprint = {2501.10634},
 primaryClass = {astro-ph.SR},
       adsurl = {https://ui.adsabs.harvard.edu/abs/2025ApJ...980..263O}
}

@ARTICLE{2022ApJS..262...13R,
       author = {{Rojas-Garc{\'\i}a}, O.~S. and {G{\'o}mez-Ruiz}, A.~I. and {Palau}, A. and {Orozco-Aguilera}, M.~T. and {Dagostino}, M. Chavez and {Kurtz}, S.~E.},
        title = "{Interstellar Complex Organic Molecules in SiO-traced Massive Outflows}",
      journal = {\apjs},
         year = 2022,
        month = sep,
       volume = {262},
       number = {1},
          eid = {13},
        pages = {13},
          doi = {10.3847/1538-4365/ac81cb},
archivePrefix = {arXiv},
       eprint = {2207.09426},
 primaryClass = {astro-ph.GA},
       adsurl = {https://ui.adsabs.harvard.edu/abs/2022ApJS..262...13R}
}

@ARTICLE{2022ApJ...933L..35F,
       author = {{Feng}, S. and {Liu}, H.~B. and {Caselli}, P. and {Burkhardt}, A. and {Du}, F. and {Bachiller}, R. and {Codella}, C. and {Ceccarelli}, C.},
        title = "{A Detailed Temperature Map of the Archetypal Protostellar Shocks in L1157}",
      journal = {\apjl},
         year = 2022,
        month = jul,
       volume = {933},
       number = {2},
          eid = {L35},
        pages = {L35},
          doi = {10.3847/2041-8213/ac75d7},
archivePrefix = {arXiv},
       eprint = {2206.06585},
 primaryClass = {astro-ph.GA},
       adsurl = {https://ui.adsabs.harvard.edu/abs/2022ApJ...933L..35F}
}

@ARTICLE{2020A&A...640A..75D,
       author = {{De Simone}, M. and {Codella}, C. and {Ceccarelli}, C. and {L{\'o}pez-Sepulcre}, A. and {Witzel}, A. and {Neri}, R. and {Balucani}, N. and {Caselli}, P. and {Favre}, C. and {Fontani}, F. and {Lefloch}, B. and {Ospina-Zamudio}, J. and {Pineda}, J.~E. and {Taquet}, V.},
        title = "{Seeds of Life in Space (SOLIS). X. Interstellar complex organic molecules in the NGC 1333 IRAS 4A outflows}",
      journal = {\aap},
         year = 2020,
        month = aug,
       volume = {640},
          eid = {A75},
        pages = {A75},
          doi = {10.1051/0004-6361/201937004},
archivePrefix = {arXiv},
       eprint = {2006.09925},
 primaryClass = {astro-ph.SR},
       adsurl = {https://ui.adsabs.harvard.edu/abs/2020A&A...640A..75D}
}

@ARTICLE{2024MNRAS.527.2110R,
       author = {{Rojas-Garc{\'\i}a}, O.~S. and {G{\'o}mez-Ruiz}, A.~I. and {Palau}, A. and {Orozco-Aguilera}, M.~T. and {Kurtz}, S.~E. and {Chavez Dagostino}, M.},
        title = "{Interstellar complex organic molecules towards outflows from the G351.16+0.70 (NGC 6334 V) massive protostellar system}",
      journal = {\mnras},
         year = 2024,
        month = jan,
       volume = {527},
       number = {2},
        pages = {2110-2127},
          doi = {10.1093/mnras/stad3161},
archivePrefix = {arXiv},
       eprint = {2303.02527},
 primaryClass = {astro-ph.GA},
       adsurl = {https://ui.adsabs.harvard.edu/abs/2024MNRAS.527.2110R}
}

@ARTICLE{2017MNRAS.467.2723P,
       author = {{Palau}, Aina and {Walsh}, Catherine and {S{\'a}nchez-Monge}, {\'A}lvaro and {Girart}, Josep M. and {Cesaroni}, Riccardo and {Jim{\'e}nez-Serra}, Izaskun and {Fuente}, Asunci{\'o}n and {Zapata}, Luis A. and {Neri}, Roberto},
        title = "{Complex organic molecules tracing shocks along the outflow cavity in the high-mass protostar IRAS 20126+4104}",
      journal = {\mnras},
         year = 2017,
        month = may,
       volume = {467},
       number = {3},
        pages = {2723-2752},
          doi = {10.1093/mnras/stx004},
archivePrefix = {arXiv},
       eprint = {1701.04802},
 primaryClass = {astro-ph.SR},
       adsurl = {https://ui.adsabs.harvard.edu/abs/2017MNRAS.467.2723P}
}

@ARTICLE{2023ApJ...949...89X,
       author = {{Xie}, Jinjin and {Li}, Juan and {Wang}, Junzhi and {Liu}, Shu and {Yang}, Kai and {Quan}, Donghui and {Zheng}, Siqi and {Li}, Yuqiang and {Wu}, Jingwen and {Duan}, Yan and {Li}, Di},
        title = "{Imaging Molecular Outflow in Massive Star-forming Regions with HNCO Lines}",
      journal = {\apj},
         year = 2023,
        month = jun,
       volume = {949},
       number = {2},
          eid = {89},
        pages = {89},
          doi = {10.3847/1538-4357/acc83f},
archivePrefix = {arXiv},
       eprint = {2303.14866},
 primaryClass = {astro-ph.GA},
       adsurl = {https://ui.adsabs.harvard.edu/abs/2023ApJ...949...89X}
}

@ARTICLE{2022A&A...659A..23C,
       author = {{Costa Silva}, A.~R. and {Fedriani}, R. and {Tan}, J.~C. and {Caratti o Garatti}, A. and {Ramsay}, S. and {Rosero}, V. and {Cosentino}, G. and {Gorai}, P. and {Leurini}, S.},
        title = "{NIR jets from a clustered region of massive star formation. Morphology and composition in the IRAS 18264-1152 region}",
      journal = {\aap},
         year = 2022,
        month = mar,
       volume = {659},
          eid = {A23},
        pages = {A23},
          doi = {10.1051/0004-6361/202142412},
archivePrefix = {arXiv},
       eprint = {2112.04463},
 primaryClass = {astro-ph.SR},
       adsurl = {https://ui.adsabs.harvard.edu/abs/2022A&A...659A..23C}
}

@ARTICLE{2024AJ....167..285X,
       author = {{Xu}, Yani and {Wang}, Junzhi and {Liu}, Shu and {Li}, Juan and {Li}, Yuqiang and {Luo}, Rui and {Ou}, Chao and {Zheng}, Siqi and {Liu}, Yijia},
        title = "{Dense Outflowing Molecular Gas in Massive Star-forming Regions}",
      journal = {\aj},
         year = 2024,
        month = jun,
       volume = {167},
       number = {6},
          eid = {285},
        pages = {285},
          doi = {10.3847/1538-3881/ad47c4},
archivePrefix = {arXiv},
       eprint = {2406.08935},
 primaryClass = {astro-ph.GA},
       adsurl = {https://ui.adsabs.harvard.edu/abs/2024AJ....167..285X}
}

@ARTICLE{2024A&A...681A.104B,
       author = {{Busch}, Laura A. and {Belloche}, Arnaud and {Garrod}, Robin T. and {M{\"u}ller}, Holger S.~P. and {Menten}, Karl M.},
        title = "{Shocking Sgr B2 (N1) with its own outflow. A new perspective on segregation between O- and N-bearing molecules}",
      journal = {\aap},
         year = 2024,
        month = jan,
       volume = {681},
          eid = {A104},
        pages = {A104},
          doi = {10.1051/0004-6361/202347256},
archivePrefix = {arXiv},
       eprint = {2310.11339},
 primaryClass = {astro-ph.GA},
       adsurl = {https://ui.adsabs.harvard.edu/abs/2024A&A...681A.104B}
}

@ARTICLE{2022A&A...665A..96B,
       author = {{Busch}, Laura A. and {Belloche}, Arnaud and {Garrod}, Robin T. and {M{\"u}ller}, Holger S.~P. and {Menten}, Karl M.},
        title = "{Resolving desorption of complex organic molecules in a hot core. Transition from non-thermal to thermal desorption or two-step thermal desorption?}",
      journal = {\aap},
         year = 2022,
        month = sep,
       volume = {665},
          eid = {A96},
        pages = {A96},
          doi = {10.1051/0004-6361/202243383},
archivePrefix = {arXiv},
       eprint = {2206.11174},
 primaryClass = {astro-ph.GA},
       adsurl = {https://ui.adsabs.harvard.edu/abs/2022A&A...665A..96B}
}

@ARTICLE{2020ApJ...901...31L,
       author = {{Liu}, Hong-Li and {Sanhueza}, Patricio and {Liu}, Tie and {Zavagno}, Annie and {Tang}, Xin-Di and {Wu}, Yuefang and {Zhang}, Siju},
        title = "{Chemistry of Protostellar Clumps in the High-mass, Star-forming Filamentary Infrared Dark Cloud G034.43+00.24}",
      journal = {\apj},
         year = {2020b},
        month = sep,
       volume = {901},
       number = {1},
          eid = {31},
        pages = {31},
          doi = {10.3847/1538-4357/abadfe},
       adsurl = {https://ui.adsabs.harvard.edu/abs/2020ApJ...901...31L}
}

@ARTICLE{2020A&ARv..28....1L,
       author = {{Lee}, Chin-Fei},
        title = "{Molecular jets from low-mass young protostellar objects}",
      journal = {\aapr},
         year = 2020,
        month = mar,
       volume = {28},
       number = {1},
          eid = {1},
        pages = {1},
          doi = {10.1007/s00159-020-0123-7},
archivePrefix = {arXiv},
       eprint = {2002.05823},
 primaryClass = {astro-ph.GA},
       adsurl = {https://ui.adsabs.harvard.edu/abs/2020A&ARv..28....1L}
}

@ARTICLE{Liu2020b,
       author = {{Liu}, Tie and {Evans}, Neal J. and {Kim}, Kee-Tae and {Goldsmith}, Paul F. and {Liu}, Sheng-Yuan and {Zhang}, Qizhou and {Tatematsu}, Ken'ichi and {Wang}, Ke and {Juvela}, Mika and {Bronfman}, Leonardo and {Cunningham}, Maria R. and {Garay}, Guido and {Hirota}, Tomoya and {Lee}, Jeong-Eun and {Kang}, Sung-Ju and {Li}, Di and {Li}, Pak-Shing and {Mardones}, Diego and {Qin}, Sheng-Li and {Ristorcelli}, Isabelle and {Tej}, Anandmayee and {Toth}, L. Viktor and {Wu}, Jing-Wen and {Wu}, Yue-Fang and {Yi}, Hee-weon and {Yun}, Hyeong-Sik and {Liu}, Hong-Li and {Peng}, Ya-Ping and {Li}, Juan and {Li}, Shang-Huo and {Lee}, Chang Won and {Shen}, Zhi-Qiang and {Baug}, Tapas and {Wang}, Jun-Zhi and {Zhang}, Yong and {Issac}, Namitha and {Zhu}, Feng-Yao and {Luo}, Qiu-Yi and {Soam}, Archana and {Liu}, Xun-Chuan and {Xu}, Feng-Wei and {Wang}, Yu and {Zhang}, Chao and {Ren}, Zhiyuan and {Zhang}, Chao},
        title = "{ATOMS: ALMA Three-millimeter Observations of Massive Star-forming regions - I. Survey description and a first look at G9.62+0.19}",
      journal = {\mnras},
         year = 2020,
        month = aug,
       volume = {496},
       unumber = {3},
        pages = {2790-2820},
          doi = {10.1093/mnras/staa1577},
archivePrefix = {arXiv},
       eprint = {2006.01549},
 primaryClass = {astro-ph.GA},
       adsurl = {https://ui.adsabs.harvard.edu/abs/2020MNRAS.496.2790L}
}

@ARTICLE{2021MNRAS.505.2801L,
       author = {{Liu}, Hong-Li and {Liu}, Tie and {Evans}, Neal J., II and {Wang}, Ke and {Garay}, Guido and {Qin}, Sheng-Li and {Li}, Shanghuo and {Stutz}, Amelia and {Goldsmith}, Paul F. and {Liu}, Sheng-Yuan and {Tej}, Anandmayee and {Zhang}, Qizhou and {Juvela}, Mika and {Li}, Di and {Wang}, Jun-Zhi and {Bronfman}, Leonardo and {Ren}, Zhiyuan and {Wu}, Yue-Fang and {Kim}, Kee-Tae and {Lee}, Chang Won and {Tatematsu}, Ken'ichi and {Cunningham}, Maria R. and {Liu}, Xun-Chuan and {Wu}, Jing-Wen and {Hirota}, Tomoya and {Lee}, Jeong-Eun and {Li}, Pak-Shing and {Kang}, Sung-Ju and {Mardones}, Diego and {Ristorcelli}, Isabelle and {Zhang}, Yong and {Luo}, Qiu-Yi and {Toth}, L. Viktor and {Yi}, Hee-weon and {Yun}, Hyeong-Sik and {Peng}, Ya-Ping and {Li}, Juan and {Zhu}, Feng-Yao and {Shen}, Zhi-Qiang and {Baug}, Tapas and {Dewangan}, L.~K. and {Chakali}, Eswaraiah and {Liu}, Rong and {Xu}, Feng-Wei and {Wang}, Yu and {Zhang}, Chao and {Li}, Jinzeng and {Zhang}, Chao and {Zhou}, Jianwen and {Tang}, Mengyao and {Xue}, Qiaowei and {Issac}, Namitha and {Soam}, Archana and {{\'A}lvarez-Guti{\'e}rrez}, Rodrigo H.},
        title = "{ATOMS: ALMA three-millimeter observations of massive star-forming regions - III. Catalogues of candidate hot molecular cores and hyper/ultra compact H II regions}",
      journal = {\mnras},
         year = 2021,
        month = aug,
       volume = {505},
       number = {2},
        pages = {2801-2818},
          doi = {10.1093/mnras/stab1352},
archivePrefix = {arXiv},
       eprint = {2105.03554},
 primaryClass = {astro-ph.GA},
       adsurl = {https://ui.adsabs.harvard.edu/abs/2021MNRAS.505.2801L}
}

@ARTICLE{2022MNRAS.511.3463Q,
       author = {{Qin}, Sheng-Li and {Liu}, Tie and {Liu}, Xunchuan and {Goldsmith}, Paul F. and {Li}, Di and {Zhang}, Qizhou and {Liu}, Hong-Li and {Wu}, Yuefang and {Bronfman}, Leonardo and {Juvela}, Mika and {Lee}, Chang Won and {Garay}, Guido and {Zhang}, Yong and {He}, Jinhua and {Hsu}, Shih-Ying and {Shen}, Zhi-Qiang and {Lee}, Jeong-Eun and {Wang}, Ke and {Tang}, Ningyu and {Tang}, Mengyao and {Zhang}, Chao and {Yue}, Yinghua and {Xue}, Qiaowei and {Li}, Shanghuo and {Peng}, Yaping and {Dutta}, Somnath and {Ge}, Jixing and {Xu}, Fengwei and {Chen}, Long-Fei and {Baug}, Tapas and {Dewangan}, Lokesh and {Tej}, Anandmayee},
        title = "{ATOMS: ALMA Three-millimeter Observations of Massive Star-forming regions - VIII. A search for hot cores by using C$_{2}$H$_{5}$CN, CH$_{3}$OCHO, and CH$_{3}$OH lines}",
      journal = {\mnras},
         year = 2022,
        month = apr,
       volume = {511},
       number = {3},
        pages = {3463-3476},
          doi = {10.1093/mnras/stac219},
archivePrefix = {arXiv},
       eprint = {2201.10044},
 primaryClass = {astro-ph.GA},
       adsurl = {https://ui.adsabs.harvard.edu/abs/2022MNRAS.511.3463Q}
}

@ARTICLE{2023MNRAS.520.3259X,
       author = {{Xu}, Fengwei and {Wang}, Ke and {Liu}, Tie and {Goldsmith}, Paul F. and {Zhang}, Qizhou and {Juvela}, Mika and {Liu}, Hong-Li and {Qin}, Sheng-Li and {Li}, Guang-Xing and {Tej}, Anandmayee and {Garay}, Guido and {Bronfman}, Leonardo and {Li}, Shanghuo and {Wu}, Yue-Fang and {G{\'o}mez}, Gilberto C. and {V{\'a}zquez-Semadeni}, Enrique and {Tatematsu}, Ken'ichi and {Ren}, Zhiyuan and {Zhang}, Yong and {Toth}, L. Viktor and {Liu}, Xunchuan and {Yue}, Nannan and {Zhang}, Siju and {Baug}, Tapas and {Issac}, Namitha and {Stutz}, Amelia M. and {Liu}, Meizhu and {Fuller}, Gary A. and {Tang}, Mengyao and {Zhang}, Chao and {Dewangan}, Lokesh and {Lee}, Chang Won and {Zhou}, Jianwen and {Xie}, Jinjin and {Jiao}, Wenyu and {Wang}, Chao and {Liu}, Rong and {Luo}, Qiuyi and {Soam}, Archana and {Eswaraiah}, Chakali},
        title = "{ATOMS: ALMA Three-millimeter Observations of Massive Star-forming regions - XV. Steady accretion from global collapse to core feeding in massive hub-filament system SDC335}",
      journal = {\mnras},
         year = 2023,
        month = apr,
       volume = {520},
       number = {3},
        pages = {3259-3285},
          doi = {10.1093/mnras/stad012},
archivePrefix = {arXiv},
       eprint = {2301.01895},
 primaryClass = {astro-ph.GA},
       adsurl = {https://ui.adsabs.harvard.edu/abs/2023MNRAS.520.3259X}
}

@ARTICLE{2024RAA....24b5009L,
       author = {{Liu}, Xunchuan and {Liu}, Tie and {Zhu}, Lei and {Garay}, Guido and {Liu}, Hong-Li and {Goldsmith}, Paul and {Evans}, Neal and {Kim}, Kee-Tae and {Liu}, Sheng-Yuan and {Xu}, Fengwei and {Lu}, Xing and {Tej}, Anandmayee and {Mai}, Xiaofeng and {Bronfman}, Leonardo and {Li}, Shanghuo and {Mardones}, Diego and {Stutz}, Amelia and {Tatematsu}, Ken'ichi and {Wang}, Ke and {Zhang}, Qizhou and {Qin}, Sheng-Li and {Zhou}, Jianwen and {Luo}, Qiuyi and {Zhang}, Siju and {Cheng}, Yu and {He}, Jinhua and {Gu}, Qilao and {Li}, Ziyang and {Zhang}, Zhenying and {Zhang}, Suinan and {Saha}, Anindya and {Dewangan}, Lokesh and {Sanhueza}, Patricio and {Shen}, Zhiqiang},
        title = "{The ALMA-QUARKS Survey. I. Survey Description and Data Reduction}",
      journal = {Research in Astronomy and Astrophysics},
         year = 2024,
        month = feb,
       volume = {24},
       number = {2},
          eid = {025009},
        pages = {025009},
          doi = {10.1088/1674-4527/ad0d5c},
archivePrefix = {arXiv},
       eprint = {2311.08651},
 primaryClass = {astro-ph.GA},
       adsurl = {https://ui.adsabs.harvard.edu/abs/2024RAA....24b5009L}
}

@ARTICLE{Xu2024Quarks2,
       author = {{Xu}, Fengwei and {Wang}, Ke and {Liu}, Tie and {Zhu}, Lei and {Garay}, Guido and {Liu}, Xunchuan and {Goldsmith}, Paul and {Zhang}, Qizhou and {Sanhueza}, Patricio and {Qin}, Shengli and {He}, Jinhua and {Juvela}, Mika and {Tej}, Anandmayee and {Liu}, Hongli and {Li}, Shanghuo and {Morii}, Kaho and {Zhang}, Siju and {Zhou}, Jianwen and {Stutz}, Amelia and {Evans}, Neal J. and {Kim}, Kee-Tae and {Liu}, Shengyuan and {Mardones}, Diego and {Li}, Guangxing and {Bronfman}, Leonardo and {Tatematsu}, Ken'ichi and {Lee}, Chang Won and {Lu}, Xing and {Mai}, Xiaofeng and {Jiao}, Sihan and {Chibueze}, James O. and {Su}, Keyun and {T{\'o}th}, Viktor L.},
        title = "{The ALMA-QUARKS Survey. II. The ACA 1.3 mm Continuum Source Catalog and the Assembly of Dense Gas in Massive Star-Forming Clumps}",
      journal = {Research in Astronomy and Astrophysics},
         year = 2024,
        month = jun,
       volume = {24},
       number = {6},
          eid = {065011},
        pages = {065011},
          doi = {10.1088/1674-4527/ad3dc3},
archivePrefix = {arXiv},
       eprint = {2404.02275},
 primaryClass = {astro-ph.GA},
       adsurl = {https://ui.adsabs.harvard.edu/abs/2024RAA....24f5011X}
}

@ARTICLE{Xu2024Assemble1,
       author = {{Xu}, Fengwei and {Wang}, Ke and {Liu}, Tie and {Tang}, Mengyao and {Evans}, II, Neal J. and {Palau}, Aina and {Morii}, Kaho and {He}, Jinhua and {Sanhueza}, Patricio and {Liu}, Hong-Li and {Stutz}, Amelia and {Zhang}, Qizhou and {Chen}, Xi and {Li}, Pak Shing and {G{\'o}mez}, Gilberto C. and {V{\'a}zquez-Semadeni}, Enrique and {Li}, Shanghuo and {Mai}, Xiaofeng and {Lu}, Xing and {Liu}, Meizhu and {Chen}, Li and {Li}, Chuanshou and {Shi}, Hongqiong and {Ren}, Zhiyuan and {Li}, Di and {Garay}, Guido and {Bronfman}, Leonardo and {Dewangan}, Lokesh and {Juvela}, Mika and {Lee}, Chang Won and {Zhang}, S. and {Yue}, Nannan and {Wang}, Chao and {Ge}, Yifei and {Jiao}, Wenyu and {Luo}, Qiuyi and {Zhou}, J. -W. and {Tatematsu}, Ken'ichi and {Chibueze}, James O. and {Su}, Keyun and {Sun}, Shenglan and {Ristorcelli}, I. and {Toth}, L. Viktor},
        title = "{The ALMA Survey of Star Formation and Evolution in Massive Protoclusters with Blue Profiles (ASSEMBLE): Core Growth, Cluster Contraction, and Primordial Mass Segregation}",
      journal = {\apjs},
         year = 2024,
        month = jan,
       volume = {270},
       number = {1},
          eid = {9},
        pages = {9},
          doi = {10.3847/1538-4365/acfee5},
archivePrefix = {arXiv},
       eprint = {2309.14684},
 primaryClass = {astro-ph.GA},
       adsurl = {https://ui.adsabs.harvard.edu/abs/2024ApJS..270....9X}
}

@article{Hoque_2025,
doi = {10.3847/1538-4357/add928},
url = {https://dx.doi.org/10.3847/1538-4357/add928},
year = {2025},
month = {jul},
publisher = {The American Astronomical Society},
volume = {987},
number = {2},
pages = {197},
author = {Hoque, Ariful and Baug, Tapas and Dewangan, Lokesh K. and Juvela, Mika and Tej, Anandmayee and Goldsmith, Paul F. and García, Pablo and Stutz, Amelia M. and Liu, Tie and Lee, Chang Won and Xu, Fengwei and Sanhueza, Patricio and Bhadari, N. K. and Tatematsu, K. and Liu, Xunchuan and Liu, Hong-Li and Zhang, Yong and Tang, Xindi and Garay, Guido and Wang, Ke and Zhang, Siju and Tóth, L. Viktor and Nazeer, Hafiz and Hwang, Jihye and Gorai, Prasanta and Bronfman, Leonardo and Das, Swagat Ranjan and Sinha, Tirthendu},
title = {The ALMA-ATOMS Survey: Exploring Protostellar Outflows in HC3N},
journal = {The Astrophysical Journal}
}

@ARTICLE{2025A&A...697A.190L,
       author = {{Li}, Zi-Yang and {Liu}, Xunchuan and {Liu}, Tie and {Qin}, Sheng-Li and {Goldsmith}, Paul F. and {Garc{\'\i}a}, Pablo and {Peng}, Yaping and {Chen}, Li and {Jiao}, Yunfan and {Kou}, Zhiping and {Li}, Chuanshou and {Zou}, Jiahang and {Tang}, Mengyao and {Li}, Shanghuo and {Liu}, Meizhu and {Garay}, Guido and {Xu}, Fengwei and {Jiao}, Wenyu and {Luo}, Qiu-Yi and {Zhang}, Suinan and {Gu}, Qi-Lao and {Mai}, Xiaofeng and {Zhang}, Yan-Kun and {Weng}, Jixiang and {Lee}, Chang Won and {Sanhueza}, Patricio and {Dib}, Sami and {Das}, Swagat R. and {Tang}, Xindi and {Bronfman}, Leonardo and {Gorai}, Prasanta and {Tatematsu}, Ken'ichi and {Liu}, Hong-Li and {Yang}, Dongting and {Zhang}, Zhenying and {Shen}, Xianjin},
        title = "{The ALMA-ATOMS survey: A sample of weak hot core candidates identified through line stacking}",
      journal = {\aap},
         year = 2025,
        month = may,
       volume = {697},
          eid = {A190},
        pages = {A190},
          doi = {10.1051/0004-6361/202452762},
archivePrefix = {arXiv},
       eprint = {2504.06802},
 primaryClass = {astro-ph.GA},
       adsurl = {https://ui.adsabs.harvard.edu/abs/2025A&A...697A.190L}
}

@ARTICLE{2025ApJS..280...33Y,
       author = {{Yang}, Dongting and {Liu}, Hong-Li and {Liu}, Tie and {Liu}, Xunchuan and {Xu}, Fengwei and {Qin}, Sheng-Li and {Tej}, Anandmayee and {Garay}, Guido and {Zhu}, Lei and {Mai}, Xiaofeng and {Jiao}, Wenyu and {Zhang}, Siju and {Dib}, Sami and {Stutz}, Amelia M. and {Palau}, Aina and {Sanhueza}, Patricio and {Zavagno}, Annie and {Yang}, A.~Y. and {Tang}, Xindi and {Tang}, Mengyao and {Zhang}, Yichen and {Garc{\'\i}a}, Pablo and {Zhang}, Tianwei and {Saha}, Anindya and {Li}, Shanghuo and {Goldsmith}, Paul F. and {Bronfman}, Leonardo and {Lee}, Chang Won and {Taniguchi}, Kotomi and {Das}, Swagat Ranjan and {Gorai}, Prasanta and {Hoque}, Ariful and {Chen}, Li and {Kou}, Zhiping and {Zhou}, Jianjun and {Zhang}, Yankun and {T{\'o}th}, L. Viktor and {Baug}, Tapas and {Shen}, Xianjin and {Li}, Chuanshou and {Zou}, Jiahang and {Das}, Ankan and {Nazeer}, Hafiz and {Dewangan}, L.~K. and {Hwang}, Jihye and {Chibueze}, James O.},
        title = "{The ALMA-QUARKS Survey. III. Clump-to-core Fragmentation and Searches for High-mass Starless Cores}",
      journal = {\apjs},
         year = 2025,
        month = sep,
       volume = {280},
       number = {1},
          eid = {33},
        pages = {33},
          doi = {10.3847/1538-4365/adf847},
archivePrefix = {arXiv},
       eprint = {2508.03229},
 primaryClass = {astro-ph.GA},
       adsurl = {https://ui.adsabs.harvard.edu/abs/2025ApJS..280...33Y}
}

@ARTICLE{2021ApJ...909..177L,
       author = {{Lu}, Xing and {Li}, Shanghuo and {Ginsburg}, Adam and {Longmore}, Steven N. and {Kruijssen}, J.~M. Diederik and {Walker}, Daniel L. and {Feng}, Siyi and {Zhang}, Qizhou and {Battersby}, Cara and {Pillai}, Thushara and {Mills}, Elisabeth A.~C. and {Kauffmann}, Jens and {Cheng}, Yu and {Inutsuka}, Shu-ichiro},
        title = "{ALMA Observations of Massive Clouds in the Central Molecular Zone: Ubiquitous Protostellar Outflows}",
      journal = {\apj},
         year = 2021,
        month = mar,
       volume = {909},
       number = {2},
          eid = {177},
        pages = {177},
          doi = {10.3847/1538-4357/abde3c},
archivePrefix = {arXiv},
       eprint = {2101.07925},
 primaryClass = {astro-ph.GA},
       adsurl = {https://ui.adsabs.harvard.edu/abs/2021ApJ...909..177L}
}

@ARTICLE{2020ApJ...896...37F,
       author = {{Feng}, S. and {Codella}, C. and {Ceccarelli}, C. and {Caselli}, P. and {Lopez-Sepulcre}, A. and {Neri}, R. and {Fontani}, F. and {Podio}, L. and {Lefloch}, B. and {Liu}, H.~B. and {Bachiller}, R. and {Viti}, S.},
        title = "{Seeds of Life in Space (SOLIS). IX. Chemical Segregation of SO$_{2}$ and SO toward the Low-mass Protostellar Shocked Region of L1157}",
      journal = {\apj},
         year = 2020,
        month = jun,
       volume = {896},
       number = {1},
          eid = {37},
        pages = {37},
          doi = {10.3847/1538-4357/ab8813},
archivePrefix = {arXiv},
       eprint = {2005.04629},
 primaryClass = {astro-ph.GA},
       adsurl = {https://ui.adsabs.harvard.edu/abs/2020ApJ...896...37F}
}

@ARTICLE{2025MNRAS.539.2380B,
       author = {{Bouvier}, M. and {Giani}, L. and {Chahine}, L. and {L{\'o}pez-Sepulcre}, A. and {Ceccarelli}, C. and {Podio}, L.},
        title = "{Discovery of an intriguing chemically rich outflow in the OMC-2/3 filament}",
      journal = {\mnras},
         year = 2025,
        month = may,
       volume = {539},
       number = {3},
        pages = {2380-2399},
          doi = {10.1093/mnras/staf631},
archivePrefix = {arXiv},
       eprint = {2504.10133},
 primaryClass = {astro-ph.GA},
       adsurl = {https://ui.adsabs.harvard.edu/abs/2025MNRAS.539.2380B}
}

@ARTICLE{2021A&A...655A..65T,
       author = {{Tychoniec}, {\L}ukasz and {van Dishoeck}, Ewine F. and {van't Hoff}, Merel L.~R. and {van Gelder}, Martijn L. and {Tabone}, Beno{\^\i}t and {Chen}, Yuan and {Harsono}, Daniel and {Hull}, Charles L.~H. and {Hogerheijde}, Michiel R. and {Murillo}, Nadia M. and {Tobin}, John J.},
        title = "{Which molecule traces what: Chemical diagnostics of protostellar sources}",
      journal = {\aap},
         year = 2021,
        month = nov,
       volume = {655},
          eid = {A65},
        pages = {A65},
          doi = {10.1051/0004-6361/202140692},
archivePrefix = {arXiv},
       eprint = {2107.03696},
 primaryClass = {astro-ph.SR},
       adsurl = {https://ui.adsabs.harvard.edu/abs/2021A&A...655A..65T}
}

@ARTICLE{2018ARA&A..56...41M,
       author = {{Motte}, Fr{\'e}d{\'e}rique and {Bontemps}, Sylvain and {Louvet}, Fabien},
        title = "{High-Mass Star and Massive Cluster Formation in the Milky Way}",
      journal = {\araa},
         year = 2018,
        month = sep,
       volume = {56},
        pages = {41-82},
          doi = {10.1146/annurev-astro-091916-055235},
archivePrefix = {arXiv},
       eprint = {1706.00118},
 primaryClass = {astro-ph.GA},
       adsurl = {https://ui.adsabs.harvard.edu/abs/2018ARA&A..56...41M}
}

@ARTICLE{2021A&A...645A.142A,
       author = {{Avison}, A. and {Fuller}, G.~A. and {Peretto}, N. and {Duarte-Cabral}, A. and {Rosen}, A.~L. and {Traficante}, A. and {Pineda}, J.~E. and {G{\"u}sten}, R. and {Cunningham}, N.},
        title = "{Continuity of accretion from clumps to Class 0 high-mass protostars in SDC335}",
      journal = {\aap},
         year = 2021,
        month = jan,
       volume = {645},
          eid = {A142},
        pages = {A142},
          doi = {10.1051/0004-6361/201936043},
archivePrefix = {arXiv},
       eprint = {2012.08948},
 primaryClass = {astro-ph.GA},
       adsurl = {https://ui.adsabs.harvard.edu/abs/2021A&A...645A.142A}
}

@ARTICLE{1997A&A...322..296C,
       author = {{Caselli}, P. and {Hartquist}, T.~W. and {Havnes}, O.},
        title = "{Grain-grain collisions and sputtering in oblique C-type shocks.}",
      journal = {\aap},
         year = 1997,
        month = jun,
       volume = {322},
        pages = {296-301},
       adsurl = {https://ui.adsabs.harvard.edu/abs/1997A&A...322..296C}
}

@ARTICLE{2020A&A...634A..17J,
       author = {{James}, T.~A. and {Viti}, S. and {Holdship}, J. and {Jim{\'e}nez-Serra}, I.},
        title = "{Tracing shock type with chemical diagnostics. An application to L1157}",
      journal = {\aap},
         year = 2020,
        month = feb,
       volume = {634},
          eid = {A17},
        pages = {A17},
          doi = {10.1051/0004-6361/201936536},
archivePrefix = {arXiv},
       eprint = {1912.03721},
 primaryClass = {astro-ph.SR},
       adsurl = {https://ui.adsabs.harvard.edu/abs/2020A&A...634A..17J}
}

@ARTICLE{2017ApJ...839...47M,
       author = {{Miura}, Hitoshi and {Yamamoto}, Tetsuo and {Nomura}, Hideko and {Nakamoto}, Taishi and {Tanaka}, Kyoko K. and {Tanaka}, Hidekazu and {Nagasawa}, Makiko},
        title = "{Comprehensive Study of Thermal Desorption of Grain-surface Species by Accretion Shocks around Protostars}",
      journal = {\apj},
         year = 2017,
        month = apr,
       volume = {839},
       number = {1},
          eid = {47},
        pages = {47},
          doi = {10.3847/1538-4357/aa67df},
       adsurl = {https://ui.adsabs.harvard.edu/abs/2017ApJ...839...47M}
}

@ARTICLE{2011ApJ...740...14O,
       author = {{{\"O}berg}, Karin I. and {van der Marel}, Nienke and {Kristensen}, Lars E. and {van Dishoeck}, Ewine F.},
        title = "{Complex Molecules toward Low-mass Protostars: The Serpens Core}",
      journal = {\apj},
         year = 2011,
        month = oct,
       volume = {740},
       number = {1},
          eid = {14},
        pages = {14},
          doi = {10.1088/0004-637X/740/1/14},
archivePrefix = {arXiv},
       eprint = {1107.5824},
 primaryClass = {astro-ph.GA},
       adsurl = {https://ui.adsabs.harvard.edu/abs/2011ApJ...740...14O}
}

@ARTICLE{2014MNRAS.445..151M,
       author = {{Mendoza}, Edgar and {Lefloch}, B. and {L{\'o}pez-Sepulcre}, A. and {Ceccarelli}, C. and {Codella}, C. and {Boechat-Roberty}, H.~M. and {Bachiller}, R.},
        title = "{Molecules with a peptide link in protostellar shocks: a comprehensive study of L1157}",
      journal = {\mnras},
         year = 2014,
        month = nov,
       volume = {445},
       number = {1},
        pages = {151-161},
          doi = {10.1093/mnras/stu1718},
archivePrefix = {arXiv},
       eprint = {1408.4857},
 primaryClass = {astro-ph.SR},
       adsurl = {https://ui.adsabs.harvard.edu/abs/2014MNRAS.445..151M}
}

@ARTICLE{2017MNRAS.469L..73L,
       author = {{Lefloch}, Bertrand and {Ceccarelli}, C. and {Codella}, C. and {Favre}, C. and {Podio}, L. and {Vastel}, C. and {Viti}, S. and {Bachiller}, R.},
        title = "{L1157-B1, a factory of complex organic molecules in a solar-type star-forming region}",
      journal = {\mnras},
         year = 2017,
        month = jul,
       volume = {469},
       number = {1},
        pages = {L73-L77},
          doi = {10.1093/mnrasl/slx050},
archivePrefix = {arXiv},
       eprint = {1704.04646},
 primaryClass = {astro-ph.GA},
       adsurl = {https://ui.adsabs.harvard.edu/abs/2017MNRAS.469L..73L}
}

@ARTICLE{2024ApJ...976...29H,
       author = {{Hsu}, Shih-Ying and {Lee}, Chin-Fei and {Liu}, Sheng-Yuan and {Johnstone}, Doug and {Liu}, Tie and {Takahashi}, Satoko and {Bronfman}, Leonardo and {Chen}, Huei-Ru Vivien and {Dutta}, Somnath and {Eden}, David J. and {Evans}, II, Neal J. and {Hirano}, Naomi and {Juvela}, Mika and {Kuan}, Yi-Jehng and {Kwon}, Woojin and {Lee}, Chang Won and {Lee}, Jeong-Eun and {Li}, Shanghuo and {Liu}, Chun-Fan and {Liu}, Xunchuan and {Luo}, Qiuyi and {Qin}, Sheng-Li and {Sahu}, Dipen and {Sanhueza}, Patricio and {Shang}, Hsien and {Tatematsu}, Kenichi and {Yang}, Yao-Lun},
        title = "{ALMASOP. The Localized and Chemically Rich Features near the Bases of the Protostellar Jet in HOPS 87}",
      journal = {\apj},
         year = 2024,
        month = nov,
       volume = {976},
       number = {1},
          eid = {29},
        pages = {29},
          doi = {10.3847/1538-4357/ad7e25},
archivePrefix = {arXiv},
       eprint = {2409.14445},
 primaryClass = {astro-ph.SR},
       adsurl = {https://ui.adsabs.harvard.edu/abs/2024ApJ...976...29H}
}

@ARTICLE{1989ApJ...342..306H,
       author = {{Hollenbach}, David and {McKee}, Christopher F.},
        title = "{Molecule Formation and Infrared Emission in Fast Interstellar Shocks. III. Results for J Shocks in Molecular Clouds}",
      journal = {\apj},
         year = 1989,
        month = jul,
       volume = {342},
        pages = {306},
          doi = {10.1086/167595},
       adsurl = {https://ui.adsabs.harvard.edu/abs/1989ApJ...342..306H}
}

@INPROCEEDINGS{1994ASPC...58..332F,
       author = {{Flower}, David},
        title = "{Chemical Processes in Interstellar Shocks}",
    booktitle = {The First Symposium on the Infrared Cirrus and Diffuse Interstellar Clouds},
         year = 1994,
       editor = {{Cutri}, Roc M. and {Latter}, William B.},
       series = {Astronomical Society of the Pacific Conference Series},
       volume = {58},
        month = jan,
        pages = {332},
       adsurl = {https://ui.adsabs.harvard.edu/abs/1994ASPC...58..332F}
}

@ARTICLE{1995Ap&SS.233..111D,
       author = {{Draine}, B.~T.},
        title = "{Grain Destruction in Interstellar Shock Waves}",
      journal = {\apss},
         year = 1995,
        month = nov,
       volume = {233},
       number = {1-2},
        pages = {111-123},
          doi = {10.1007/BF00627339},
archivePrefix = {arXiv},
       eprint = {astro-ph/9508066},
 primaryClass = {astro-ph},
       adsurl = {https://ui.adsabs.harvard.edu/abs/1995Ap&SS.233..111D}
}

@INPROCEEDINGS{2011IAUS..280...88T,
       author = {{Tafalla}, Mario and {Bachiller}, Rafael},
        title = "{Molecules in Bipolar Outflows}",
    booktitle = {The Molecular Universe},
         year = 2011,
       editor = {{Cernicharo}, Jos{\'e} and {Bachiller}, Rafael},
       series = {IAU Symposium},
       volume = {280},
        month = dec,
        pages = {88-102},
          doi = {10.1017/S1743921311024896},
archivePrefix = {arXiv},
       eprint = {1203.2181},
 primaryClass = {astro-ph.GA},
       adsurl = {https://ui.adsabs.harvard.edu/abs/2011IAUS..280...88T}
}

@ARTICLE{2010A&A...522A..91T,
       author = {{Tafalla}, M. and {Santiago-Garc{\'\i}a}, J. and {Hacar}, A. and {Bachiller}, R.},
        title = "{A molecular survey of outflow gas: velocity-dependent shock chemistry and the peculiar composition of the EHV gas}",
      journal = {\aap},
         year = 2010,
        month = nov,
       volume = {522},
          eid = {A91},
        pages = {A91},
          doi = {10.1051/0004-6361/201015158},
archivePrefix = {arXiv},
       eprint = {1007.4549},
 primaryClass = {astro-ph.GA},
       adsurl = {https://ui.adsabs.harvard.edu/abs/2010A&A...522A..91T}
}

@ARTICLE{2017A&A...605L...3C,
       author = {{Codella}, C. and {Ceccarelli}, C. and {Caselli}, P. and {Balucani}, N. and {Barone}, V. and {Fontani}, F. and {Lefloch}, B. and {Podio}, L. and {Viti}, S. and {Feng}, S. and {Bachiller}, R. and {Bianchi}, E. and {Dulieu}, F. and {Jim{\'e}nez-Serra}, I. and {Holdship}, J. and {Neri}, R. and {Pineda}, J.~E. and {Pon}, A. and {Sims}, I. and {Spezzano}, S. and {Vasyunin}, A.~I. and {Alves}, F. and {Bizzocchi}, L. and {Bottinelli}, S. and {Caux}, E. and {Chac{\'o}n-Tanarro}, A. and {Choudhury}, R. and {Coutens}, A. and {Favre}, C. and {Hily-Blant}, P. and {Kahane}, C. and {Jaber Al-Edhari}, A. and {Laas}, J. and {L{\'o}pez-Sepulcre}, A. and {Ospina}, J. and {Oya}, Y. and {Punanova}, A. and {Puzzarini}, C. and {Quenard}, D. and {Rimola}, A. and {Sakai}, N. and {Skouteris}, D. and {Taquet}, V. and {Testi}, L. and {Theul{\'e}}, P. and {Ugliengo}, P. and {Vastel}, C. and {Vazart}, F. and {Wiesenfeld}, L. and {Yamamoto}, S.},
        title = "{Seeds of Life in Space (SOLIS). II. Formamide in protostellar shocks: Evidence for gas-phase formation}",
      journal = {\aap},
         year = 2017,
        month = sep,
       volume = {605},
          eid = {L3},
        pages = {L3},
          doi = {10.1051/0004-6361/201731249},
archivePrefix = {arXiv},
       eprint = {1708.04663},
 primaryClass = {astro-ph.EP},
       adsurl = {https://ui.adsabs.harvard.edu/abs/2017A&A...605L...3C}
}

@ARTICLE{2024MNRAS.531.2653C,
       author = {{Chahine}, Layal and {Ceccarelli}, Cecilia and {De Simone}, Marta and {Chandler}, Claire J. and {Codella}, Claudio and {Podio}, Linda and {L{\'o}pez-Sepulcre}, Ana and {Sakai}, Nami and {Loinard}, Laurent and {Bouvier}, Mathilde and {Caselli}, Paola and {Vastel}, Charlotte and {Bianchi}, Eleonora and {Cuello}, Nicol{\'a}s and {Fontani}, Francesco and {Johnstone}, Doug and {Sabatini}, Giovanni and {Hanawa}, Tomoyuki and {Zhang}, Ziwei E. and {Aikawa}, Yuri and {Busquet}, Gemma and {Caux}, Emmanuel and {Dur{\'a}n}, Aurore and {Herbst}, Eric and {M{\'e}nard}, Fran{\c{c}}ois and {Segura-Cox}, Dominique and {Svoboda}, Brian and {Balucani}, Nadia and {Charnley}, Steven and {Dulieu}, Fran{\c{c}}ois and {Evans}, Lucy and {Fedele}, Davide and {Feng}, Siyi and {Hama}, Tetsuya and {Hirota}, Tomoya and {Isella}, Andrea and {J{\'\i}menez-Serra}, Izaskun and {Lefloch}, Bertrand and {Maud}, Luke T. and {Maureira}, Mar{\'\i}a Jos{\'e} and {Miotello}, Anna and {Moellenbrock}, George and {Nomura}, Hideko and {Oba}, Yasuhiro and {Ohashi}, Satoshi and {Okoda}, Yuki and {Oya}, Yoko and {Pineda}, Jaime and {Rimola}, Albert and {Sakai}, Takeshi and {Shirley}, Yancy and {Testi}, Leonardo and {Viti}, Serena and {Watanabe}, Naoki and {Watanabe}, Yoshimasa and {Zhang}, Yichen and {Yamamoto}, Satoshi},
        title = "{Multiple chemical tracers finally unveil the intricate NGC 1333 IRAS 4A outflow system. FAUST XVI}",
      journal = {\mnras},
         year = 2024,
        month = jun,
       volume = {531},
       number = {2},
        pages = {2653-2668},
          doi = {10.1093/mnras/stae1320},
archivePrefix = {arXiv},
       eprint = {2405.12735},
 primaryClass = {astro-ph.GA},
       adsurl = {https://ui.adsabs.harvard.edu/abs/2024MNRAS.531.2653C}
}

@ARTICLE{2021ApJ...909..199O,
       author = {{Olguin}, Fernando A. and {Sanhueza}, Patricio and {Guzm{\'a}n}, Andr{\'e}s E. and {Lu}, Xing and {Saigo}, Kazuya and {Zhang}, Qizhou and {Silva}, Andrea and {Chen}, Huei-Ru Vivien and {Li}, Shanghuo and {Ohashi}, Satoshi and {Nakamura}, Fumitaka and {Sakai}, Takeshi and {Wu}, Benjamin},
        title = "{Digging into the Interior of Hot Cores with ALMA (DIHCA). I. Dissecting the High-mass Star-forming Core G335.579-0.292 MM1}",
      journal = {\apj},
         year = 2021,
        month = mar,
       volume = {909},
       number = {2},
          eid = {199},
        pages = {199},
          doi = {10.3847/1538-4357/abde3f},
archivePrefix = {arXiv},
       eprint = {2101.08284},
 primaryClass = {astro-ph.GA},
       adsurl = {https://ui.adsabs.harvard.edu/abs/2021ApJ...909..199O}
}

@ARTICLE{2021A&A...648A..45P,
       author = {{Podio}, L. and {Tabone}, B. and {Codella}, C. and {Gueth}, F. and {Maury}, A. and {Cabrit}, S. and {Lefloch}, B. and {Maret}, S. and {Belloche}, A. and {Andr{\'e}}, P. and {Anderl}, S. and {Gaudel}, M. and {Testi}, L.},
        title = "{The CALYPSO IRAM-PdBI survey of jets from Class 0 protostars. Exploring whether jets are ubiquitous in young stars}",
      journal = {\aap},
         year = 2021,
        month = apr,
       volume = {648},
          eid = {A45},
        pages = {A45},
          doi = {10.1051/0004-6361/202038429},
archivePrefix = {arXiv},
       eprint = {2012.15379},
 primaryClass = {astro-ph.SR},
       adsurl = {https://ui.adsabs.harvard.edu/abs/2021A&A...648A..45P}
}

@ARTICLE{2020A&A...636A..60T,
       author = {{Tabone}, B. and {Godard}, B. and {Pineau des For{\^e}ts}, G. and {Cabrit}, S. and {van Dishoeck}, E.~F.},
        title = "{Molecule formation in dust-poor irradiated jets. I. Stationary disk winds}",
      journal = {\aap},
         year = 2020,
        month = apr,
       volume = {636},
          eid = {A60},
        pages = {A60},
          doi = {10.1051/0004-6361/201937383},
archivePrefix = {arXiv},
       eprint = {2003.01845},
 primaryClass = {astro-ph.GA},
       adsurl = {https://ui.adsabs.harvard.edu/abs/2020A&A...636A..60T}
}

@ARTICLE{2019A&A...632A.101T,
       author = {{Tychoniec}, {\L}ukasz and {Hull}, Charles L.~H. and {Kristensen}, Lars E. and {Tobin}, John J. and {Le Gouellec}, Valentin J.~M. and {van Dishoeck}, Ewine F.},
        title = "{Chemical and kinematic structure of extremely high-velocity molecular jets in the Serpens Main star-forming region}",
      journal = {\aap},
         year = 2019,
        month = dec,
       volume = {632},
          eid = {A101},
        pages = {A101},
          doi = {10.1051/0004-6361/201935409},
archivePrefix = {arXiv},
       eprint = {1910.07857},
 primaryClass = {astro-ph.SR},
       adsurl = {https://ui.adsabs.harvard.edu/abs/2019A&A...632A.101T}
}

@ARTICLE{2009A&A...503L..13B,
       author = {{Bruderer}, S. and {Benz}, A.~O. and {Bourke}, T.~L. and {Doty}, S.~D.},
        title = "{Evidence of warm and dense material along the outflow of a high-mass YSO}",
      journal = {\aap},
         year = 2009,
        month = aug,
       volume = {503},
       number = {2},
        pages = {L13-L16},
          doi = {10.1051/0004-6361/200912620},
       adsurl = {https://ui.adsabs.harvard.edu/abs/2009A&A...503L..13B}
}

@ARTICLE{2018A&A...615A..75V,
       author = {{Visser}, Ruud and {Bruderer}, Simon and {Cazzoletti}, Paolo and {Facchini}, Stefano and {Heays}, Alan N. and {van Dishoeck}, Ewine F.},
        title = "{Nitrogen isotope fractionation in protoplanetary disks}",
      journal = {\aap},
         year = 2018,
        month = jul,
       volume = {615},
          eid = {A75},
        pages = {A75},
          doi = {10.1051/0004-6361/201731898},
archivePrefix = {arXiv},
       eprint = {1802.02841},
 primaryClass = {astro-ph.SR},
       adsurl = {https://ui.adsabs.harvard.edu/abs/2018A&A...615A..75V}
}

@ARTICLE{1980ApJ...236..182H,
       author = {{Hartquist}, T.~W. and {Dalgarno}, A. and {Oppenheimer}, M.},
        title = "{Molecular diagnostics of interstellar shocks}",
      journal = {\apj},
         year = 1980,
        month = feb,
       volume = {236},
        pages = {182-188},
          doi = {10.1086/157731},
       adsurl = {https://ui.adsabs.harvard.edu/abs/1980ApJ...236..182H}
}

@BOOK{1993dca..book.....M,
       author = {{Millar}, T.~J. and {Williams}, D.~A.},
        title = "{Dust and chemistry in astronomy}",
         year = 1993,
       adsurl = {https://ui.adsabs.harvard.edu/abs/1993dca..book.....M}
}

@ARTICLE{1992A&A...265L..49G,
       author = {{Guilloteau}, S. and {Bachiller}, R. and {Fuente}, A. and {Lucas}, R.},
        title = "{First observations of young bipolar outflows with the IRAM interferometer : 2'' resolution SiO images of the molecular jet in L 1448.}",
      journal = {\aap},
         year = 1992,
        month = nov,
       volume = {265},
        pages = {L49-L52},
       adsurl = {https://ui.adsabs.harvard.edu/abs/1992A&A...265L..49G}
}

@ARTICLE{1997A&A...325..758D,
       author = {{Dutrey}, A. and {Guilloteau}, S. and {Bachiller}, R.},
        title = "{Successive SiO shocks along the L1448 jet axis.}",
      journal = {\aap},
         year = 1997,
        month = sep,
       volume = {325},
        pages = {758-768},
       adsurl = {https://ui.adsabs.harvard.edu/abs/1997A&A...325..758D}
}

@ARTICLE{1997A&A...321..293S,
       author = {{Schilke}, P. and {Walmsley}, C.~M. and {Pineau des Forets}, G. and {Flower}, D.~R.},
        title = "{SiO production in interstellar shocks.}",
      journal = {\aap},
         year = 1997,
        month = may,
       volume = {321},
        pages = {293-304},
       adsurl = {https://ui.adsabs.harvard.edu/abs/1997A&A...321..293S}
}

@ARTICLE{2008A&A...490..695G,
       author = {{Gusdorf}, A. and {Pineau Des For{\^e}ts}, G. and {Cabrit}, S. and {Flower}, D.~R.},
        title = "{SiO line emission from interstellar jets and outflows: silicon-containing mantles and non-stationary shock waves}",
      journal = {\aap},
         year = 2008,
        month = nov,
       volume = {490},
       number = {2},
        pages = {695-706},
          doi = {10.1051/0004-6361:200810443},
       adsurl = {https://ui.adsabs.harvard.edu/abs/2008A&A...490..695G}
}

@ARTICLE{2008A&A...482..809G,
       author = {{Gusdorf}, A. and {Cabrit}, S. and {Flower}, D.~R. and {Pineau Des For{\^e}ts}, G.},
        title = "{SiO line emission from C-type shock waves: interstellar jets and outflows}",
      journal = {\aap},
         year = 2008,
        month = may,
       volume = {482},
       number = {3},
        pages = {809-829},
          doi = {10.1051/0004-6361:20078900},
archivePrefix = {arXiv},
       eprint = {0803.2791},
 primaryClass = {astro-ph},
       adsurl = {https://ui.adsabs.harvard.edu/abs/2008A&A...482..809G}
}

@ARTICLE{2015A&A...581A..85P,
       author = {{Podio}, L. and {Codella}, C. and {Gueth}, F. and {Cabrit}, S. and {Bachiller}, R. and {Gusdorf}, A. and {Lee}, C. -F. and {Lefloch}, B. and {Leurini}, S. and {Nisini}, B. and {Tafalla}, M.},
        title = "{The jet and the disk of the HH 212 low-mass protostar imaged by ALMA: SO and SO$_{2}$ emission}",
      journal = {\aap},
         year = 2015,
        month = sep,
       volume = {581},
          eid = {A85},
        pages = {A85},
          doi = {10.1051/0004-6361/201525778},
archivePrefix = {arXiv},
       eprint = {1505.05919},
 primaryClass = {astro-ph.GA},
       adsurl = {https://ui.adsabs.harvard.edu/abs/2015A&A...581A..85P}
}

@ARTICLE{2014Natur.507...78S,
       author = {{Sakai}, Nami and {Sakai}, Takeshi and {Hirota}, Tomoya and {Watanabe}, Yoshimasa and {Ceccarelli}, Cecilia and {Kahane}, Claudine and {Bottinelli}, Sandrine and {Caux}, Emmanuel and {Demyk}, Karine and {Vastel}, Charlotte and {Coutens}, Audrey and {Taquet}, Vianney and {Ohashi}, Nagayoshi and {Takakuwa}, Shigehisa and {Yen}, Hsi-Wei and {Aikawa}, Yuri and {Yamamoto}, Satoshi},
        title = "{Change in the chemical composition of infalling gas forming a disk around a protostar}",
      journal = {\nat},
         year = 2014,
        month = mar,
       volume = {507},
       number = {7490},
        pages = {78-80},
          doi = {10.1038/nature13000},
       adsurl = {https://ui.adsabs.harvard.edu/abs/2014Natur.507...78S}
}

@ARTICLE{2004A&A...415.1021J,
       author = {{J{\o}rgensen}, J.~K. and {Hogerheijde}, M.~R. and {Blake}, G.~A. and {van Dishoeck}, E.~F. and {Mundy}, L.~G. and {Sch{\"o}ier}, F.~L.},
        title = "{The impact of shocks on the chemistry of molecular clouds. High resolution images of chemical differentiation along the NGC 1333-IRAS 2A outflow}",
      journal = {\aap},
         year = 2004,
        month = mar,
       volume = {415},
        pages = {1021-1037},
          doi = {10.1051/0004-6361:20034216},
archivePrefix = {arXiv},
       eprint = {astro-ph/0311132},
 primaryClass = {astro-ph},
       adsurl = {https://ui.adsabs.harvard.edu/abs/2004A&A...415.1021J}
}

@ARTICLE{2009A&A...507L..25C,
       author = {{Codella}, C. and {Benedettini}, M. and {Beltr{\'a}n}, M.~T. and {Gueth}, F. and {Viti}, S. and {Bachiller}, R. and {Tafalla}, M. and {Cabrit}, S. and {Fuente}, A. and {Lefloch}, B.},
        title = "{Methyl cyanide as tracer of bow shocks in L1157-B1}",
      journal = {\aap},
         year = 2009,
        month = nov,
       volume = {507},
       number = {2},
        pages = {L25-L28},
          doi = {10.1051/0004-6361/200913340},
       adsurl = {https://ui.adsabs.harvard.edu/abs/2009A&A...507L..25C}
}

@ARTICLE{2010A&A...518L.112C,
       author = {{Codella}, C. and {Lefloch}, B. and {Ceccarelli}, C. and {Cernicharo}, J. and {Caux}, E. and {Lorenzani}, A. and {Viti}, S. and {Hily-Blant}, P. and {Parise}, B. and {Maret}, S. and {Nisini}, B. and {Caselli}, P. and {Cabrit}, S. and {Pagani}, L. and {Benedettini}, M. and {Boogert}, A. and {Gueth}, F. and {Melnick}, G. and {Neufeld}, D. and {Pacheco}, S. and {Salez}, M. and {Schuster}, K. and {Bacmann}, A. and {Baudry}, A. and {Bell}, T. and {Bergin}, E.~A. and {Blake}, G. and {Bottinelli}, S. and {Castets}, A. and {Comito}, C. and {Coutens}, A. and {Crimier}, N. and {Dominik}, C. and {Demyk}, K. and {Encrenaz}, P. and {Falgarone}, E. and {Fuente}, A. and {Gerin}, M. and {Goldsmith}, P. and {Helmich}, F. and {Hennebelle}, P. and {Henning}, Th. and {Herbst}, E. and {Jacq}, T. and {Kahane}, C. and {Kama}, M. and {Klotz}, A. and {Langer}, W. and {Lis}, D. and {Lord}, S. and {Pearson}, J. and {Phillips}, T. and {Saraceno}, P. and {Schilke}, P. and {Tielens}, X. and {van der Tak}, F. and {van der Wiel}, M. and {Vastel}, C. and {Wakelam}, V. and {Walters}, A. and {Wyrowski}, F. and {Yorke}, H. and {Borys}, C. and {Delorme}, Y. and {Kramer}, C. and {Larsson}, B. and {Mehdi}, I. and {Ossenkopf}, V. and {Stutzki}, J.},
        title = "{The CHESS spectral survey of star forming regions: Peering into the protostellar shock L1157-B1. I. Shock chemical complexity}",
      journal = {\aap},
         year = 2010,
        month = jul,
       volume = {518},
          eid = {L112},
        pages = {L112},
          doi = {10.1051/0004-6361/201014582},
archivePrefix = {arXiv},
       eprint = {1006.1864},
 primaryClass = {astro-ph.GA},
       adsurl = {https://ui.adsabs.harvard.edu/abs/2010A&A...518L.112C}
}

@ARTICLE{2012A&A...541L..12B,
       author = {{Bacmann}, A. and {Taquet}, V. and {Faure}, A. and {Kahane}, C. and {Ceccarelli}, C.},
        title = "{Detection of complex organic molecules in a prestellar core: a new challenge for astrochemical models}",
      journal = {\aap},
         year = 2012,
        month = may,
       volume = {541},
          eid = {L12},
        pages = {L12},
          doi = {10.1051/0004-6361/201219207},
       adsurl = {https://ui.adsabs.harvard.edu/abs/2012A&A...541L..12B}
}

@ARTICLE{2020ApJ...891...73S,
       author = {{Scibelli}, Samantha and {Shirley}, Yancy},
        title = "{Prevalence of Complex Organic Molecules in Starless and Prestellar Cores within the Taurus Molecular Cloud}",
      journal = {\apj},
         year = 2020,
        month = mar,
       volume = {891},
       number = {1},
          eid = {73},
        pages = {73},
          doi = {10.3847/1538-4357/ab7375},
archivePrefix = {arXiv},
       eprint = {2002.02469},
 primaryClass = {astro-ph.GA},
       adsurl = {https://ui.adsabs.harvard.edu/abs/2020ApJ...891...73S}
}

@ARTICLE{2004ApJ...616..638W,
       author = {{Watanabe}, Naoki and {Nagaoka}, Akihiro and {Shiraki}, Takahiro and {Kouchi}, Akira},
        title = "{Hydrogenation of CO on Pure Solid CO and CO-H$_{2}$O Mixed Ice}",
      journal = {\apj},
         year = 2004,
        month = nov,
       volume = {616},
       number = {1},
        pages = {638-642},
          doi = {10.1086/424815},
       adsurl = {https://ui.adsabs.harvard.edu/abs/2004ApJ...616..638W}
}

@ARTICLE{2009A&A...504..891O,
       author = {{{\"O}berg}, K.~I. and {Garrod}, R.~T. and {van Dishoeck}, E.~F. and {Linnartz}, H.},
        title = "{Formation rates of complex organics in UV irradiated CH\_3OH-rich ices. I. Experiments}",
      journal = {\aap},
         year = 2009,
        month = sep,
       volume = {504},
       number = {3},
        pages = {891-913},
          doi = {10.1051/0004-6361/200912559},
archivePrefix = {arXiv},
       eprint = {0908.1169},
 primaryClass = {astro-ph.GA},
       adsurl = {https://ui.adsabs.harvard.edu/abs/2009A&A...504..891O}
}

@ARTICLE{2020A&A...639A..87V,
       author = {{van Gelder}, M.~L. and {Tabone}, B. and {Tychoniec}, {\L}. and {van Dishoeck}, E.~F. and {Beuther}, H. and {Boogert}, A.~C.~A. and {Caratti o Garatti}, A. and {Klaassen}, P.~D. and {Linnartz}, H. and {M{\"u}ller}, H.~S.~P. and {Taquet}, V.},
        title = "{Complex organic molecules in low-mass protostars on Solar System scales. I. Oxygen-bearing species}",
      journal = {\aap},
         year = 2020,
        month = jul,
       volume = {639},
          eid = {A87},
        pages = {A87},
          doi = {10.1051/0004-6361/202037758},
archivePrefix = {arXiv},
       eprint = {2005.06784},
 primaryClass = {astro-ph.SR},
       adsurl = {https://ui.adsabs.harvard.edu/abs/2020A&A...639A..87V}
}

@ARTICLE{2021A&A...650A.150N,
       author = {{Nazari}, P. and {van Gelder}, M.~L. and {van Dishoeck}, E.~F. and {Tabone}, B. and {van't Hoff}, M.~L.~R. and {Ligterink}, N.~F.~W. and {Beuther}, H. and {Boogert}, A.~C.~A. and {Caratti o Garatti}, A. and {Klaassen}, P.~D. and {Linnartz}, H. and {Taquet}, V. and {Tychoniec}, {\L}.},
        title = "{Complex organic molecules in low-mass protostars on Solar System scales. II. Nitrogen-bearing species}",
      journal = {\aap},
         year = 2021,
        month = jun,
       volume = {650},
          eid = {A150},
        pages = {A150},
          doi = {10.1051/0004-6361/202039996},
archivePrefix = {arXiv},
       eprint = {2104.03326},
 primaryClass = {astro-ph.GA},
       adsurl = {https://ui.adsabs.harvard.edu/abs/2021A&A...650A.150N}
}

@ARTICLE{2021MNRAS.508.2964A,
       author = {{Anderson}, Michael and {Peretto}, Nicolas and {Ragan}, Sarah E. and {Rigby}, Andrew J. and {Avison}, Adam and {Duarte-Cabral}, Ana and {Fuller}, Gary A. and {Shirley}, Yancy L. and {Traficante}, Alessio and {Williams}, Gwenllian M.},
        title = "{An ALMA study of hub-filament systems - I. On the clump mass concentration within the most massive cores}",
      journal = {\mnras},
         year = 2021,
        month = dec,
       volume = {508},
       number = {2},
        pages = {2964-2978},
          doi = {10.1093/mnras/stab2674},
archivePrefix = {arXiv},
       eprint = {2109.07489},
 primaryClass = {astro-ph.GA},
       adsurl = {https://ui.adsabs.harvard.edu/abs/2021MNRAS.508.2964A}
}
\bibliographystyle{aasjournalv7}
\clearpage


\appendix

\renewcommand\thefigure{\Alph{section}\arabic{figure}}
\renewcommand\thetable{\Alph{section}\arabic{table}}

\section{List of detected molecular lines along the jet} \label{sec:A}
\restartappendixnumbering

\begin{table}[!ht]
\caption{Identified Molecular Transitions in the 1.3 mm and 3 mm Bands}
\label{tab:transitions}
\centering
\footnotesize
\begin{tabular}{lcccr}
\hline\hline
\colhead{Species} & \colhead{Rest Freq.} & \colhead{Quantum No.} & \colhead{$E_u/k$} & \colhead{$\log A_{ij}$} \\
 & \colhead{(MHz)} & & \colhead{(K)} & \colhead{(s$^{-1}$)} \\
\hline
CO & 230538.0 & 2$-$1 & 16.60 & $-6.1605$ \\
\textsuperscript{13}CO & 220398.7 & 2$-$1 & 15.87 & $-6.2165$ \\
C\textsuperscript{18}O & 219560.4 & 2$-$1 & 15.81 & $-6.2210$ \\
\hline
SiO & 86847.0 & 2$-$1 & 6.25 & $-4.5335$ \\
SiO & 217104.9 & 5$-$4 & 31.26 & $-3.2843$ \\
\hline
CS & 97981.0 & 2$-$1 & 7.05 & $-4.7763$ \\
\textsuperscript{13}CS & 231220.6 & 5$-$4 & 22.19 & $-3.6008$ \\
SO & 99299.9 & $3_2$–$2_1$ & 9.23 & $-4.9488$ \\
SO & 100029.6 & $4_5$-$4_4$ & 38.57 & $-5.96577$ \\
SO & 219949.4 & $6_5$–$5_4$ & 34.98 & $-3.8745$ \\
OCS & 231060.9 & 19$-$18 & 110.9 & $-4.4463$ \\
\hline
CCH & 87316.9 & 1$-$0 & 4.19 & $-5.8161$ \\
HCN & 88631.6 & 1$-$0 & 4.25 & $-4.6171$ \\
H\textsuperscript{13}CN & 86339.9 & 1$-$0 & 4.14 & $-4.6526$ \\
DCN & 217238.5 & 3$-$2 & 20.85 & $-3.3396$ \\
HCO$^+$ & 89188.5 & 1$-$0 & 4.28 & $-4.3781$ \\
H\textsuperscript{13}CO$^+$ & 86754.3 & 1$-$0 & 4.16 & $-4.1635$ \\
\hline
HC$_3$N & 100076.4 & 11$-$10 & 28.82 & $-4.1114$ \\
HC$_3$N & 218324.7 & 24$-$23 & 130.98 & $-3.0830$ \\
HNCO & 219798.3 & $10_{0,10}$–$9_{0,9}$ & 58.02 & $-3.8329$ \\
\hline
H$_2$CO & 101333.0 & $6_{1,5}$–$6_{1,6}$ & 87.56 & $-5.8038$ \\
H$_2$CO & 218222.2 & $3_{0,3}$–$2_{0,2}$ & 20.96 & $-3.5501$ \\
H$_2$CO & 218475.6 & $3_{2,2}$–$2_{2,1}$ & 68.09 & $-3.8037$ \\
H$_2$CO & 218760.1 & $3_{2,1}$–$2_{2,0}$ & 68.11 & $-3.8021$ \\
H$_2^{13}$CO & 219908.5 & $3_{1,2}$–$2_{1,1}$ & 32.94 & $-3.5911$ \\
\hline
CH$_3$OH (A) & 97582.8 & $2_{-1,1}$–$1_{-1,0}$ & 21.56 & $-5.5807$ \\
CH$_3$OH (E) & 216945.5 & $5_{-1,4}$–$4_{-2,3}$ & 55.87 & $-4.9159$ \\
CH$_3$OH (E) & 218440.1 & $4_{-2,3}$–$3_{-1,2}$ & 45.46 & $-4.3292$ \\
CH$_3$OH (E) & 220078.6 & $8_{-0,8}$–$7_{-1,6}$ & 96.61 & $-4.5993$ \\
\hline
CH$_3$CN & 220730.3 & $12_2$–$11_2$ & 97.44 & $-3.0465$ \\
CH$_3$CN & 220743.0 & $12_1$–$11_1$ & 76.01 & $-3.0372$ \\
CH$_3$CN & 220747.3 & $12_0$–$11_0$ & 68.67 & $-3.0342$ \\
\hline
CH$_3$CHO (E) & 98863.3 & $5_{1,4}$–$4_{1,3}$ & 16.59 & $-4.5082$ \\
CH$_3$CHO (A) & 98900.9 & $5_{1,4}$–$4_{1,3}$ & 16.51 & $-4.5077$ \\
\hline
CH$_3$OCHO (A+E) & 216966 & $20-19$ & 111.5 & $-3.815$ \\
\hline\hline
\end{tabular}
\tablecomments{
\mf~and \mecn~are marginally detected.
}
\end{table}

Table~\ref{tab:transitions} summarizes all molecular transitions detected at the bright knots of the outflow. The corresponding spectra are shown in Fig.~\ref{fig:spectra-3} and \ref{fig:spectra-1.3}, overlaid with cumulative integrated intensity curves.


\begin{sidewaysfigure*}
  \centering
  \includegraphics[page=1,width=\textheight]{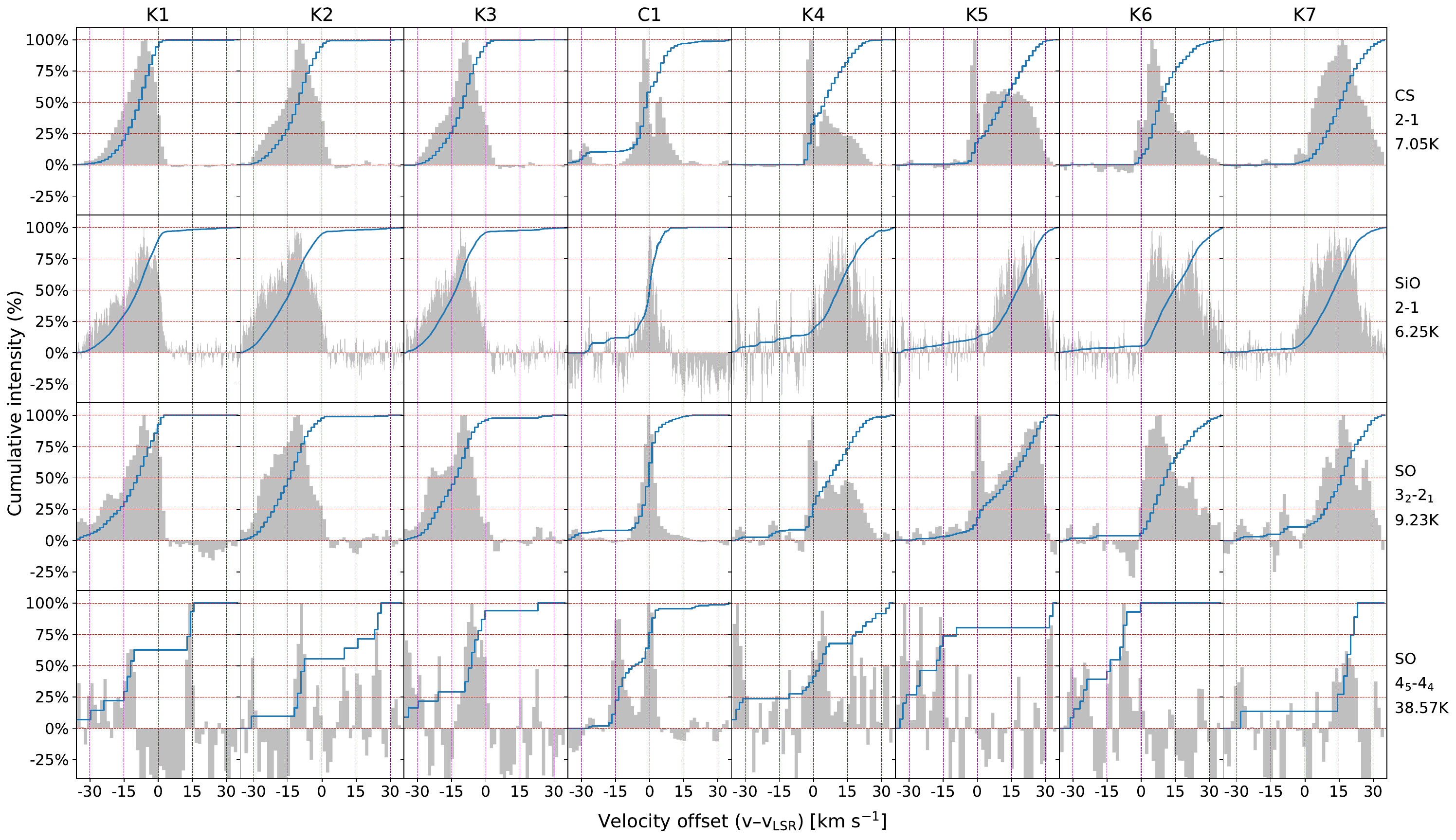}
  \caption{(a) Identified molecular transitions in the 3-mm Band.}
  \label{fig:spectra-3}
\end{sidewaysfigure*}

\begin{sidewaysfigure*}\ContinuedFloat
  \centering
  \includegraphics[page=2,width=\textheight]{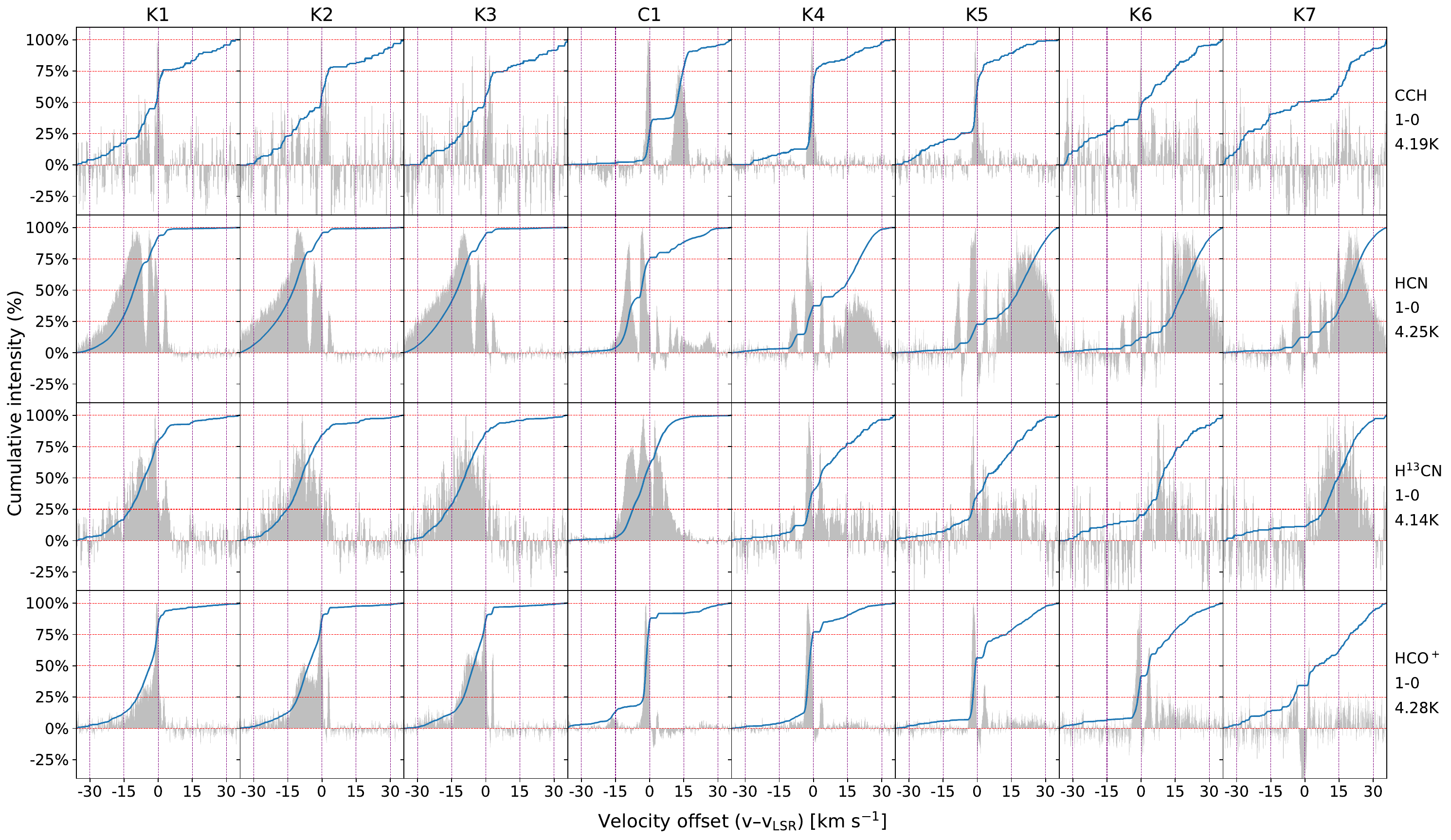}
  \caption{(b) Continued.}
\end{sidewaysfigure*}

\begin{sidewaysfigure*}\ContinuedFloat
  \centering
  \includegraphics[page=3,width=\textheight]{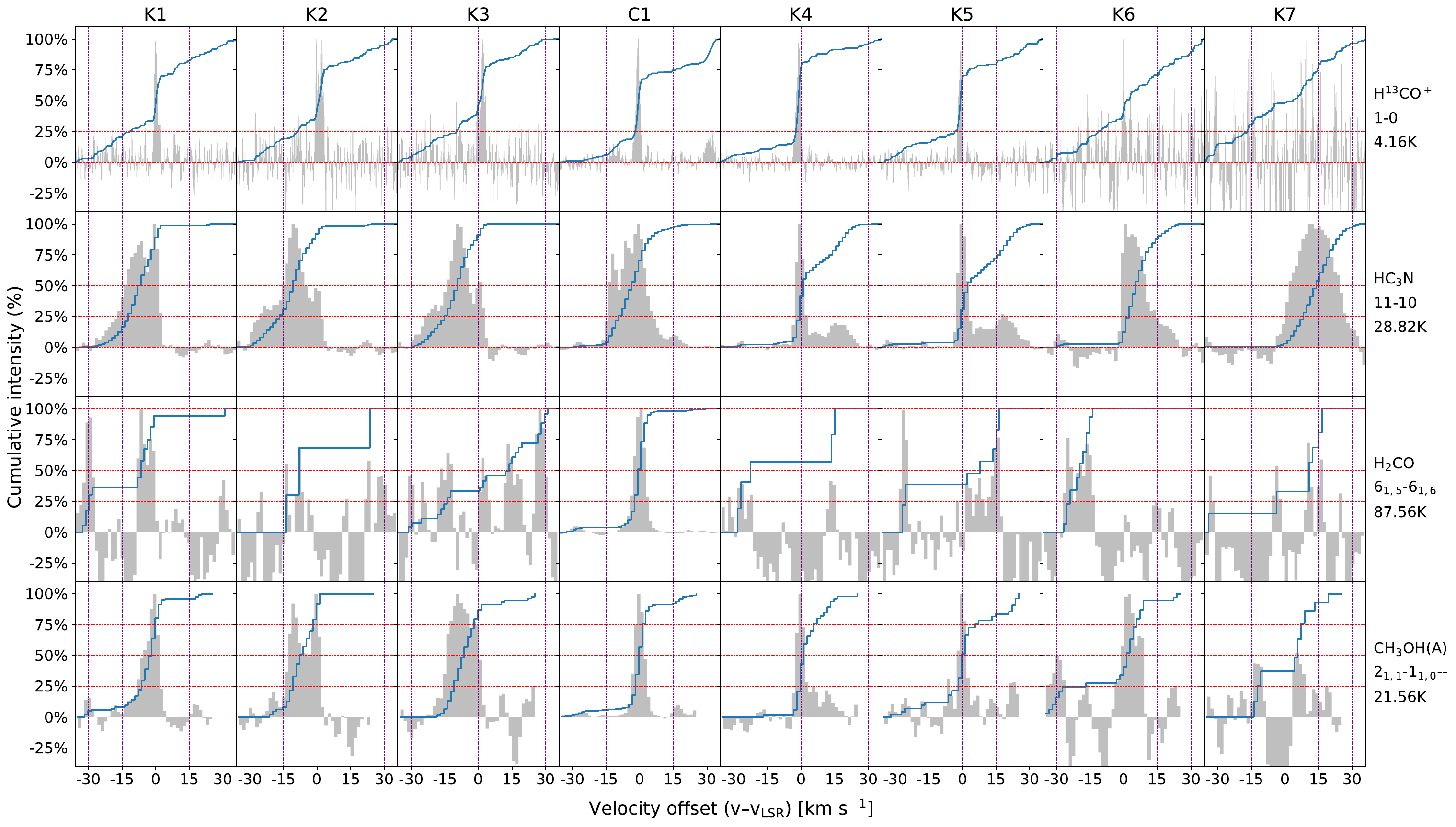}
  \caption{(c) Continued.}
\end{sidewaysfigure*}

\begin{sidewaysfigure*}\ContinuedFloat
  \centering
  \includegraphics[page=4,width=\textheight]{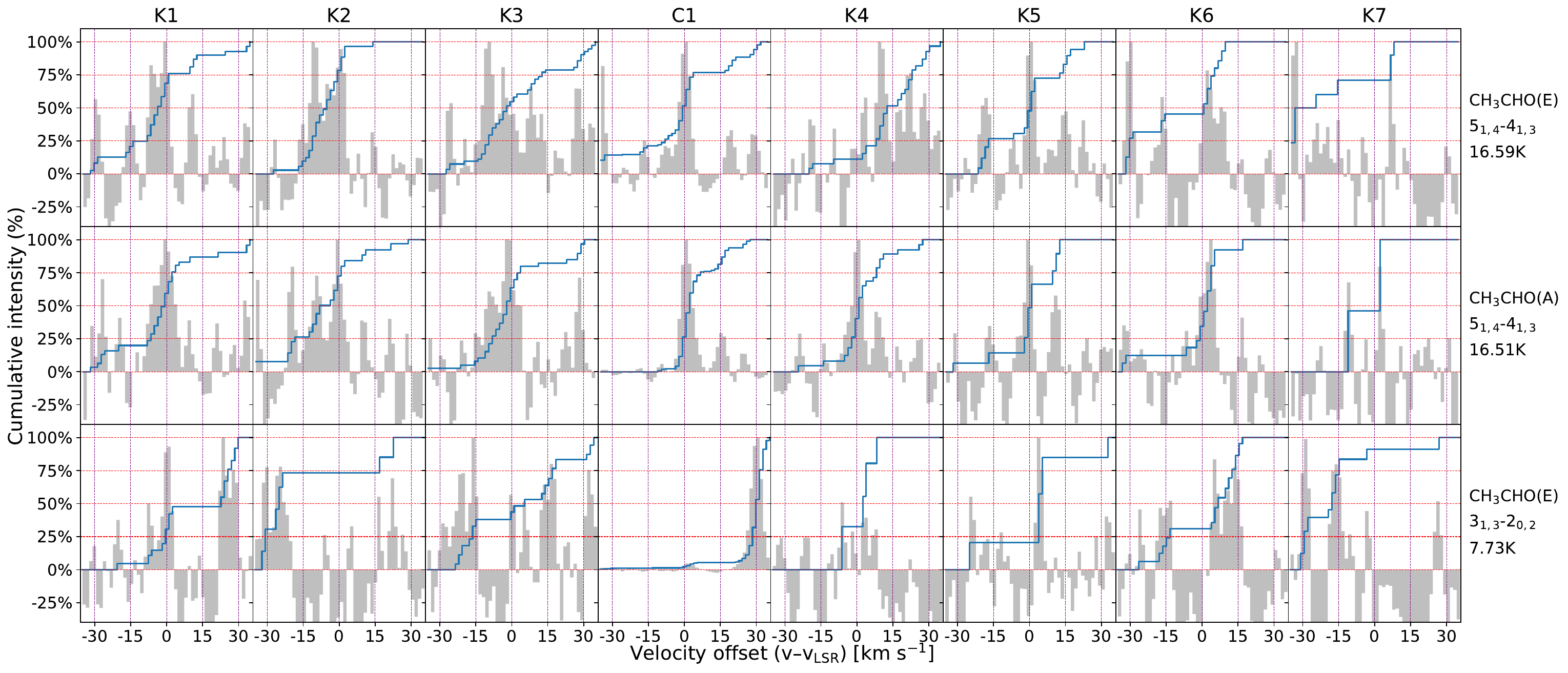}
  \caption{(d) Continued.}
\end{sidewaysfigure*}


\begin{sidewaysfigure*}[p!]
\centering
\includegraphics[page=1,width=\textheight]{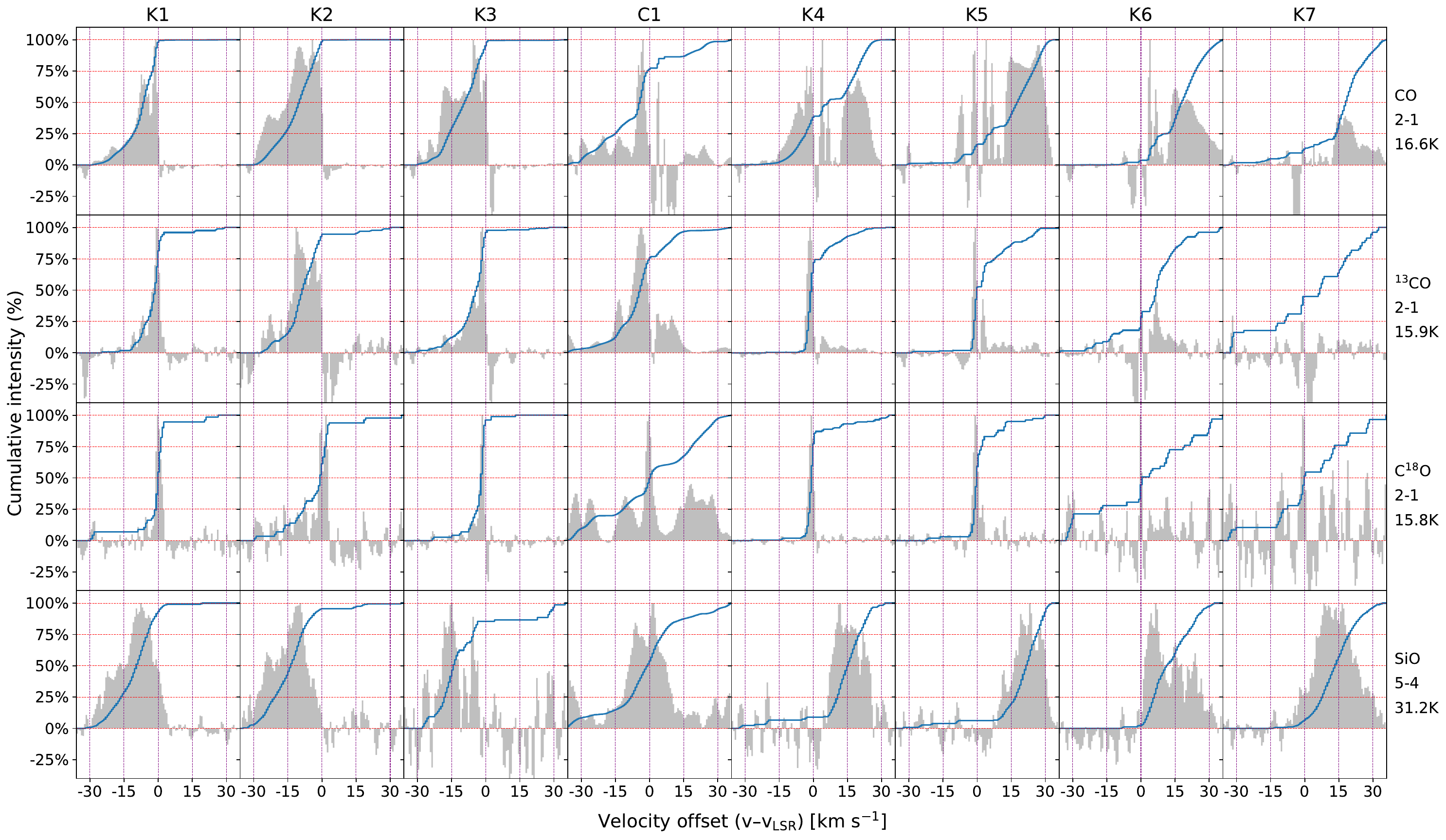}
\caption{(a) Identified molecular transitions in the 1.3-mm Band.}
\label{fig:spectra-1.3}
\end{sidewaysfigure*}

\begin{sidewaysfigure*}[p!]\ContinuedFloat
\centering
\includegraphics[page=2,width=\textheight]{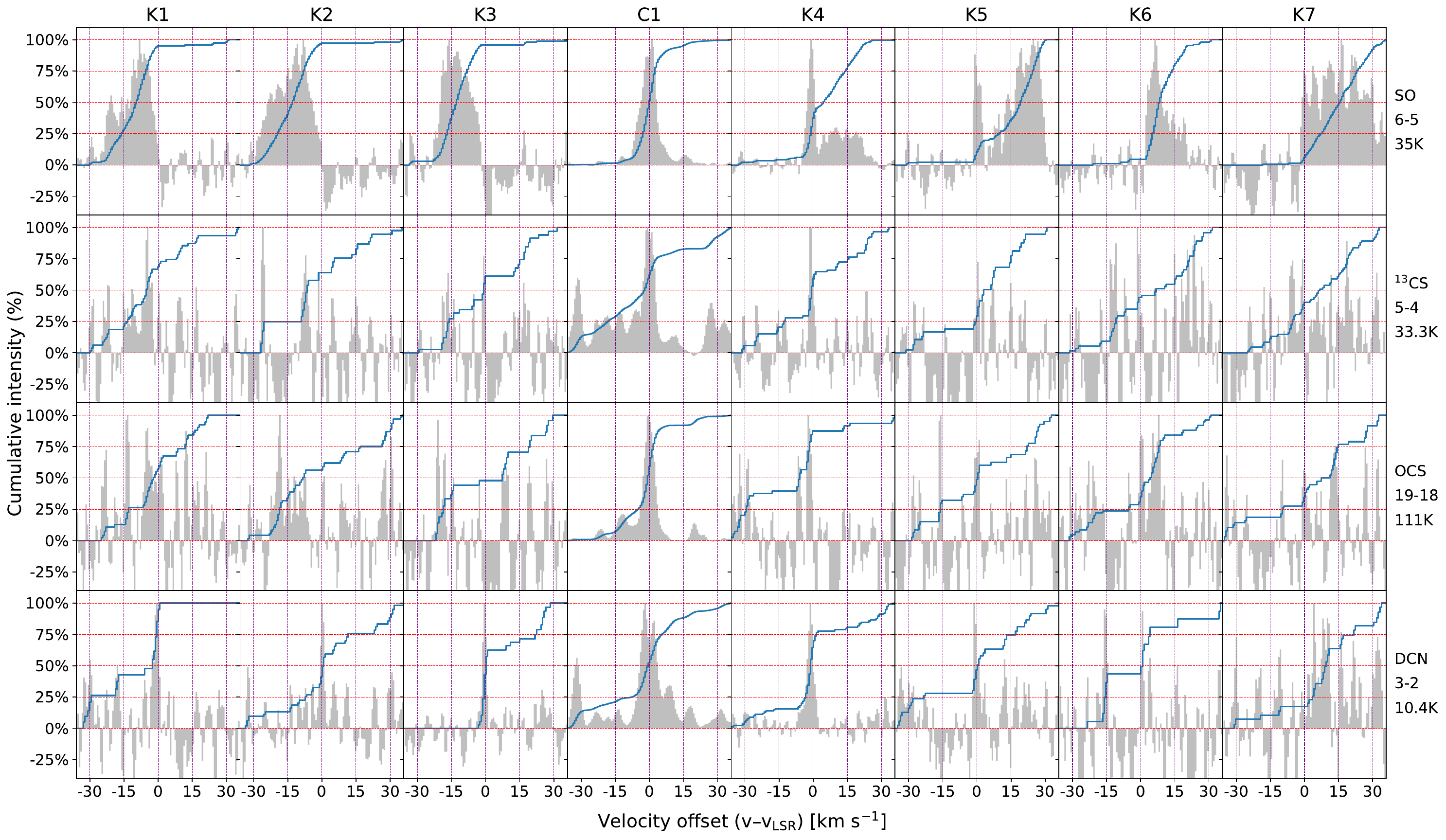}
\caption{(b) Continued.}
\end{sidewaysfigure*}

\begin{sidewaysfigure*}[p!]\ContinuedFloat
\centering
\includegraphics[page=3,width=\textheight]{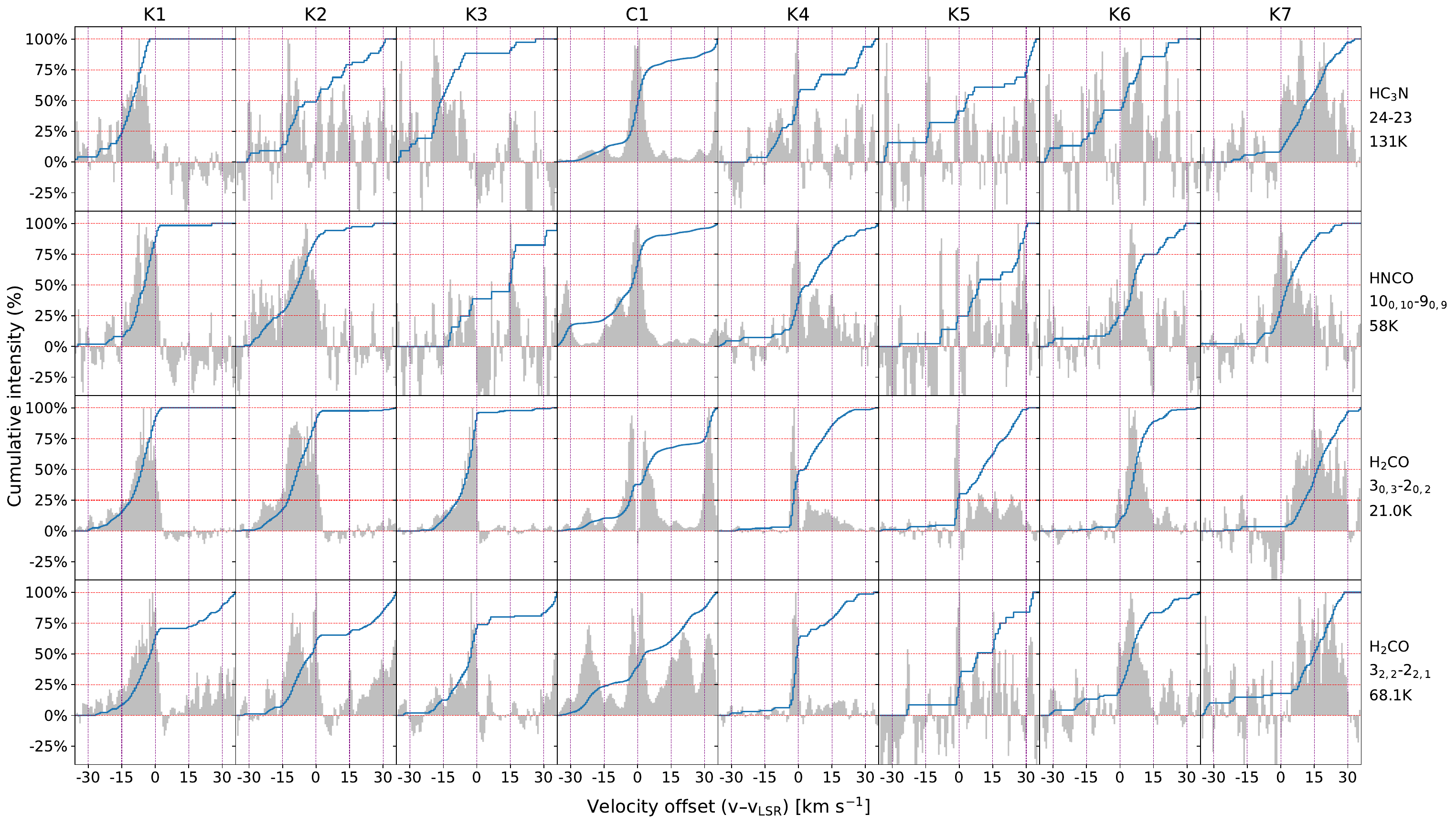}
\caption{(c) Continued.}
\end{sidewaysfigure*}

\begin{sidewaysfigure*}[p!]\ContinuedFloat
\centering
\includegraphics[page=4,width=\textheight]{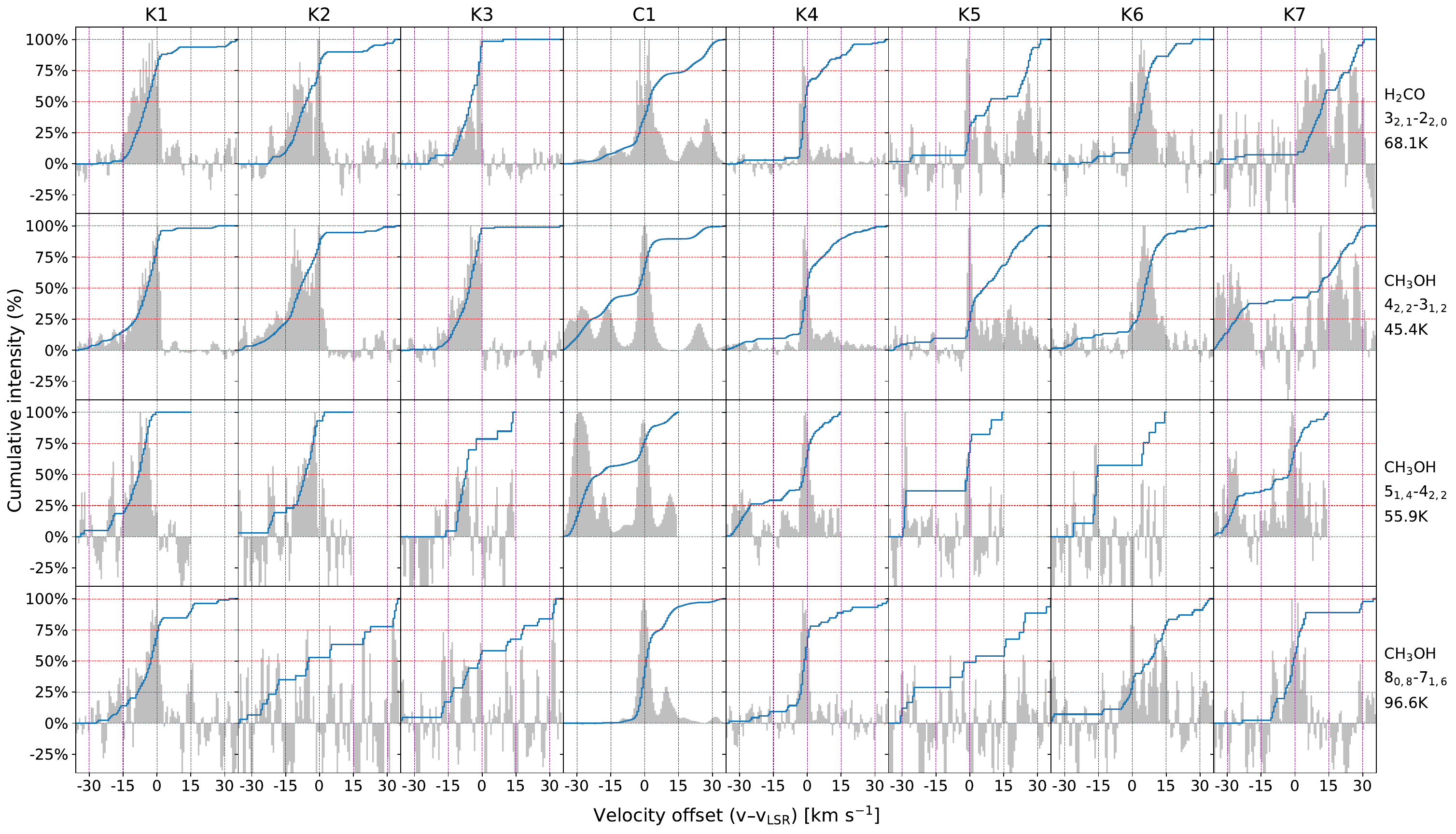}
\caption{(d) Continued.}
\end{sidewaysfigure*}

\begin{sidewaysfigure*}[p!]\ContinuedFloat
\centering
\includegraphics[page=5,width=\textheight]{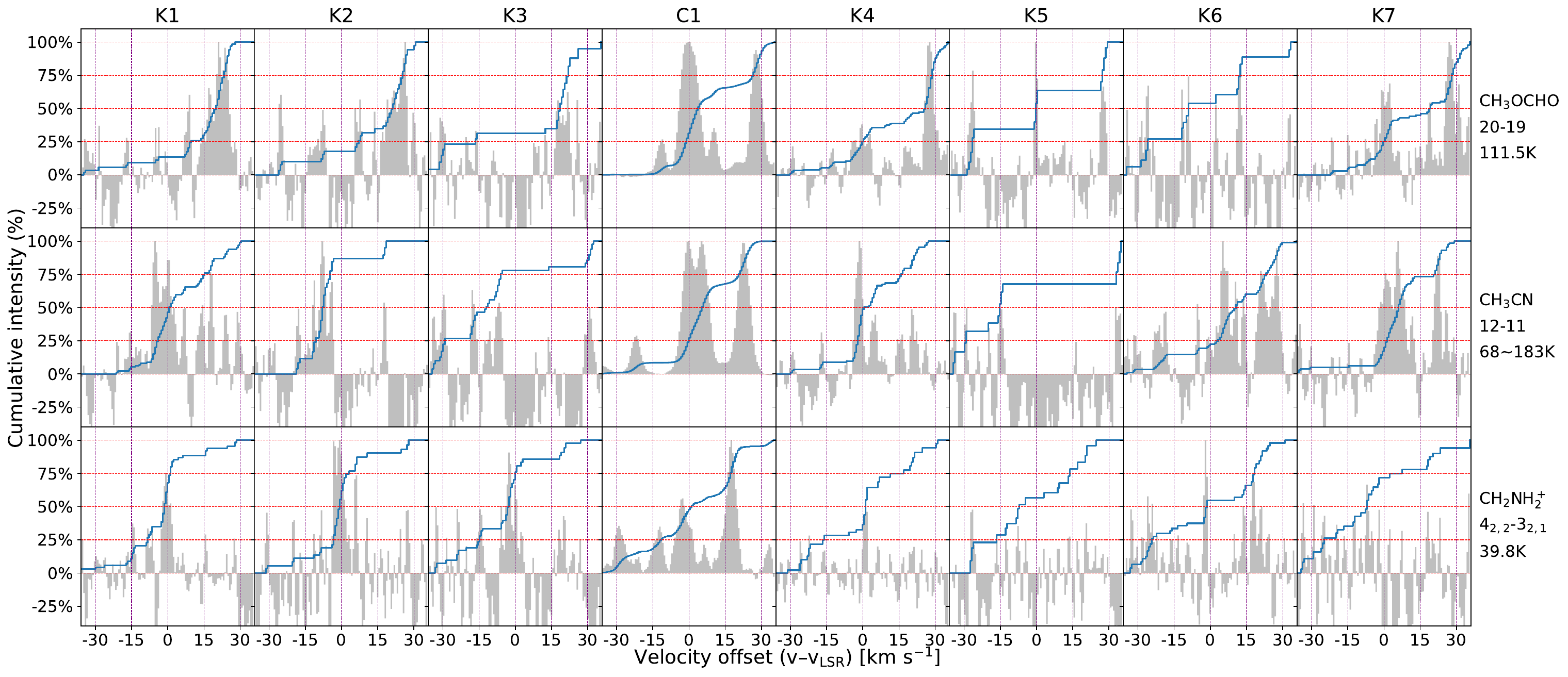}
\caption{(e) Continued. The U line in Fig.~\ref{fig:pca_model} and Fig.~\ref{fig:kendall}
here is marked as CH$_2$NH$_2^{+*}$.}
\end{sidewaysfigure*}

\clearpage
\restartappendixnumbering
\section{Supplementary Position--Velocity Analysis} \label{sec:B}

\begin{figure*}[ht]
\centering
\includegraphics[width=1\textwidth]{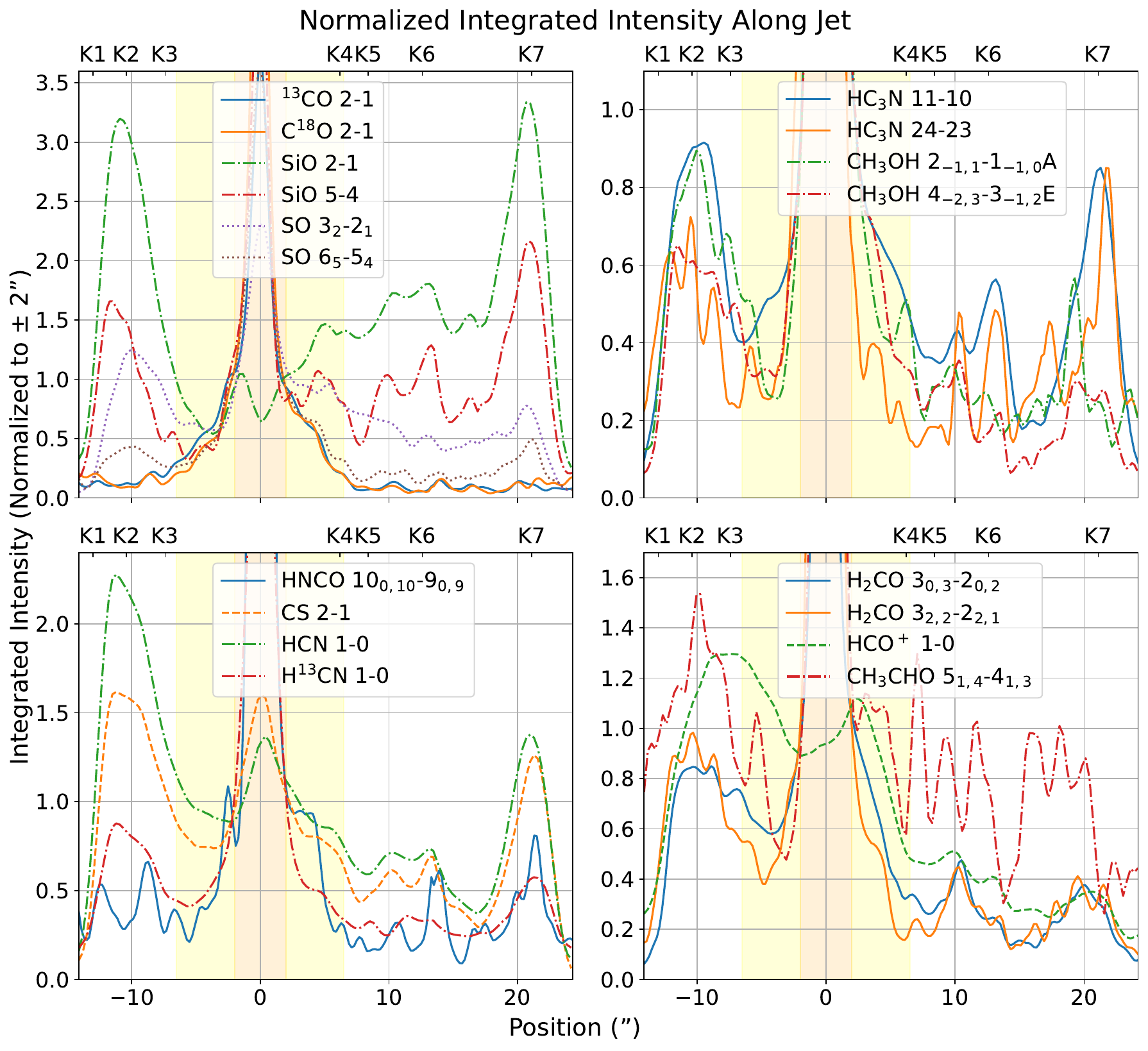}
\caption{
Integrated intensities of selected molecular lines along the jet axis of SDC335, plotted as a function of projected distance from the central hot core (C1). 
All intensities are normalized to their values at $+2\arcsec$ to highlight relative enhancements or suppressions along the flow. 
Positive offsets correspond to the northern (red-shifted) lobe, and negative offsets to the southern (blue-shifted) lobe.
}
\label{fig:B1}
\end{figure*}

Fig.~\ref{fig:B1} presents the integrated intensity along the outflow axis defined in Fig.~\ref{fig:cont}(a), with a slit width of $1''$ and velocity range from $-36$ to $+36$~km~s$^{-1}$. Based on the spatial distributions of the continuum, CCH, C$^{18}$O and other molecules (see Fig.~\ref{fig:cont}(b) and Fig.~\ref{fig:moment0}), we consider the inner $2''$ to be dominated by hot-core emission. To minimize this contamination, all values are divided by the reference intensity at $2''$ from the core; regions beyond $2''$ are referred to as the ``pure'' outflow in the following discussion. The integrated intensities of $^{13}$CO and C$^{18}$O decrease monotonically with distance, whereas other molecules show oscillatory enhancements.

\begin{figure*}[t]
\centering
\includegraphics[width=1\textwidth]{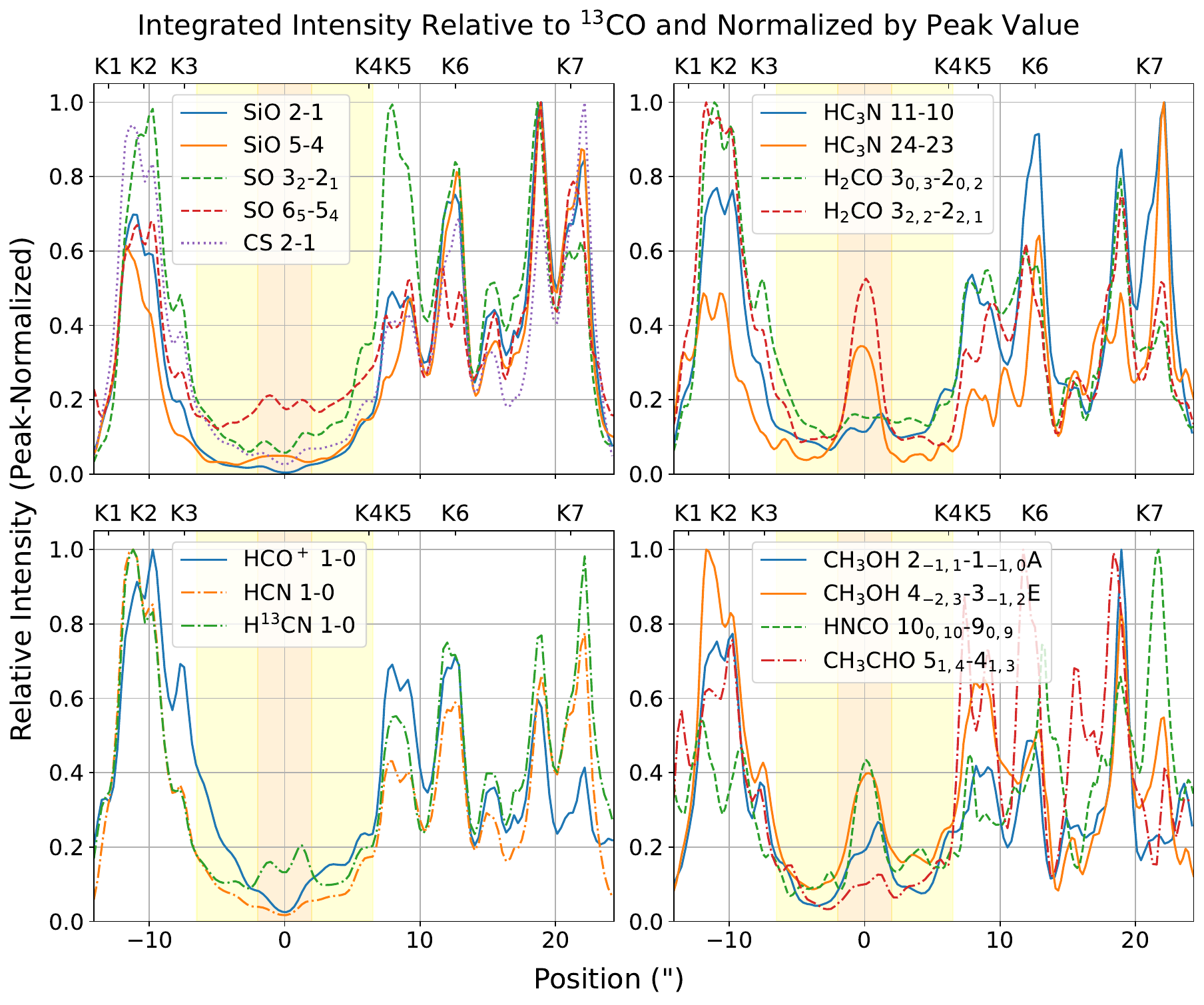}
\caption{
Relative abundance proxies of selected molecules along the jet axis, computed as the ratio of their integrated intensities to that of $^{13}$CO at each position. 
Each line is normalized to its own maximum value across the spatial extent to emphasize the distribution pattern rather than absolute abundance contrast. 
This highlights molecule-specific spatial differentiation along the outflow.
}
\label{fig:ni-13co}
\end{figure*}

\begin{figure*}[ht]
\centering
\includegraphics[width=1\textwidth]{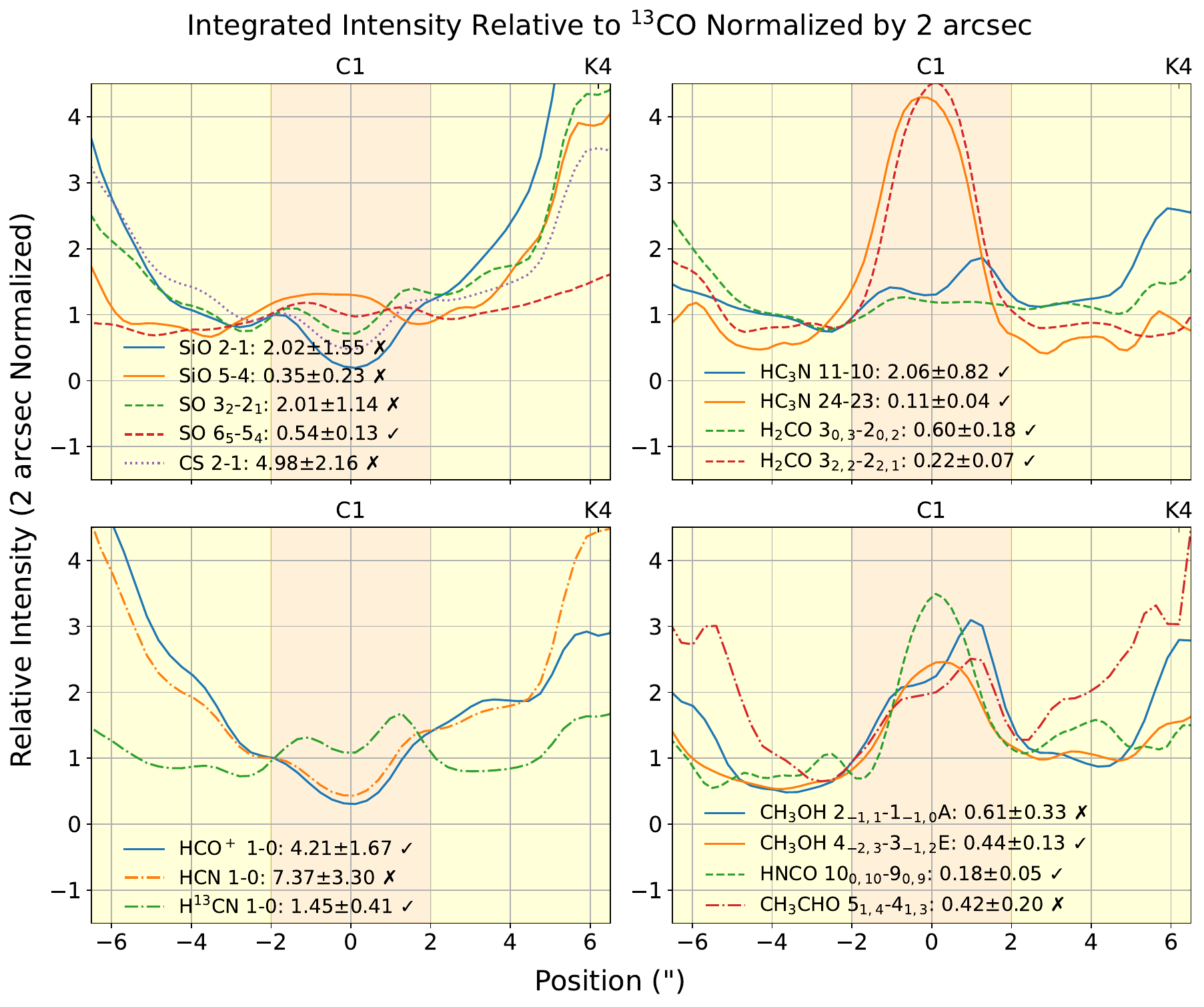}
\caption{
An enlarged view of the weak-shock region (light yellow background) in Figure~\ref{fig:ni-13co}, showing that some transitions, such as H$_2$CO $3_{0,3}-2_{0,2}$ and SO $6_{5}-5_{4}$, exhibit relatively little variation in intensity within this range, whereas other transitions show significant intensity changes.
}
\label{fig:ni-13co-static}
\end{figure*}

Because the outflow expands with distance, absolute integrated intensities are not direct tracers of abundance variations. Fig.~\ref{fig:ni-13co} therefore shows the ratios of each molecule to the local $^{13}$CO intensity, normalized at each position, to reveal abundance gradients. Compared with Fig.~\ref{fig:B1}, the relative abundances display mirror symmetry with respect to the hot core, particularly within $\pm 14''$, consistent with the integrated maps (Fig.~\ref{fig:moment0}) and PVDs (Fig.~\ref{fig:pvds}). The middle outflow ($2''$–$6.5''$) exhibits systematically lower abundances, highlighted with a pale yellow background. Fig.~\ref{fig:ni-13co-static} further zooms into shock–weak regions: after re-normalization at $2''$, the vertical axis shows the abundance relative to the $2''$ value. The mean and standard deviation of ``abundance factors relative to $^{13}$CO'' are adopted to assess significance, as annotated in the figure. SO~6$_5$–5$_4$, HC$_3$N, both H$_2$CO transitions, CH$_3$OH-E, H$^{13}$CN, and HNCO remain nearly constant, suggesting weak shock sensitivity; by contrast, SiO, SO~3$_2$–2$_2$, CS, HCO$^+$, HCN, CH$_3$OH-A, and CH$_3$CHO show various levels of enhancement, pointing to distinct formation pathways.

\begin{figure*}[t]
\hspace*{0.07\textwidth}
\includegraphics[width=0.85\textwidth]{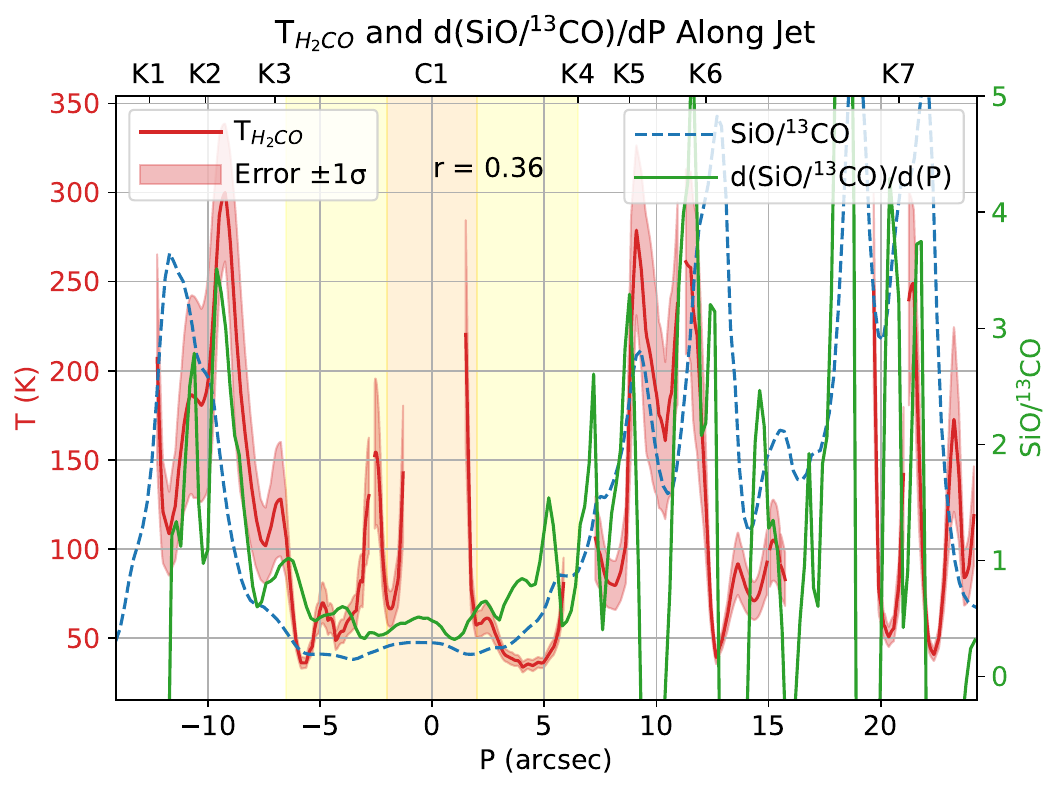}
\caption{
Comparison between the rotational temperature ($T_{\mathrm{rot}}$) derived from \hcho~and the spatial derivative of the SiO-to-$^{13}$CO integrated intensity ratio along the outflow axis of SDC335.
The horizontal axis represents the projected distance from the central hot core, while the two vertical axes show $T_{\mathrm{rot}}$ (in K, left axis) and $d$(SiO/$^{13}$CO)/$d$P where P is position (right axis).
The central light orange background indicates the extent of the hot core and its warm envelope, while the flanking light yellow regions mark areas in the outflow where some molecules do not show significant weakening relative to $^{13}$CO.
Regions with elevated SiO 5--4 gradients coincide with local temperature peaks, indicating strong shock heating and enhanced molecular desorption.
}
\label{fig:trot-sio}
\end{figure*}

Fig.~\ref{fig:trot-sio} shows the variation rate of SiO~5$_4$–4$_3$ intensity compared with the kinetic temperature $T_{\rm kin}$ derived from H$_2$CO~3–2 transitions. The kinetic temperature is calculated using the non-LTE radiative transfer code RADEX \citep{2007A&A...468..627V}. We assume the volume density $10^5$--$10^6$~cm$-3$ of the outflow, and using minimum $\chi^2$ to optimize the line ratio between transitions of H$_2$CO~3–2. This method has a good constrain of temperature below 300~K otherwise the temperature can be sensitive to the volume density. 
Bright knots correspond to high $T_{\rm kin}$ and steep SiO gradients, indicating that jet impacts compress and heat the outflowing shell, leading to the release of icy species into the gas phase.

\begin{figure*}[ht!]
\centering
\gridline{
  \fig{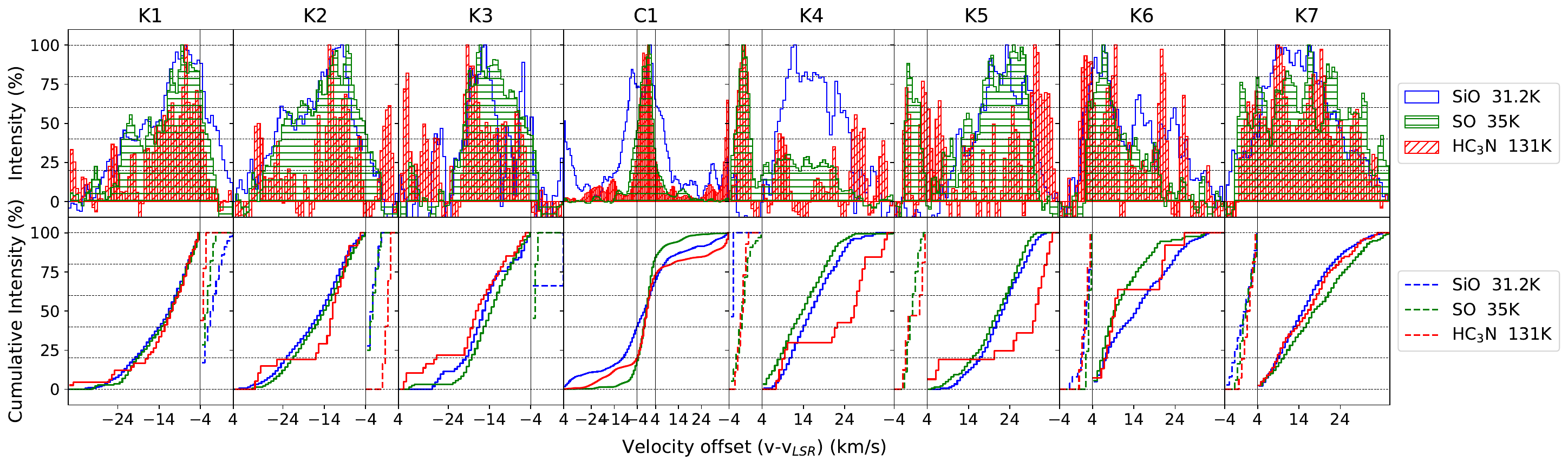}{1\textwidth}{(a) High-velocity components in the 1.3 mm band;}
}
\vspace{-0.1cm}
\gridline{
  \fig{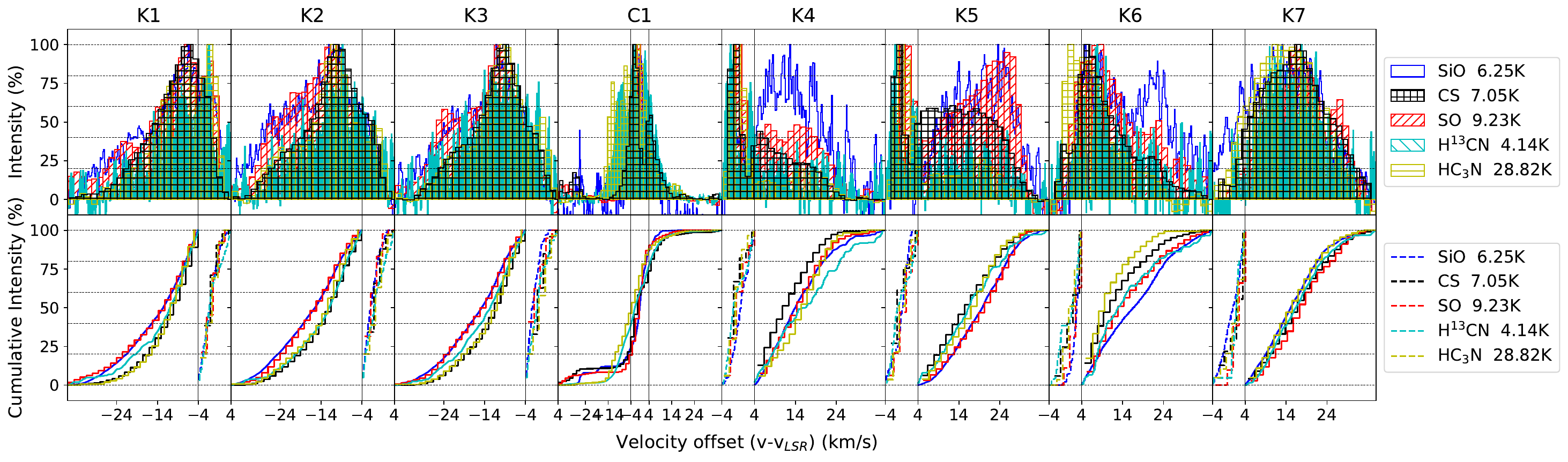}{1\textwidth}{(b) High-velocity components in the 3 mm band;}
}
\vspace{-0.1cm}
\gridline{
  \fig{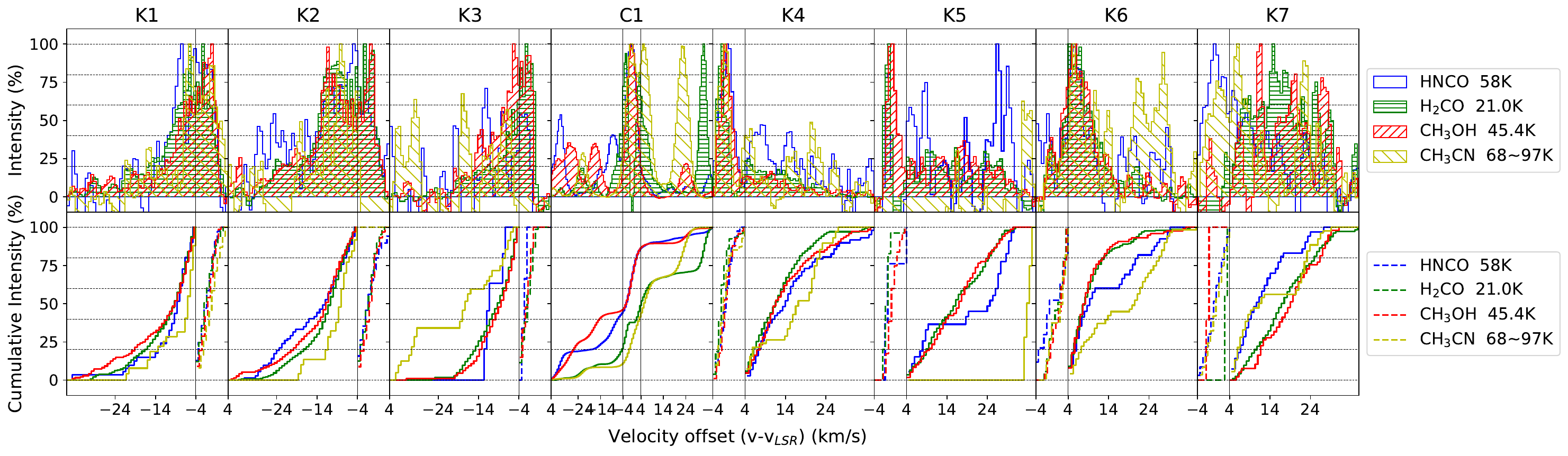}{1\textwidth}{(c) Low-velocity components in the 1.3 mm band; }
}
\vspace{-0.1cm}
\gridline{
  \fig{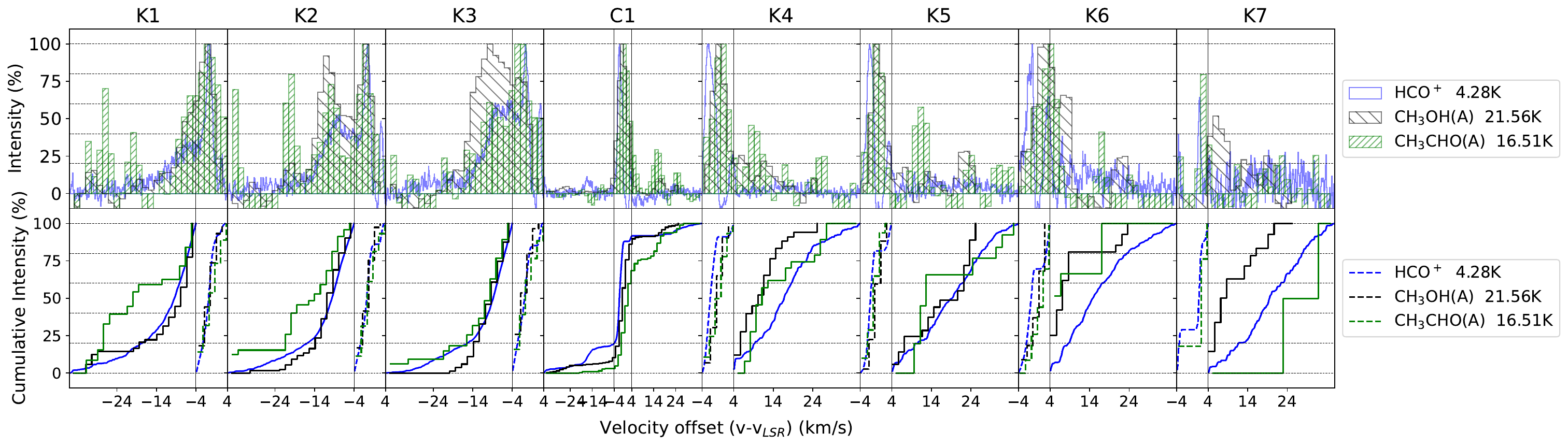}{1\textwidth}{(d) Low-velocity components in the 3 mm band.}
}
\caption{
The cumulative curves quantify the total line flux for each source, aiding comparison between knots and the central core.
}
\label{fig:spectra1}
\end{figure*}

Fig.~\ref{fig:spectra1} compares selected line profiles in the 1.3~mm and 3~mm bands. The cumulative intensity at the bottom separates systemic and non-systemic velocity components to highlight differences. 
\begin{itemize}
    \item (a) High-velocity molecules at 1.3~mm: SiO, SO, HC$_3$N. For HC$_3$N the S/N is low at knots K2, K3, K5, and K6, but near-Gaussian at C1.  
    \item (b) High-velocity molecules at 3~mm: SiO, CS, SO, HCN, HC$_3$N. Overall profiles resemble those at 1.3~mm, but the cumulative plots differ significantly: SiO and SO dominate $|v| > 12$~km~s$^{-1}$, while CS and HC$_3$N are weaker; H$^{13}$CN lies in between.  
    \item (c) Low-velocity molecules at 1.3~mm: HNCO, H$_2$CO, CH$_3$OH-E, CH$_3$CN. The three adjacent CH$_3$CN 12$_k$–11$_k$ lines show absorption at knots K3–K5, indicating excitation temperatures lower than the outflow background. HNCO has insufficient S/N at some knots. These species are dominated by systemic velocity emission (see Fig.~\ref{fig:moment0}), with weaker high- and low-velocity components.  
    \item (d) Low-velocity molecules at 3~mm: HCO$^+$, CH$_3$OH-A, CH$_3$CHO. CH$_3$OH-A ($E_u = 21.1$~K) is more concentrated at systemic velocity compared with CH$_3$OH-E ($E_u = 45.5$~K), and shows stronger emission in the northern (redshifted) lobe. Whether this A/E difference reflects intrinsic nuclear-spin properties remains to be confirmed.  
\end{itemize}


\clearpage
\restartappendixnumbering
\section{Channel Maps} \label{sec:channelmap}

We also produced channel maps of CO, $^{13}$CO, SiO, SO, H$_2$CO, and CH$_3$OH using the combined ACA+TM1+TM2 data with angular resolution of $0.34''$. To suppress random noise in single channels, we integrated over $\pm 1.3$~km~s$^{-1}$ centered on each velocity (this is also why these data were not used for quantitative analysis). The maps show that the high-velocity jet is located along the outflow axis, while the low-velocity component forms cavity-like structures. The maximum velocities reach $\pm 36$~km~s$^{-1}$ for SiO and $\pm 33$~km~s$^{-1}$ for SO. Components with $|v| > 12$~km~s$^{-1}$ are concentrated near the axis, whereas slower components attach to the cavity walls. H$_2$CO and CH$_3$OH follow similar trends, with maximum velocities of $\pm 27$~km~s$^{-1}$ and $\pm 24$~km~s$^{-1}$, respectively. Several smaller-scale jets in other directions are also discernible in these maps.

\begin{sidewaysfigure*}
\centering
\includegraphics[width=0.85\textwidth]{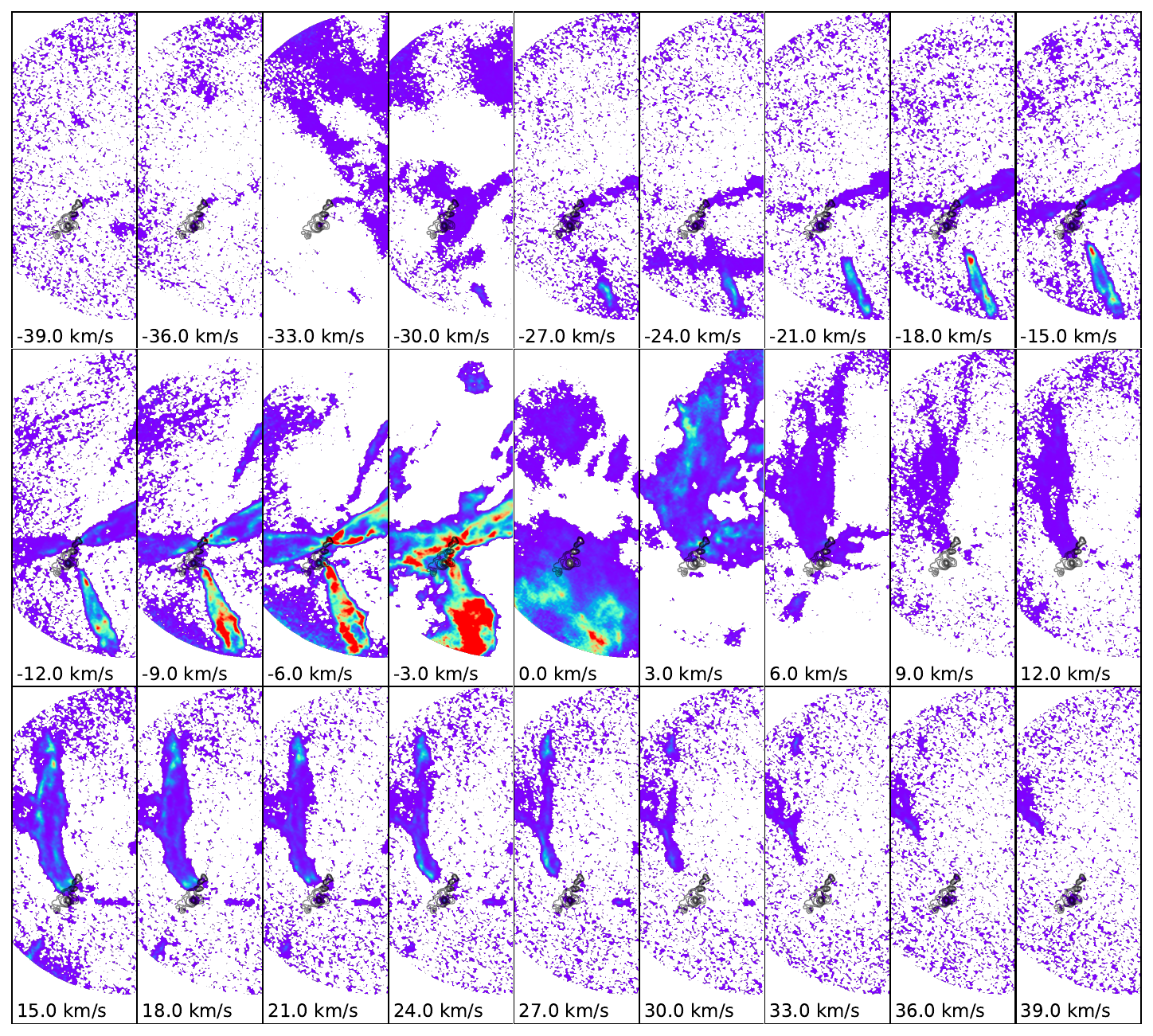}
\caption{
CO 2-1 channel map.
}
\label{fig:channelCO}
\end{sidewaysfigure*}

\begin{sidewaysfigure*}
\centering
\includegraphics[width=0.85\textwidth]{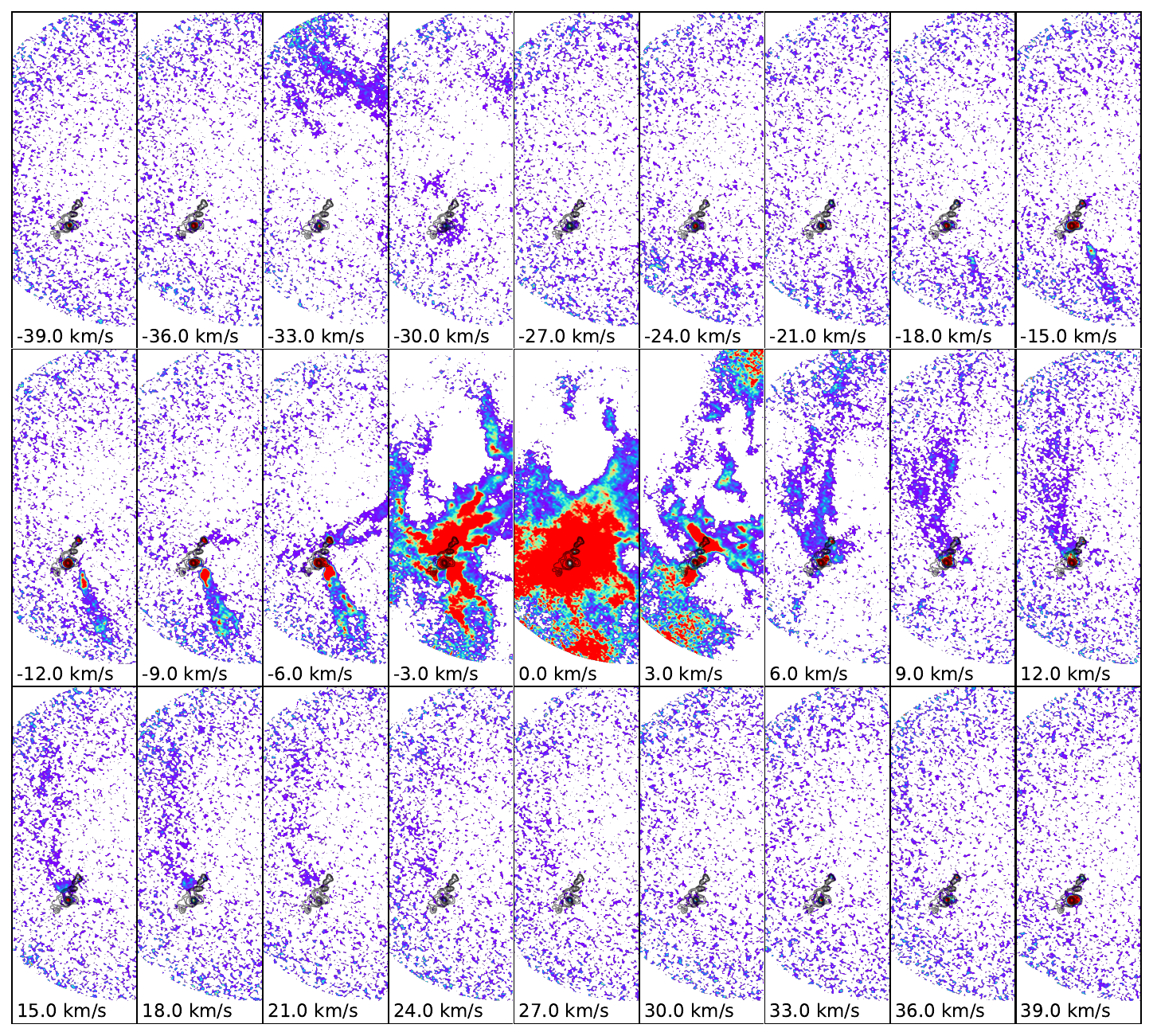}
\caption{
$^{13}$CO 2-1 channel map.
}
\label{fig:channel13CO}
\end{sidewaysfigure*}

\begin{sidewaysfigure*}
\centering
\includegraphics[width=0.85\textwidth]{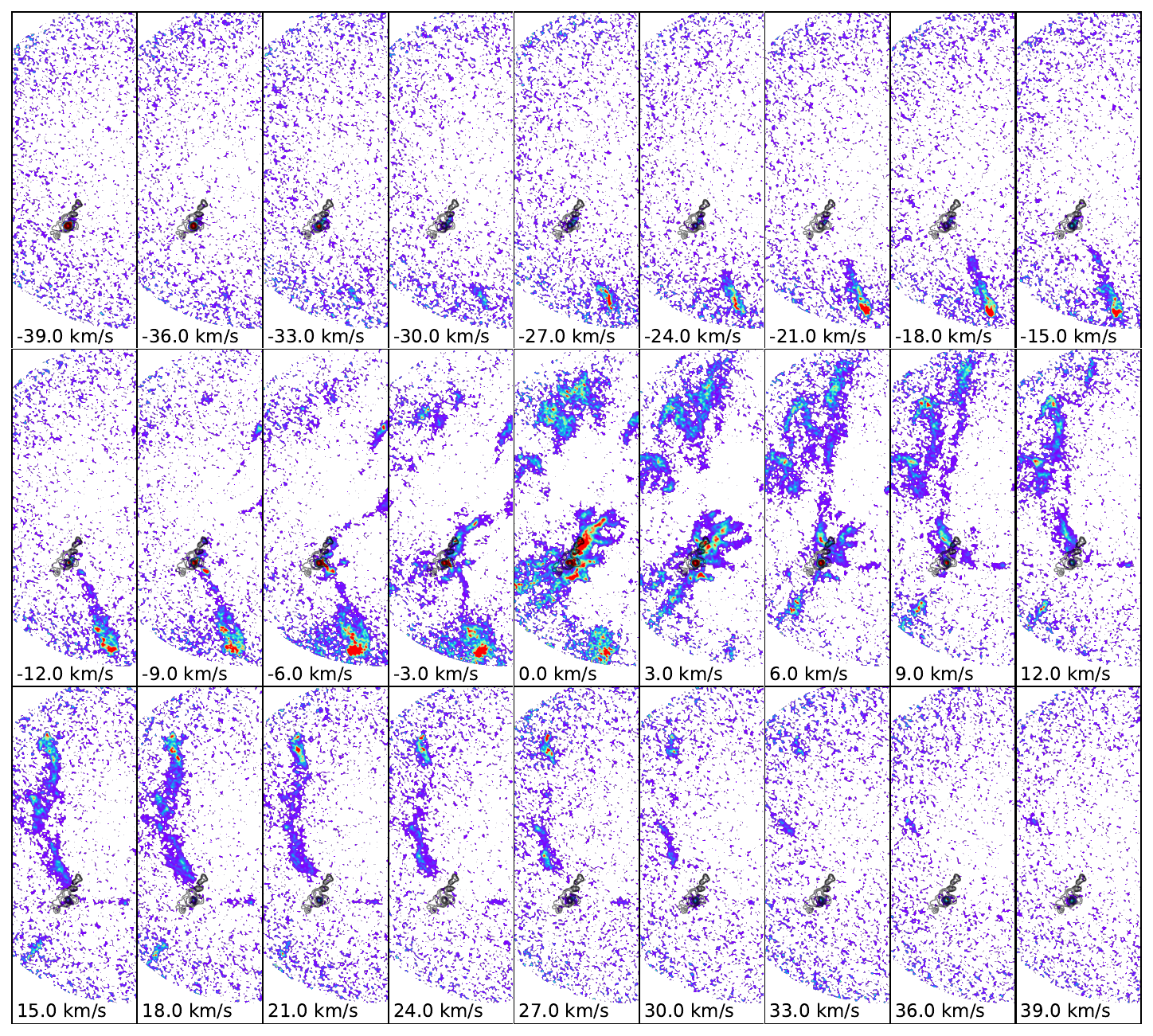}
\caption{
SiO 5-4 channel map.
}
\label{fig:channelSiO}
\end{sidewaysfigure*}

\begin{sidewaysfigure*}
\centering
\includegraphics[width=0.85\textwidth]{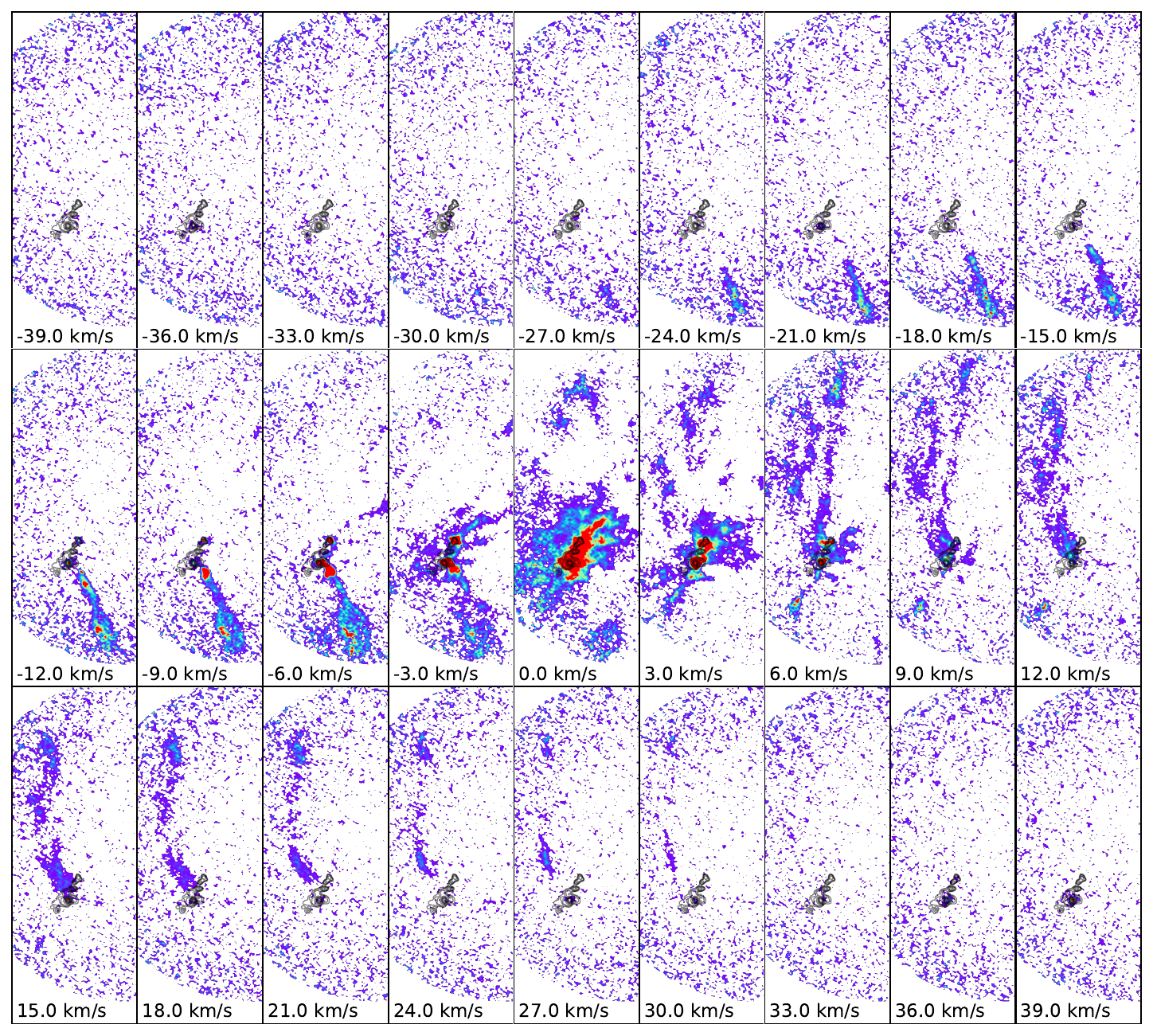}
\caption{
SO $6_5$–$5_4$ channel map.
}
\label{fig:channelSO}
\end{sidewaysfigure*}

\begin{sidewaysfigure*}
\centering
\includegraphics[width=0.85\textwidth]{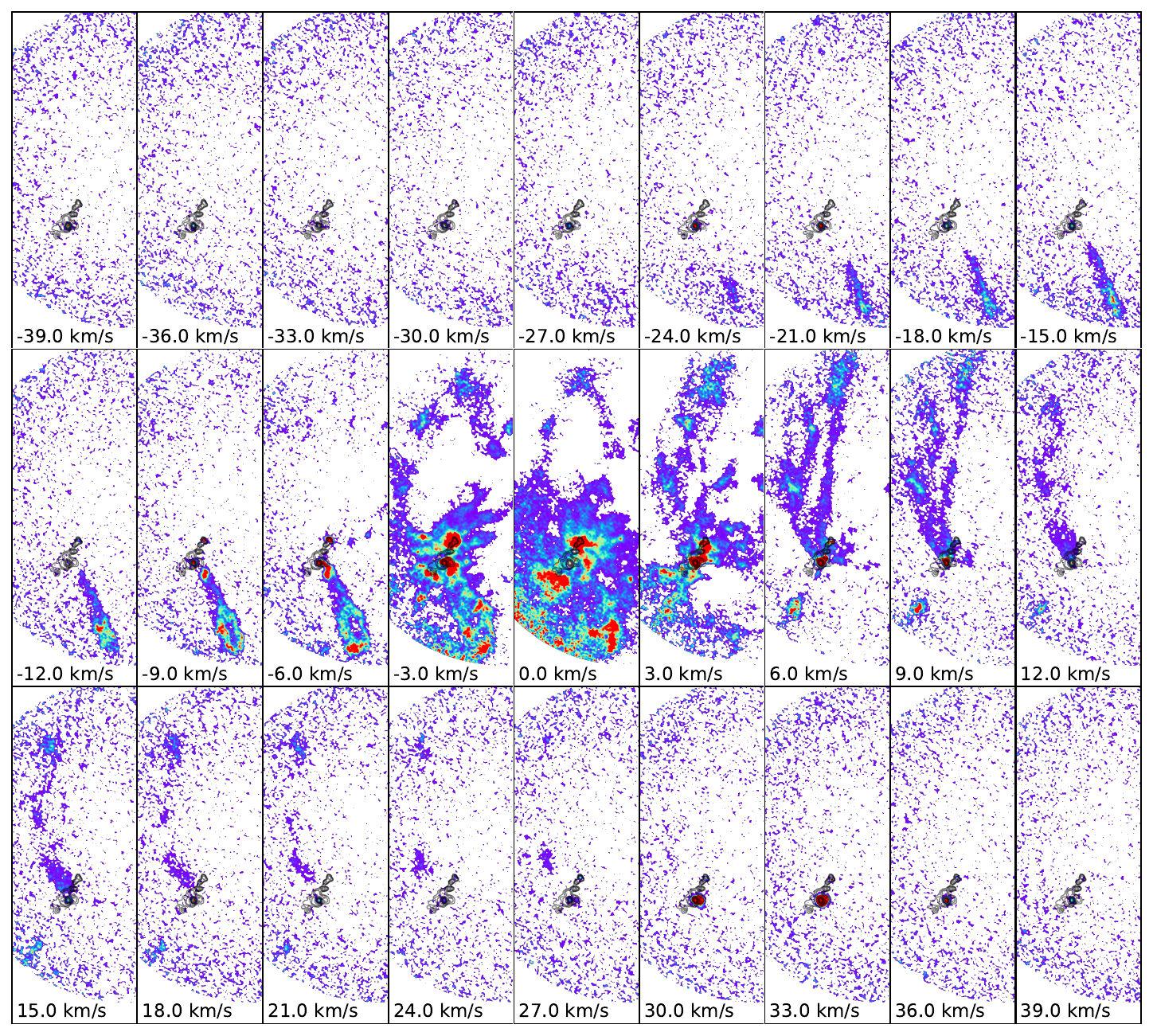}
\caption{
\hcho~$3_{0,3}$–$2_{0,2}$ channel map.
}
\label{fig:channelH2CO}
\end{sidewaysfigure*}

\begin{sidewaysfigure*}
\centering
\includegraphics[width=0.85\textwidth]{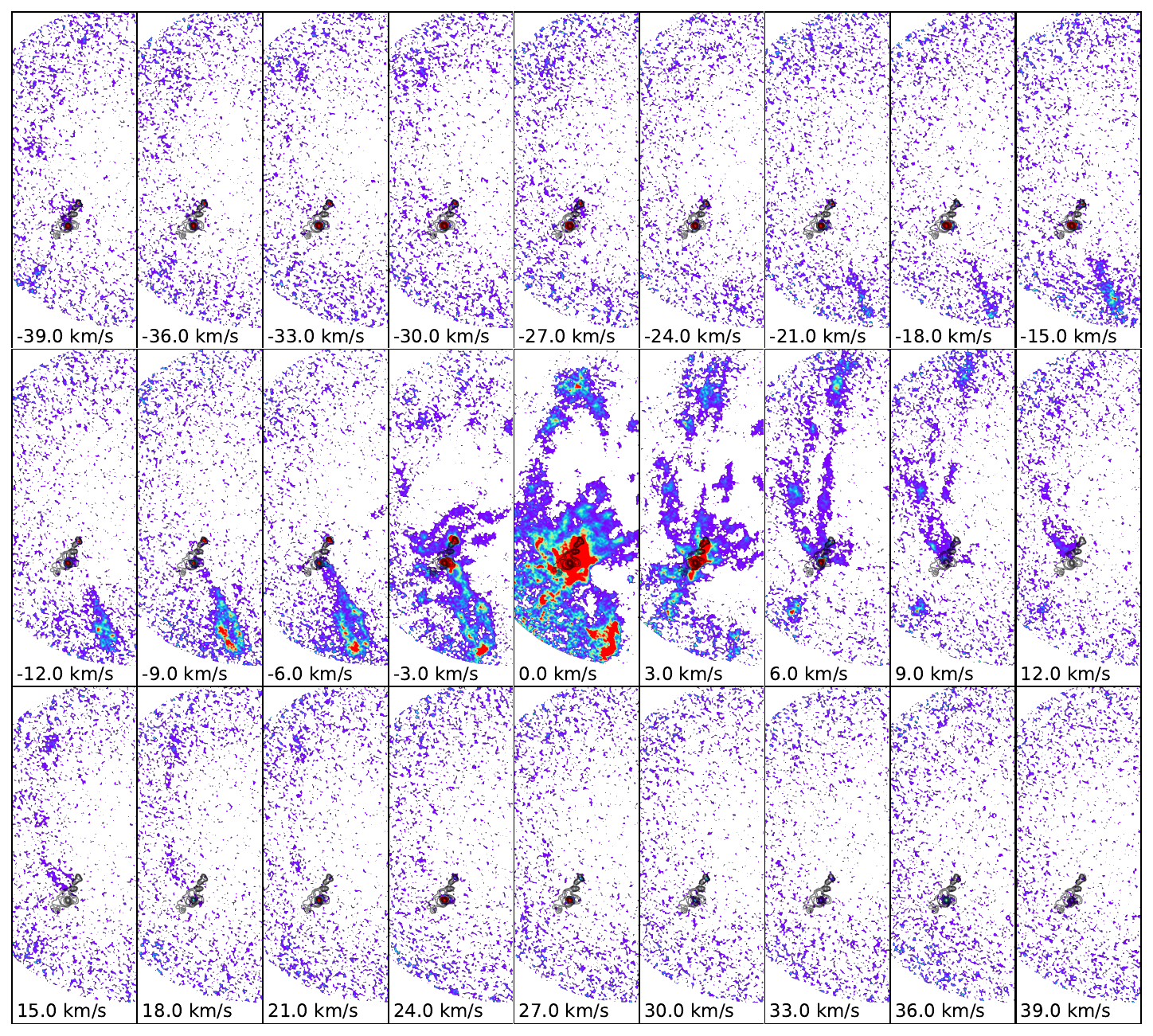}
\caption{
\meoh~$4_{-2,3}$–$3_{-1,2}$ channel map.
}
\label{fig:channelMeOH}
\end{sidewaysfigure*}





\end{document}